\documentclass[a4paper,fleqn,usenatbib]{mnras}

\usepackage[T1]{fontenc}
\usepackage{ae,aecompl}

\usepackage{graphicx}	
\usepackage{amsmath}	
\usepackage{amssymb}	

\usepackage{multirow}
\usepackage{lineno}
\usepackage{verbatim}
\usepackage[export]{adjustbox}
\usepackage{upgreek}

\modulolinenumbers[10]

\def\mujybm {${\rm \mu}$Jy\,bm$^{-1}$}
\newcommand{\undertext}[2] {$\underbrace{\textrm{#1}}_{\textrm{#2}}$}

\title[X-galaxy in the lens]{ Role of the companion lensing galaxy in
  the \emph{CLASS} gravitational lens B1152+199 }

\author[M. Zhang]{
M. Zhang,$^{1,2}$\thanks{E-mail: mgnahz@gmail.com} Q. Yuan,$^{1,3}$ J.-Y. Liu,$^{1,3}$ L. Zhang$^{4}$
\\
$^{1}$Xinjiang Astronomical Observatory, Chinese Academy of Sciences, 150 Science 1-Street, Urumqi 831001, China \\
$^{2}$Key Laboratory for Radio Astronomy, Chinese Academy of Sciences, 2 West Beijing Road, Nanjing 210008, China \\
$^{3}$University of Chinese Academy of Sciences, 19A Yuquan Road, Beijing 100049, China \\
$^{4}$College of Big Data and Information Engineering, Guizhou University, Guiyang 550025, China
}

\date{10 January 2022}

\pubyear{2022}

\begin{document}
\label{firstpage}
\pagerange{\pageref{firstpage}--\pageref{lastpage}}
\maketitle

\begin{abstract}
  We reinvestigate the Cosmic Lens All-Sky Survey (\emph{CLASS})
  gravitational lens B1152+199 using archived Hubble Space Telescope
  (\emph{HST}) data and Very Long Baseline Interferometry
  (\emph{VLBI}) data. A consistent luminosity ratio within effective
  radius between the host galaxy and the X-galaxy is measured from
  \emph{HST} tri-band images, which leads to a mass ratio between the
  two galaxies as $r_b\sim 2$. To determine the role of the X-galaxy
  in the lens system, we modelled the dual-lens system with
  constraints from the \emph{VLBI}-resolved jet components and the
  \emph{HST} images. The 8.4-GHz global-\emph{VLBI} data currently
  provide the most stringent constraints to the mass model, especially
  to the radial power-law slope. The optimized models for this
  two-image three-component radio lens favour a
  steeper-than-isothermal inner slope. The jet bending in image B was
  also investigated and it turned out to be rather a misalignment than
  a curvature. The goodness of fit indicates that the role of the
  X-galaxy is crucial in the lens system if three pairs of resolved
  jet components are to be fitted. When we imported the optimal model
  from radio constraints to optical modelling with the \emph{HST}
  tri-band data, the optimization kept the consistency of the optimal
  model and successfully reproduced the features observed in the
  \emph{HST} images. This implies that the diffuse emission discovered
  in the \emph{HST} images is actually a detection of the secondary
  lensing effects from the companion lens.
 
\end{abstract}




\section{Introduction}

Gravitational lensing is an effect of light deflection caused by
gravitational fields due to the distribution of lensing
masses. According to the different lensing strengths, it can be
categorized as strong or weak regimes, where the multiple images or
distorted shapes of background galaxies are produced respectively. The
strong gravitational lensing provides a dynamics-independent method to
infer substructures of mass distribution in lensing galaxies,
regardless of whether they are dark or luminous. While the Lambda cold dark
matter (\emph{$\Lambda$CDM}) cosmological model has enjoyed great
success in predicting large-scale structures in the cosmic microwave
background (\emph{CMB}) observations~\citep{hinshaw.07.apjs}, $N$-body
simulations in this paradigm tend to over-predict the amount of
substructures in smoothed dark matter halos~\citep{navarro.96.apj}
that is yet to be confirmed by observational evidence, especially at
galactic scales. Recently, there has been some progress that a large
number of dwarf galaxies in the Milky Way have been detected with the
Sloan Digital Sky Survey (\emph{SDSS}) and the Dark Energy Survey
(\emph{DES}) \citep{bechtol.15.apj}. The galactic-scale hydrodynamical
simulations have also been done to investigate the subhalo
existence~\citep{despali.16.mn,despali.17.mn}, as well the
line-of-sight structures~\citep{gilman.18.mn,despali.18.mn}. The
tension between \emph{CDM} model and observations seems mitigated with
more newfound Milky Way dwarfs. However, the underdetected
substructures in extragalactic early-type galaxies still leave the
discrepancy. 

Generally, substructures can be inferred from mass modelling, provided
the discrete lensing mass can be verified.  The direct observable
phenomena include the flux-ratio
anomalies~\citep{mao.98.mn4,bradac.02.a&a,keeton.03.apj,biggs.04.mn,
  keeton.06.apj} and surface brightness
anomalies~\citep{vegetti.12.nat}, as well as the time-delay anomalies
which is more difficult to detect. The flux-ratio anomaly happens in
the event of a caustic crossing in a quad-image system, which violates
the cusp or fold relations, while the surface brightness anomaly takes
place on the arc or Einstein ring. It should be noticed that there is
evidence of galactic-scale discs may account for the flux-ratio
anomalies from both the observational and the theoretical
perspectives~\citep{hsueh.16.mn,hsueh.17.mn,hsueh.18.mn,gilman.18.mn}.
It is also possible to observe morphological distortions due the
presence of a substructure, i.e. the gravitational
millilensing~\citep{wambsganss.92.apjl, metcalf.02.apj}.  Then,
indirect inference can be drawn about the lensing mass, either via a
mass or potential reconstruction with non-parametric methods, such as
extended structures (arcs and rings) reconstruction via Maximum
Likelihood (\emph{ML}) methods~\citep[e.g.]{koopmans.05.mn} and
point-like structure ensemble modelling~\citep{saha.97.mn}. There has
been extensive work on non-parametric gravitational potential
reconstruction from
lensing~\citep{vegetti.09.mn,vegetti.10.mn,vegetti.12.nat,
  vegetti.14.mn,vegetti.18.mn,ritondale.19.mn}.

There have been several luminous cases in which faint dwarf galaxies
have been found, and whereby models were fit to the observables: one
is the anomalous flux-ratio in B2045+265~\citep{mckean.07.mn}, another
is the single perturbing mass (object X) model in
MG0414+0534~\citep{ros.00.a&a,kochanek.06.glsw2,macleod.13.apj,stacey.18.mn},
and yet another possible case is
MG2016+112~\citep{koopmans.02.mn,more.09.mn,spingola.19.a&a}. To be
noticed, there are still interesting anomaly cases like
B1422+231~\citep{patnaik.92.mn} in which the perturbing object has not
been identified. Additionally, \citet{cohn.04.apj} have shown that
image multiplicities in lenses were mostly produced by
substructure-perturbed systems, and many previously discovered simple
lens system were detected with satellites~\citep{ros.00.a&a,
  more.09.mn, macleod.13.apj,spingola.18.mn}.

The gravitational lens B1152+199 was discovered through the
\emph{CLASS} survey~\citep{myers.99.aj} and
investigated~\citep{rusin.02.mn, metcalf.02.apj} as a primary
dual-image lens with a possible substructure
inference. Higher-frequency radio interferometric observations have
also been studied recently~\citep{asadi.20.mn} to testify the bending
significance of the jet. In \citeauthor{rusin.02.mn}'s investigation,
there is a faint blob identified as the X galaxy west to the lens
galaxy. Due to the inadequate number of observables, there is a
degeneracy between the mass profile and the perturber mass. Therefore,
no significant effect of the X galaxy on the model could be addressed.

In this paper, we reinvestigate the optical and radio interferometric
observations of B1152+199 to date and present the data analysis and
lens modellings with regard to the impacts of detected arc-shaped
diffuse emission and multi-component images. The role of the satellite
galaxy in the lensing configuration is prudently examined, which was
not shown in previous literature. Section 2 analyses the \emph{HST}
and \emph{VLBI} data and raises the question about the diffuse
emission and the role of the X-galaxy. Then the answers to the
question are inferred by the mass modelling with both point and image
constraints in Section 3. Section 4 discusses the observation and
modelling results and their implications. Section 5 summarizes our
findings and proposes future investigations.

\section{Data}

\subsection{\emph{HST} observations and data reduction}\label{sec:hst}

Since the lensing galaxy has not been proved to be radio-loud with
interferometric detections, the optical observations then become very
useful to provide extra constraints on the lens position and mass
profile, as well as to spectroscopically determine the redshifts of
images and lenses. To study the lensing configuration of B1152+199,
there were several HST observations carried out including imaging with
the Wide Field and Planetary Camera 2 (\emph{WFPC2}) using $V$-band
(F555W) and $I$-band (F814W) filters in 2000 (8248, PI: Wilkinson),
and imaging with the Near Infrared Camera and Multi-Object
Spectrometer (\emph{NICMOS}) using $H$-band (F160W) filter in 2004
(9744, PI: Kochanek). The lensing galaxy (G) and a satellite galaxy
(X) were detected with \emph{HST/WFPC2} $I$-band image of this
dual-image (A and B) system, and the redshifts of the lens and the
source were measured with the Keck II telescope as $z_l$=0.439 and
$z_s$=1.019 respectively~\citep{rusin.02.mn}. The CfA-Arizona Space
Telescope LEns Survey of gravitational lenses
(\emph{CASTLES})~\citep{kochanek.99.aip} has maintained an up-to-date
catalogue of the \emph{HST} optical lenses with introductory
measurements including B1152+199.

The mass ratio between the lensing galaxy and the X-galaxy is
conventionally derived from the Faber-Jackson relation ($\sigma\propto
L^{1/4}$) and the mass-velocity dispersion relation
($b\propto\sigma^2$), together with the magnitude relation
($F_1/F_2\simeq2.512^{\Delta m}$), as done
by~\citet{rusin.02.mn}. Since the photometry is essential to speculate
the mass scale of the X-galaxy, we re-measured the \emph{WFPC2}
$V$/$I$-band and \emph{NICMOS} $H$-band photometries of both the lens
galaxy and the X-galaxy with the drizzled image data from Hubble
Legacy Archive~\footnote{
Based on observations made with the NASA/ESA Hubble Space Telescope, and obtained from the Hubble Legacy Archive, which is a collaboration between the Space Telescope Science Institute (STScI/NASA), the Space Telescope European Coordinating Facility (ST-ECF/ESA) and the Canadian Astronomy Data Centre (CADC/NRC/CSA).}. Both observations
had a long total exposure time ($V$/$I$-band: 2000~sec, $H$-band:
5311~sec), but the integration times are not as long as the total the
exposure time, since the \emph{HST} produces multiple frames for
removing cosmic rays, bad pixels and so that the point spread function
(\emph{PSF}) can be better sampled from dithered observations. We
found it is difficult to remove the point spread function (\emph{PSF})
effect in the image merely by fitting the bright point source with the
convolution kernel. This is in part due to the dynamic range
limitations from having a coarsely sampled \emph{PSF}, as well as the
\emph{PSF} used will likely not match the colour of the lensed
emission, and therefore, the luminosity weighted width of the
\emph{PSF} as a function of frequency will be poorly known. This means
there is a mismatch between the \emph{PSF} and the point-source
emission from quasars.

\begin{figure*}
   \centering
   \includegraphics[width=3.9cm]{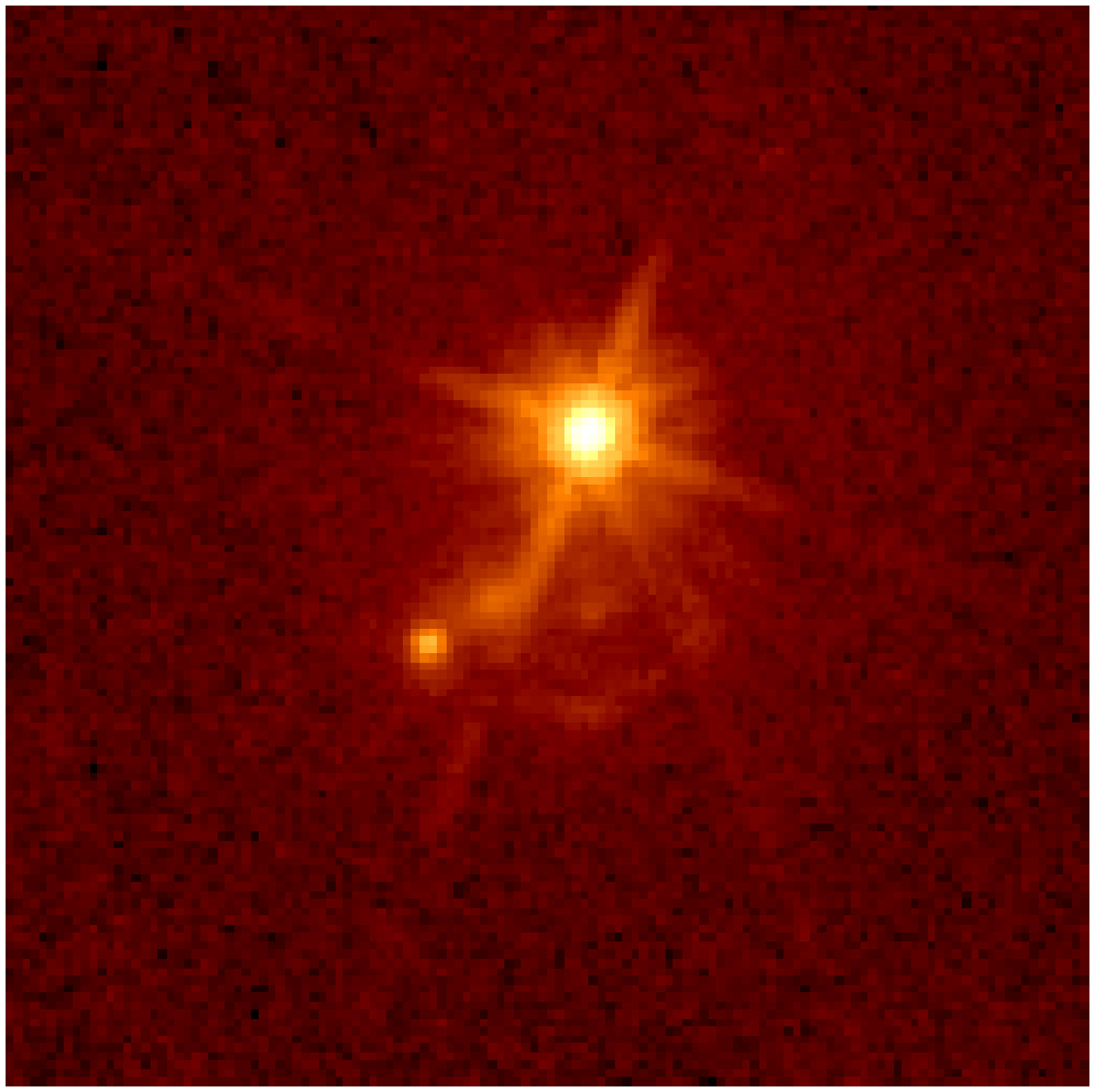}
   \includegraphics[width=3.9cm]{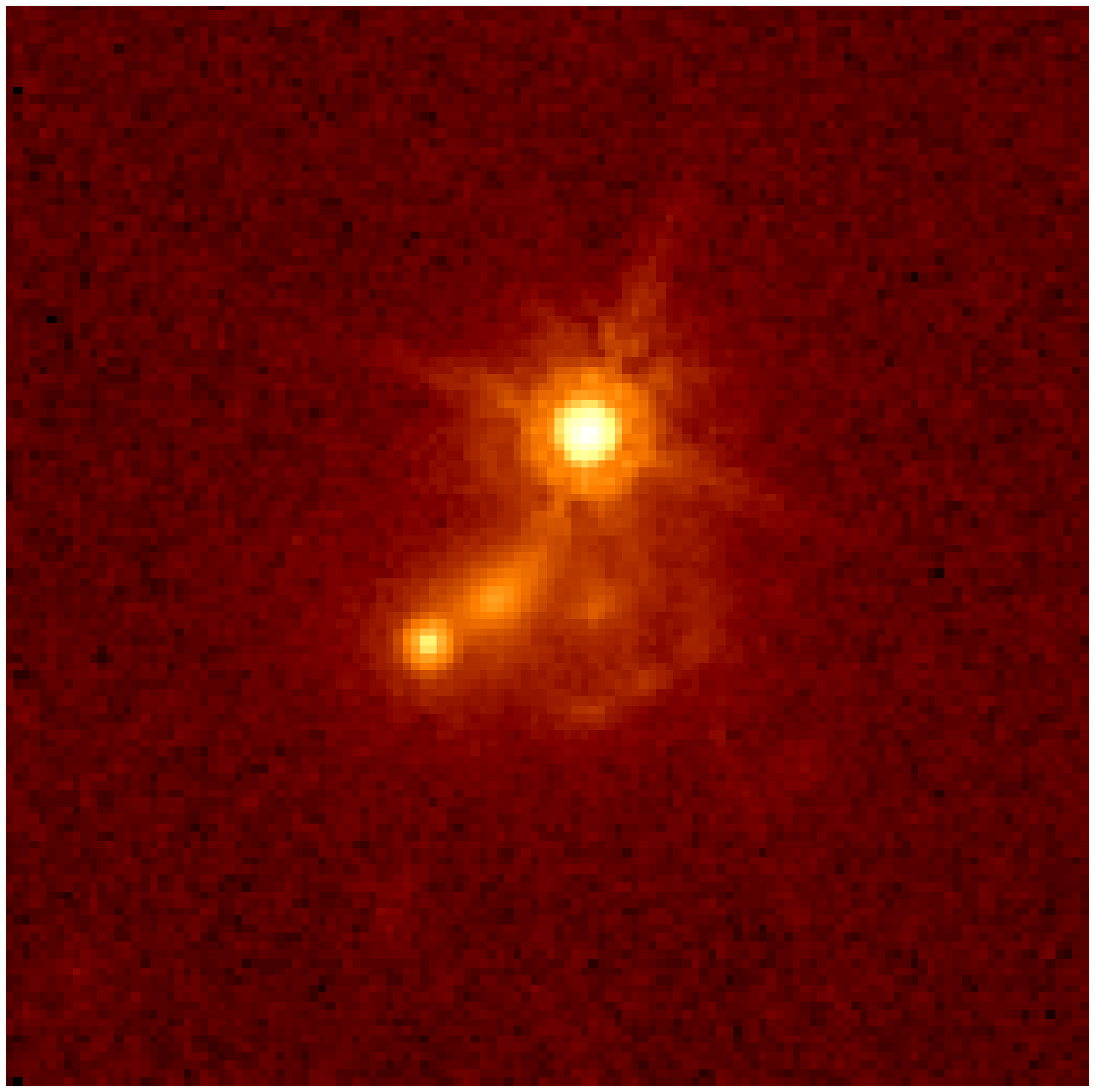}
   \includegraphics[width=3.9cm]{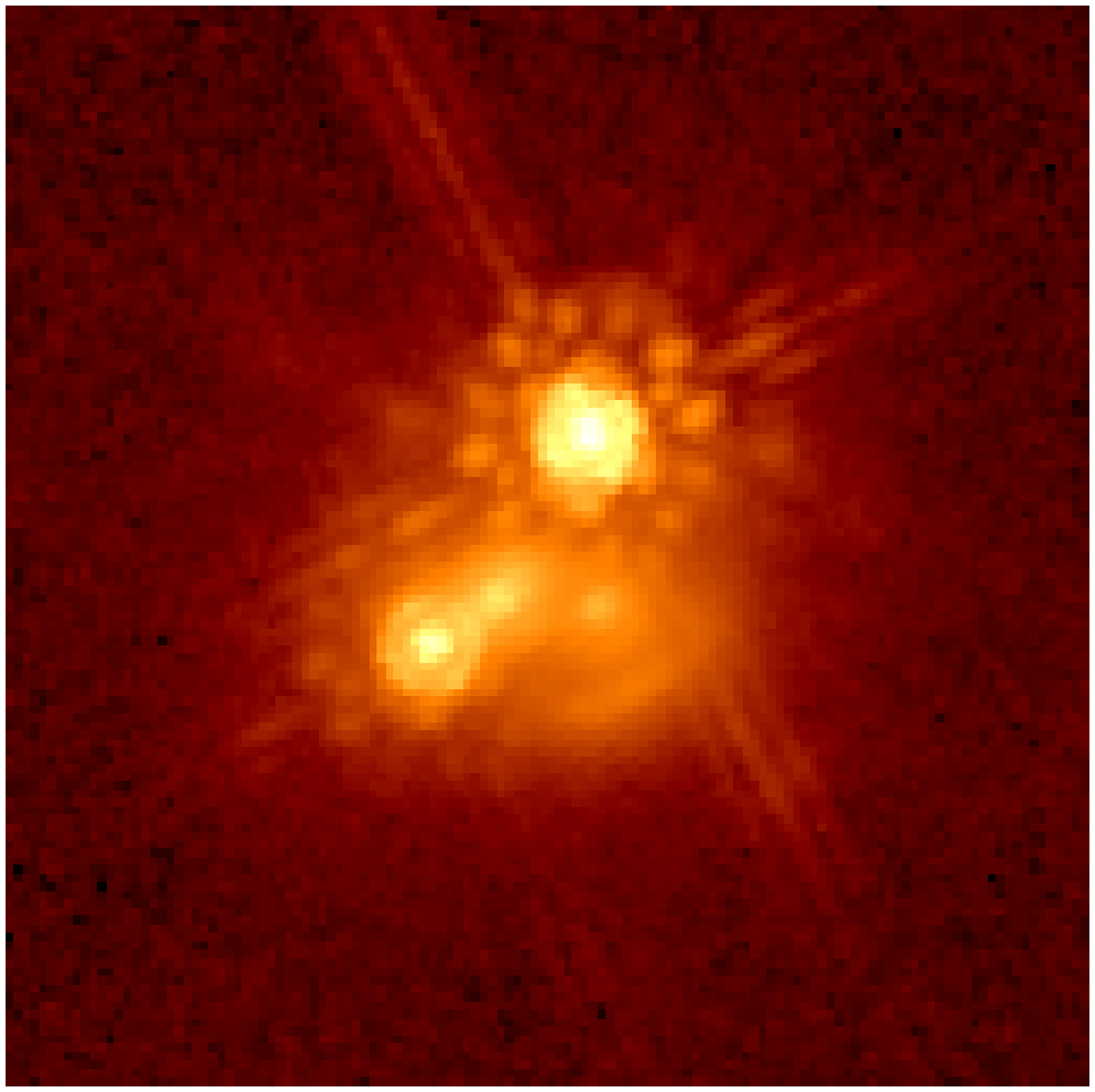}
   \includegraphics[width=3.9cm]{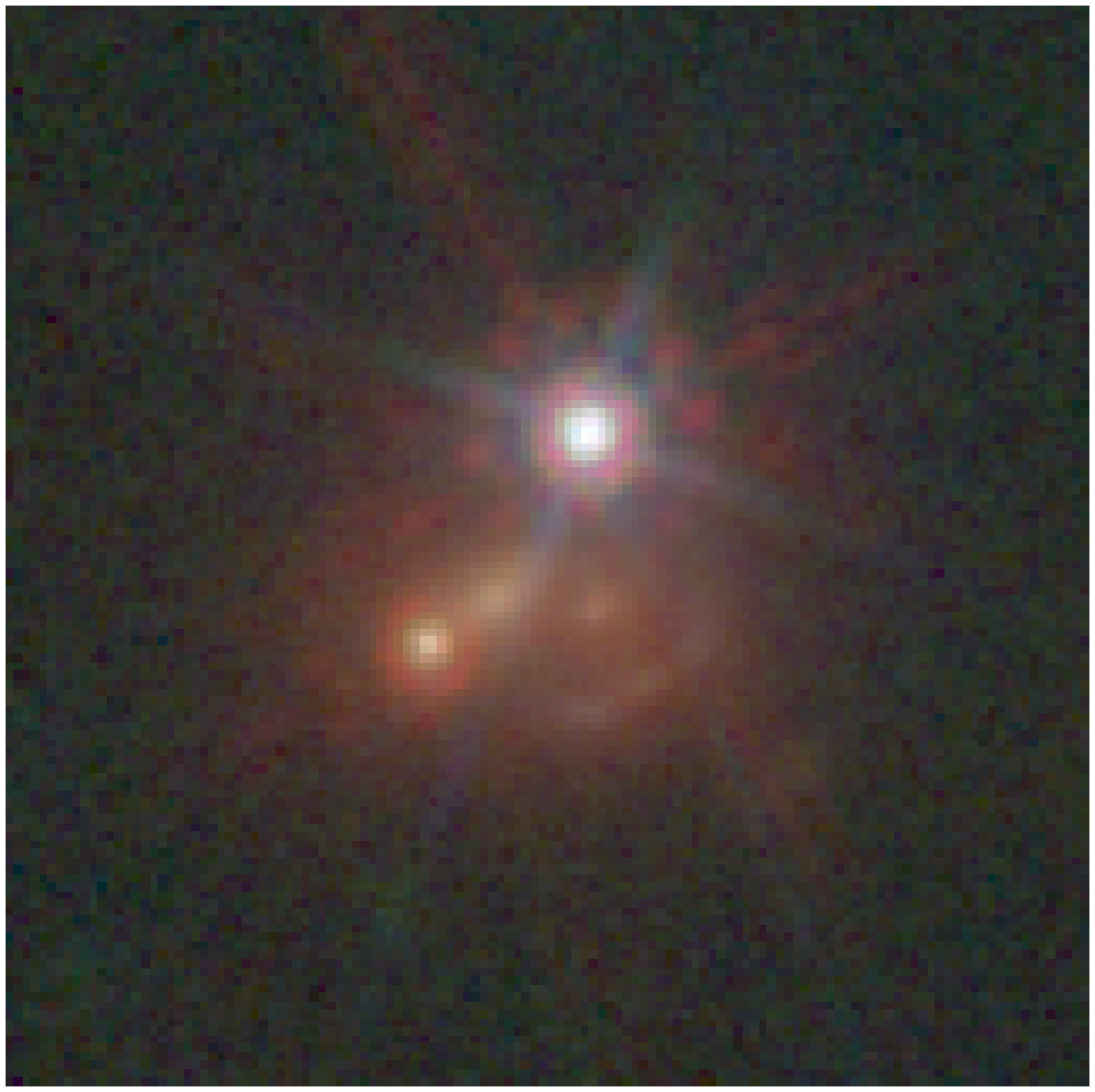}
   \caption[RGB images]{The \emph{RGB} images of tri-band \emph{HST}
     optical observations. The first three panels from left to right
     are the \emph{HST/WFPC2} $V$-band, $I$-band and \emph{HST/NICMOS}
     $H$-band images, which are assigned to the B, G and R channels,
     respectively, according to their wavelengths. The last panel
     is the composed three-channel \emph{RGB} image. The hue is set to
     a black-body profile with a logarithmic scale. }
   \label{fig:rgb}
\end{figure*}

\begin{figure*}
   \centering
   \includegraphics[width=4cm]{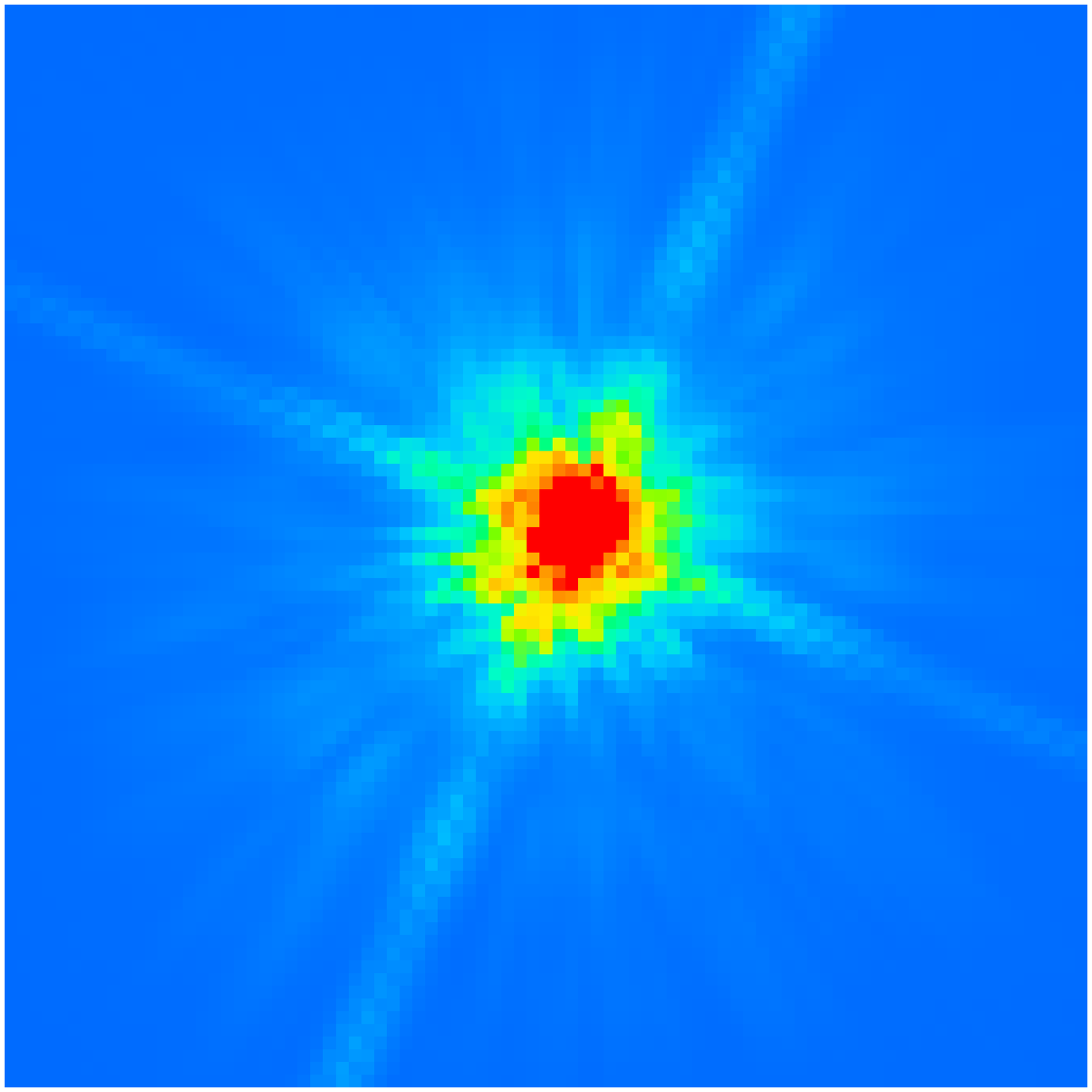}
   \includegraphics[width=4cm]{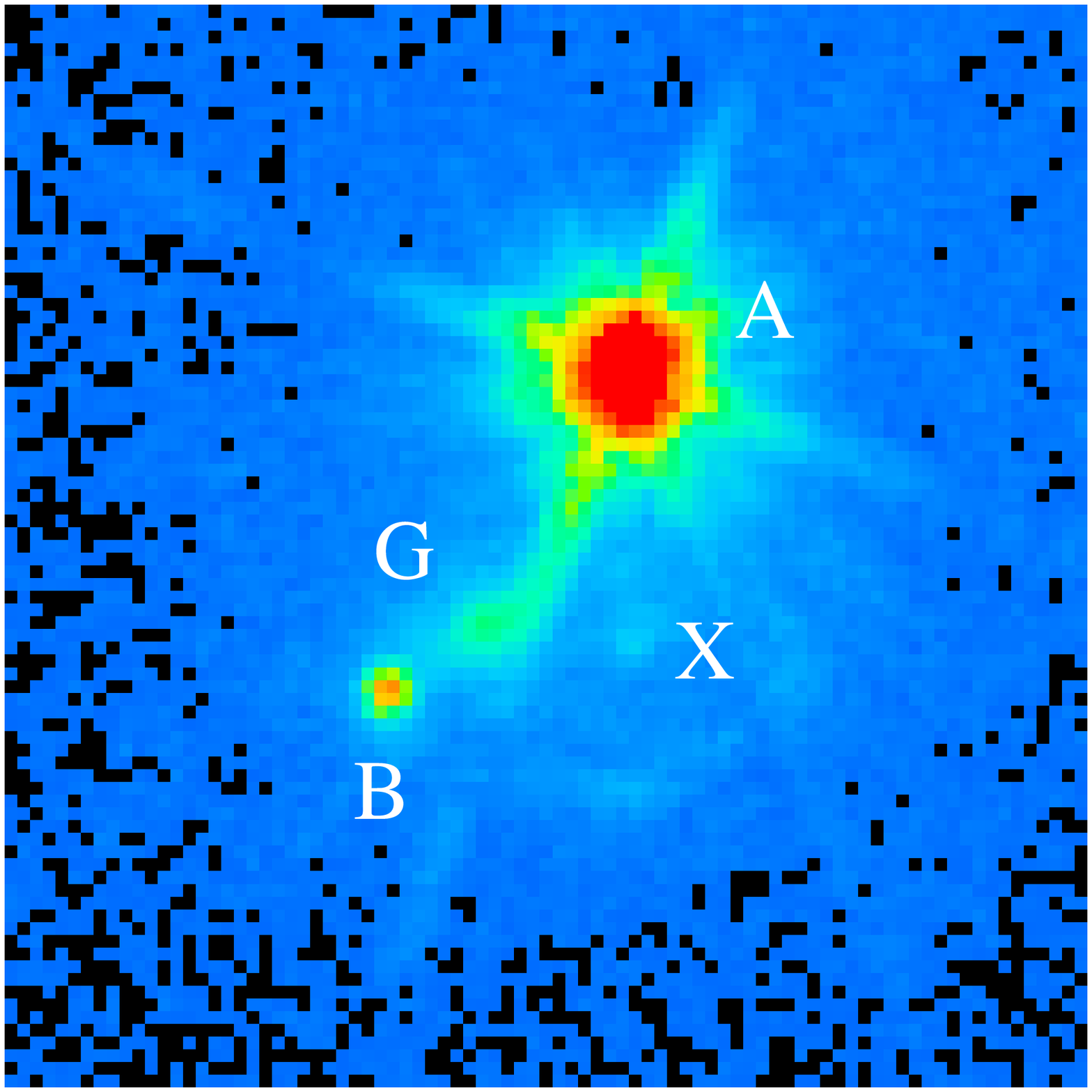}
   \includegraphics[width=4cm]{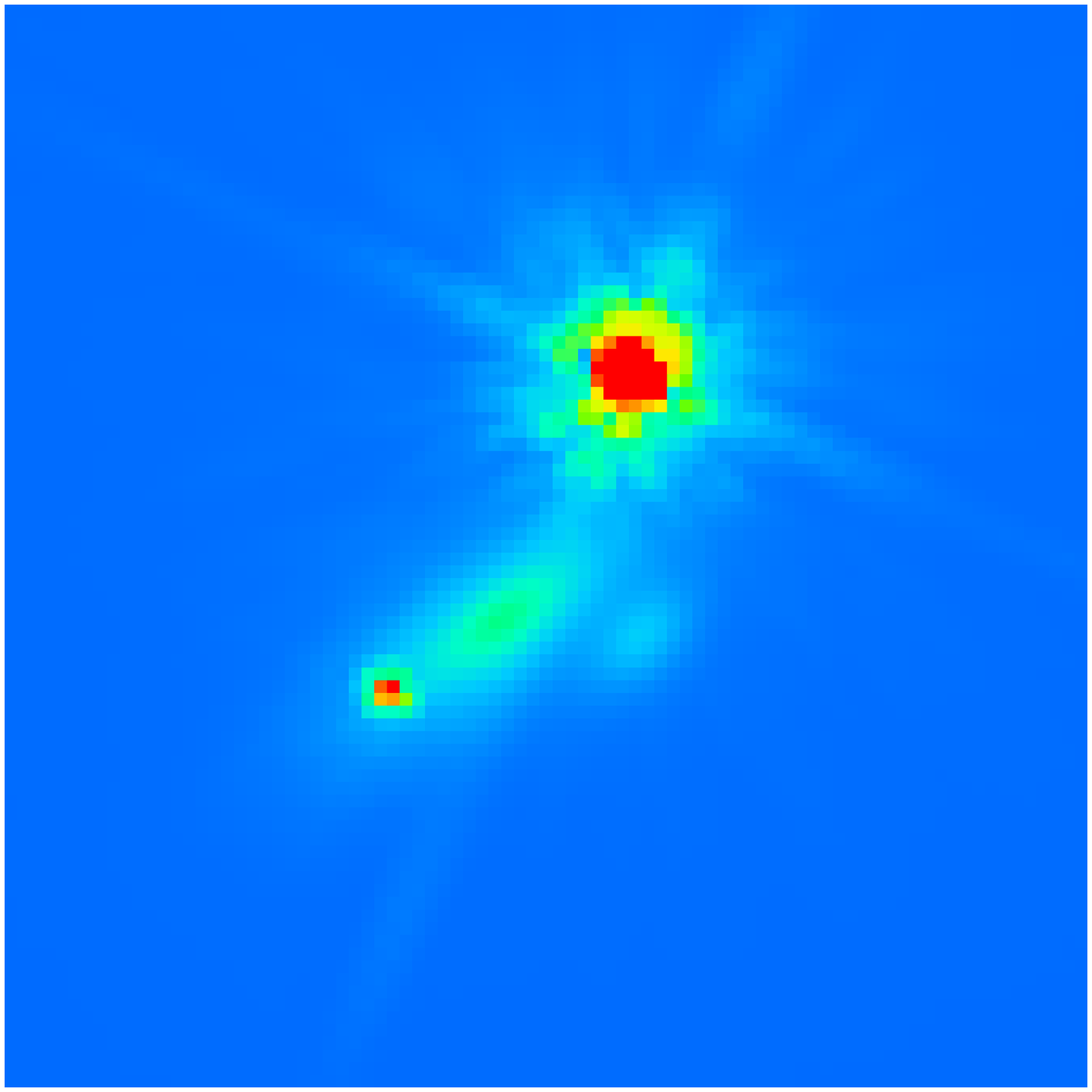}
   \includegraphics[width=4cm]{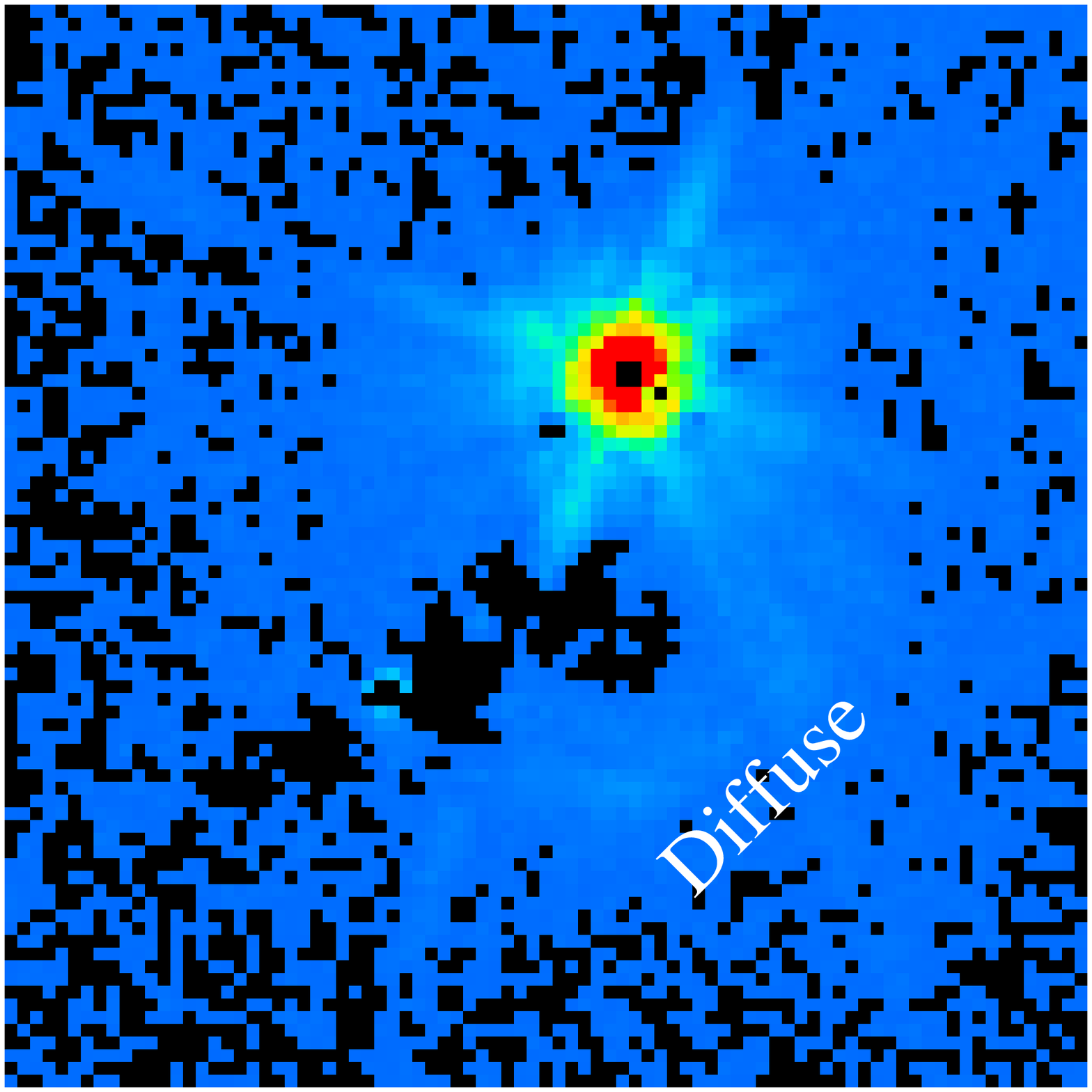} \\
   \includegraphics[width=4cm]{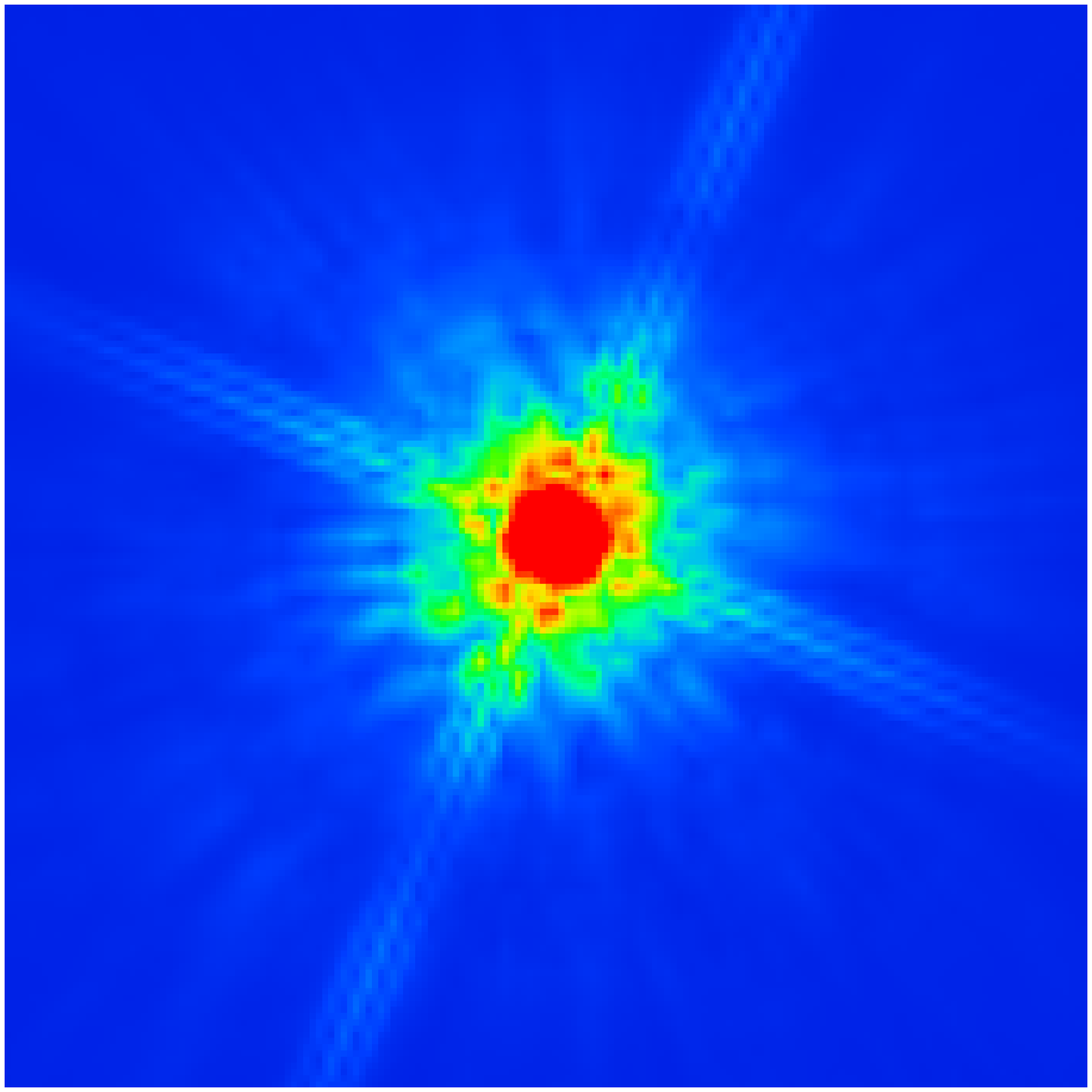}
   \includegraphics[width=4cm]{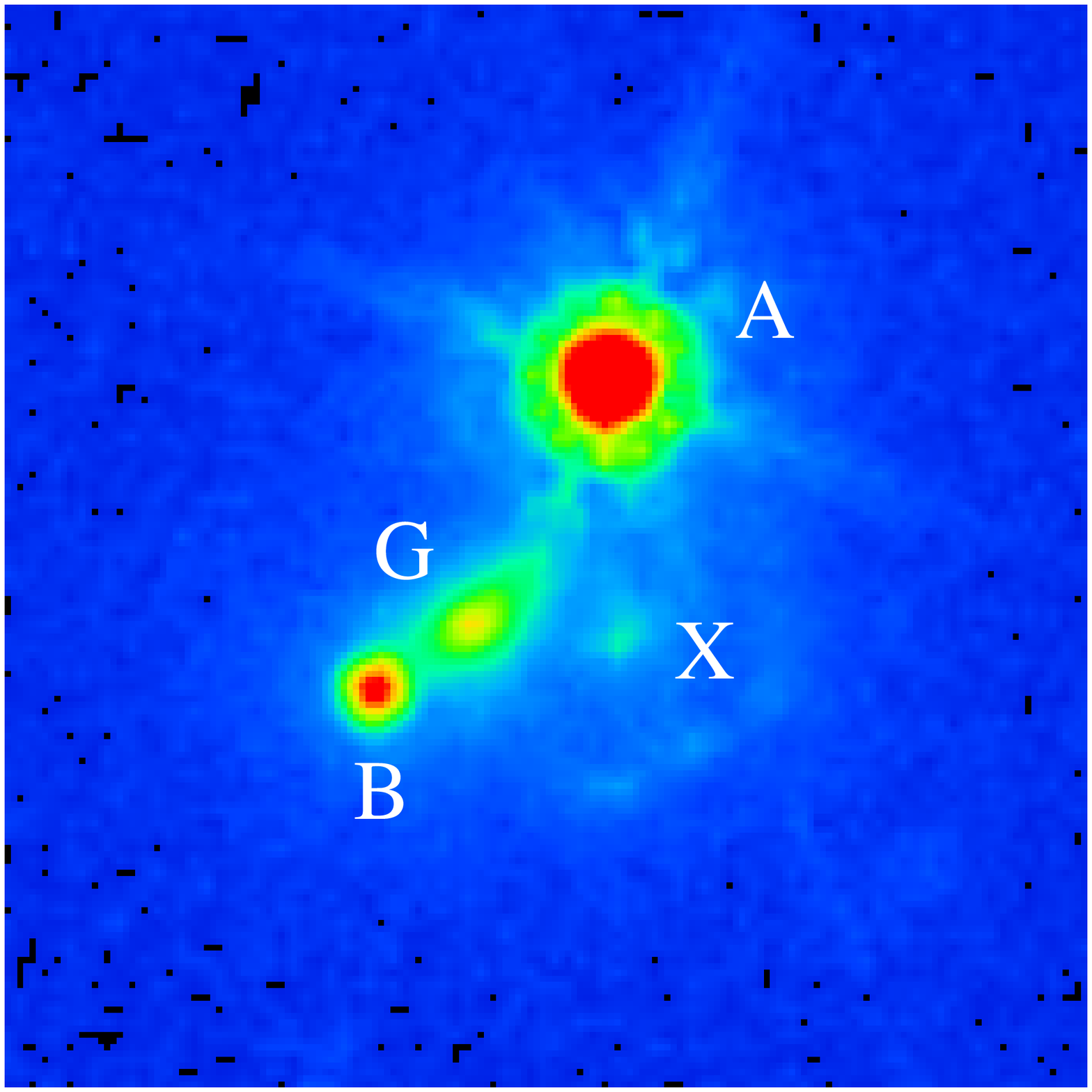}
   \includegraphics[width=4cm]{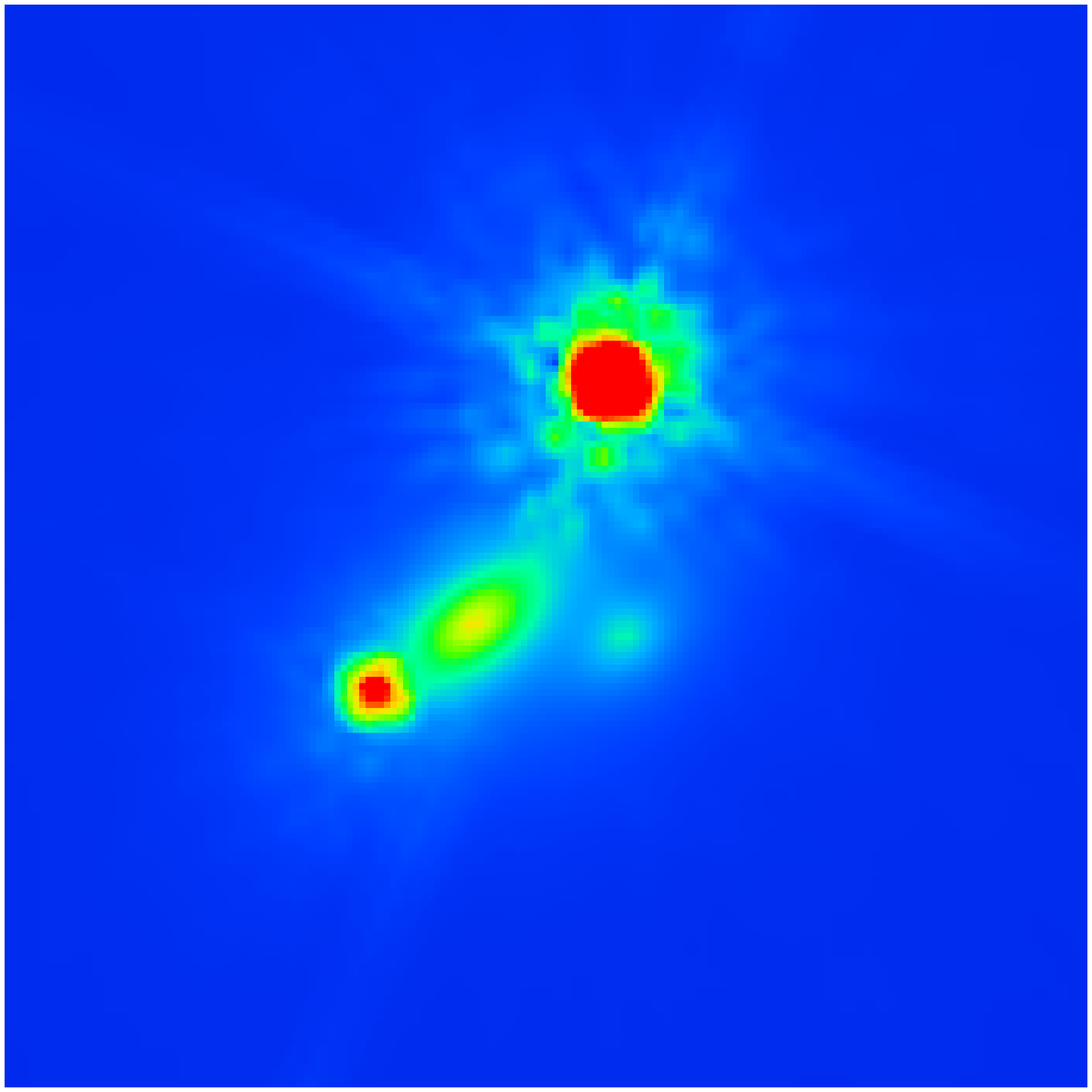}
   \includegraphics[width=4cm]{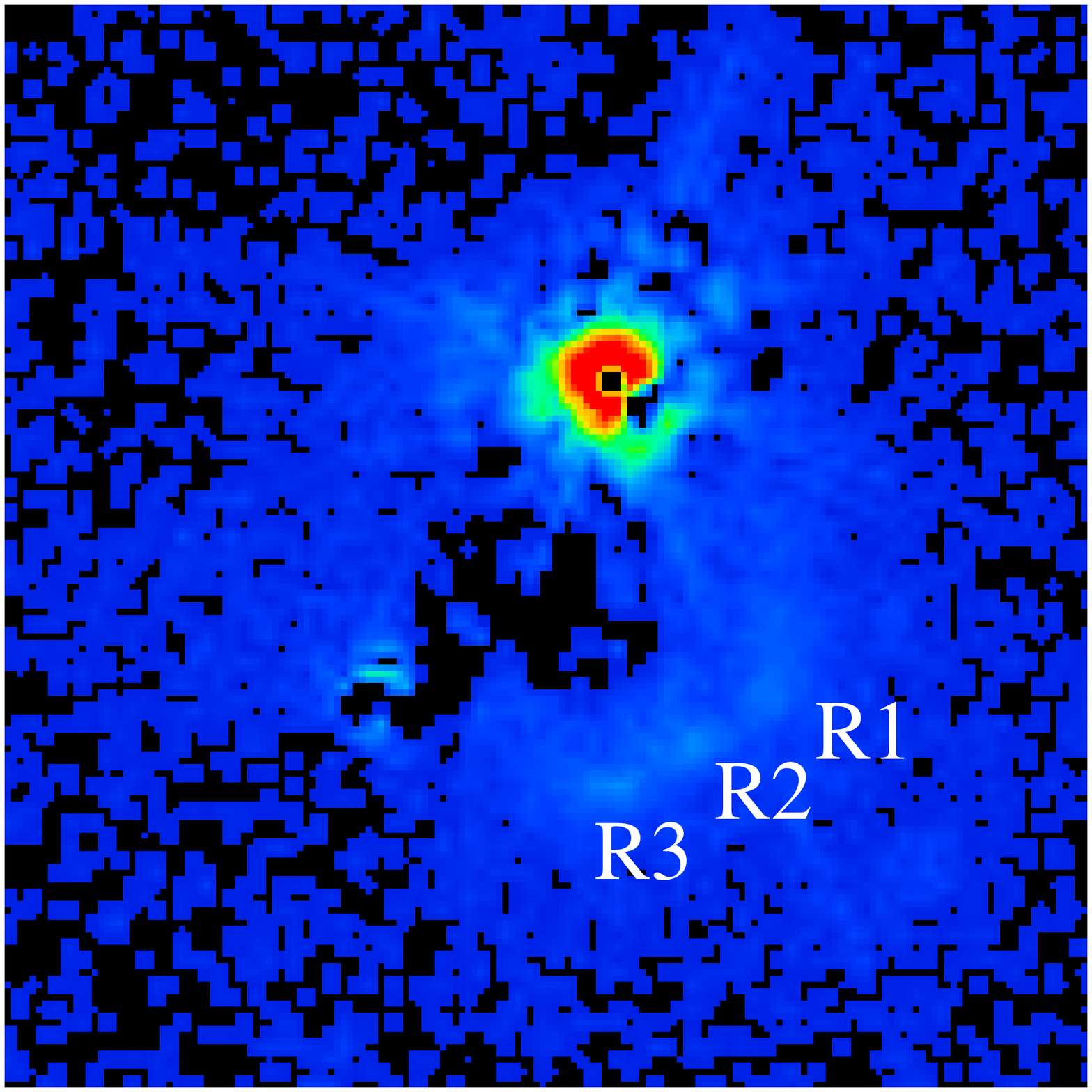} \\
   \includegraphics[width=4cm]{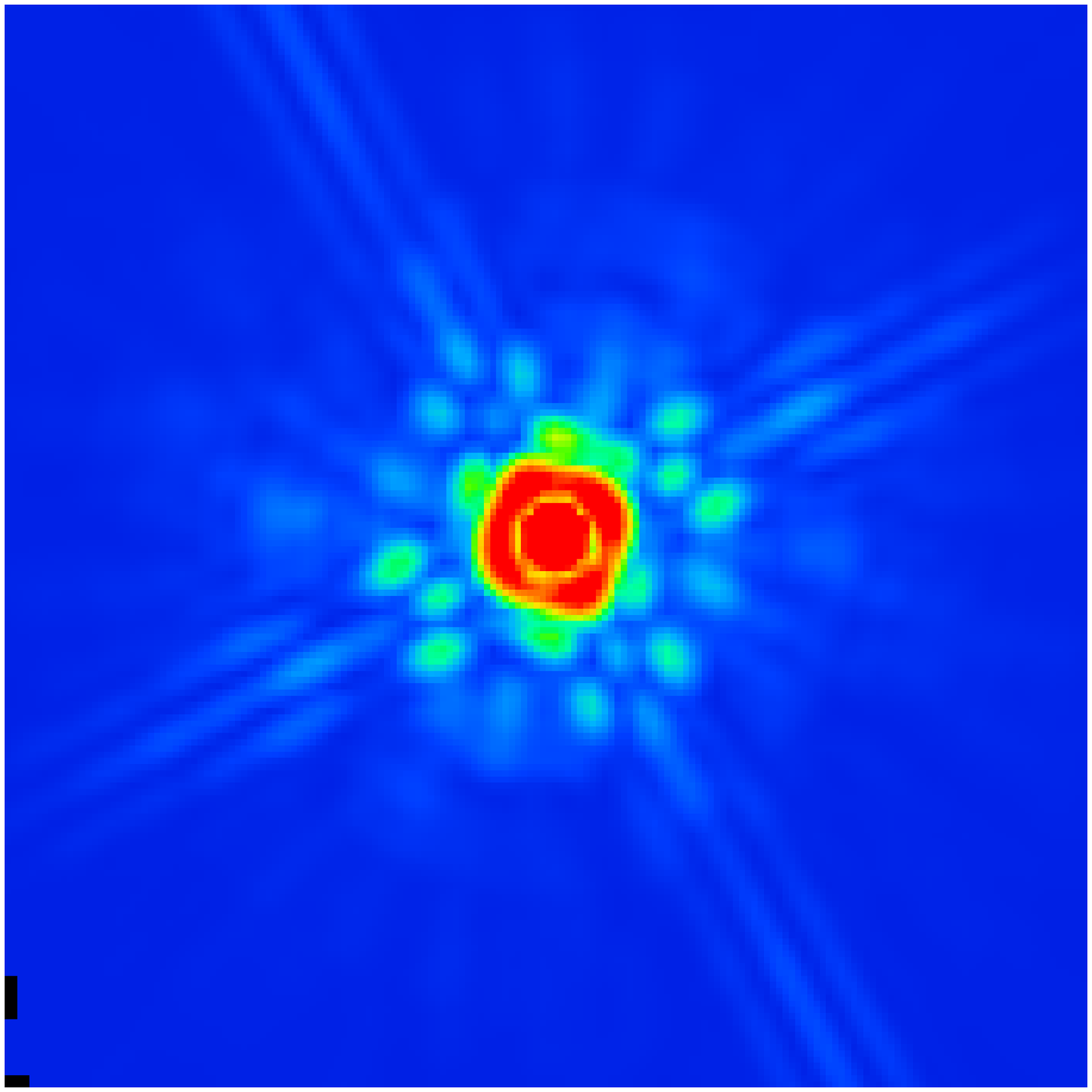}
   \includegraphics[width=4cm]{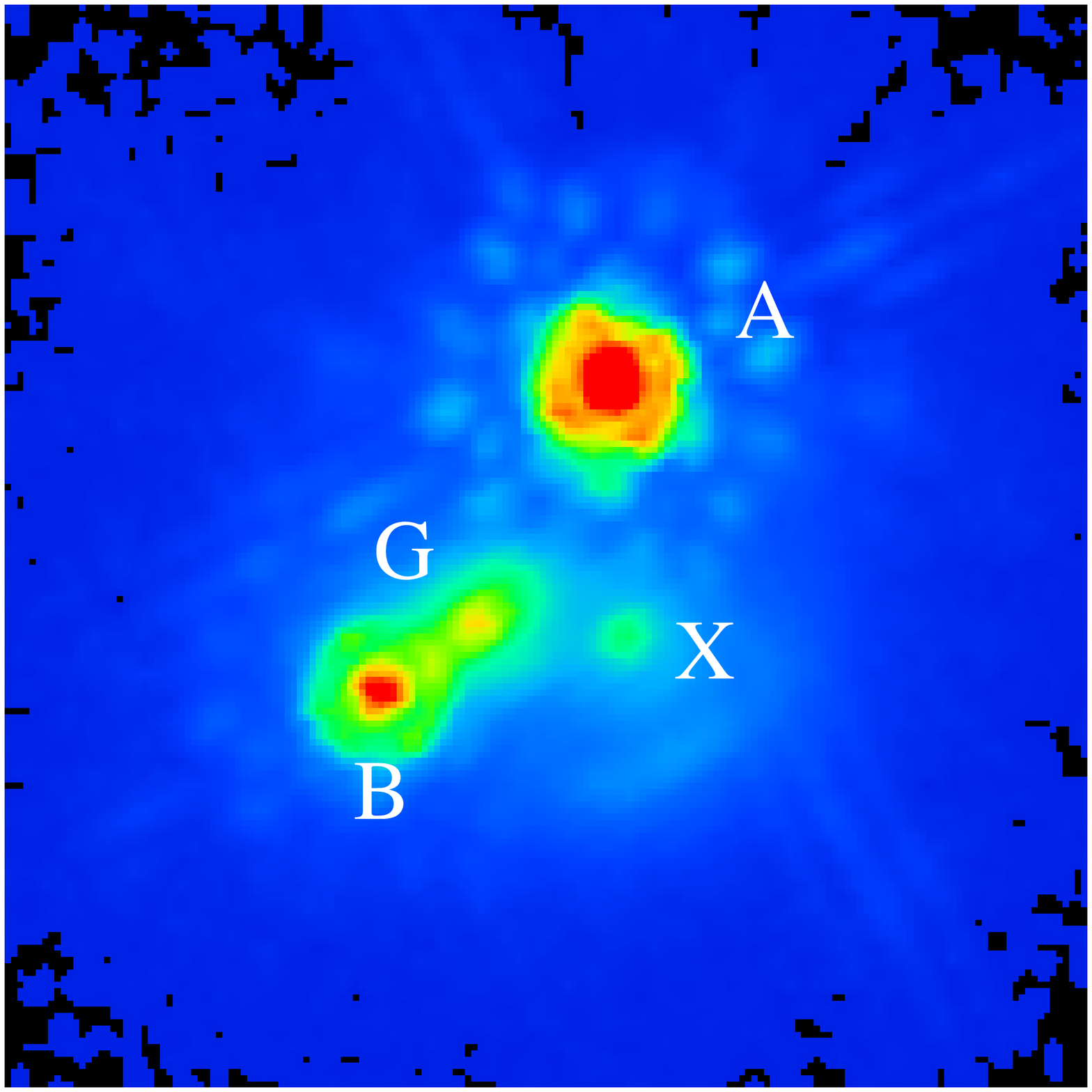}
   \includegraphics[width=4cm]{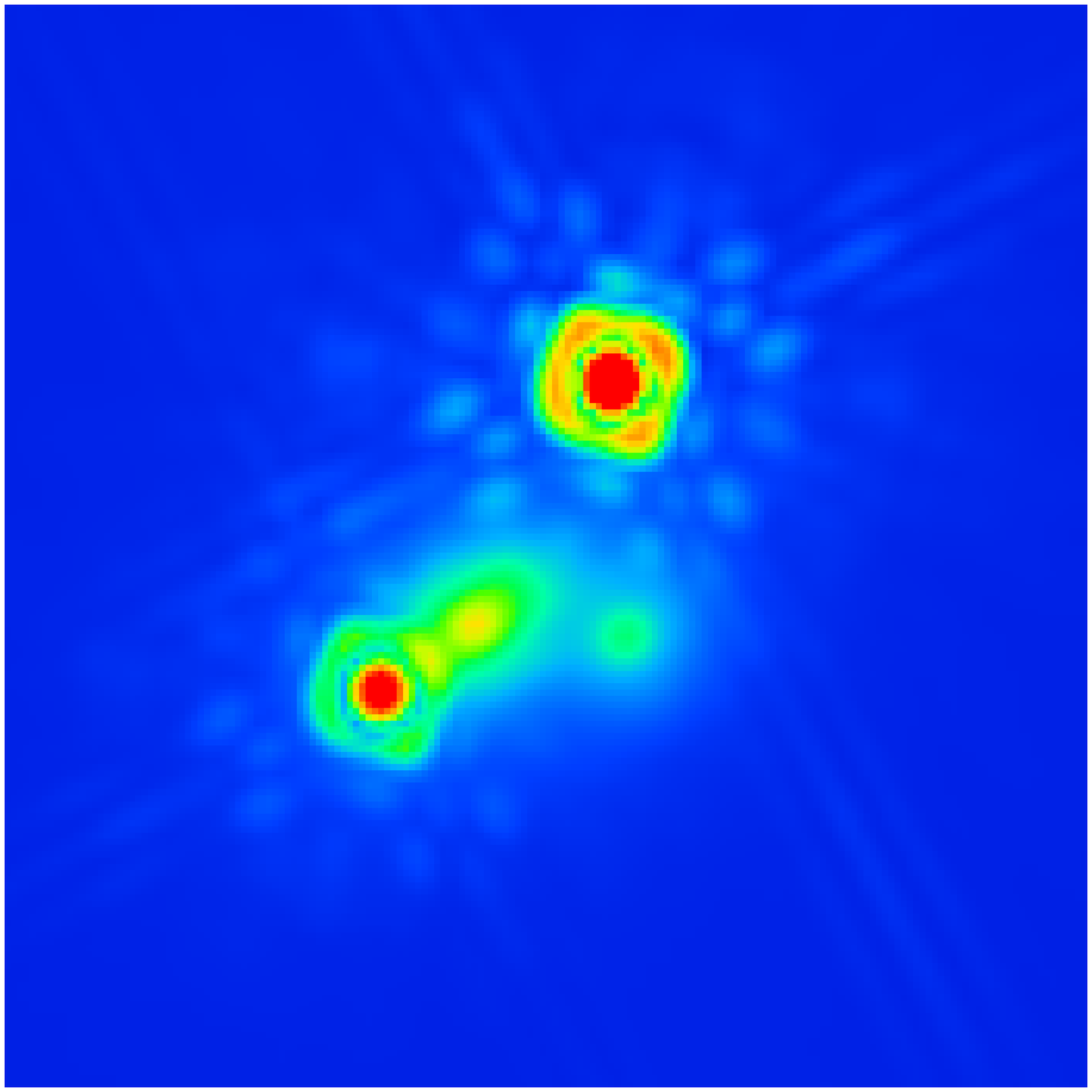}
   \includegraphics[width=4cm]{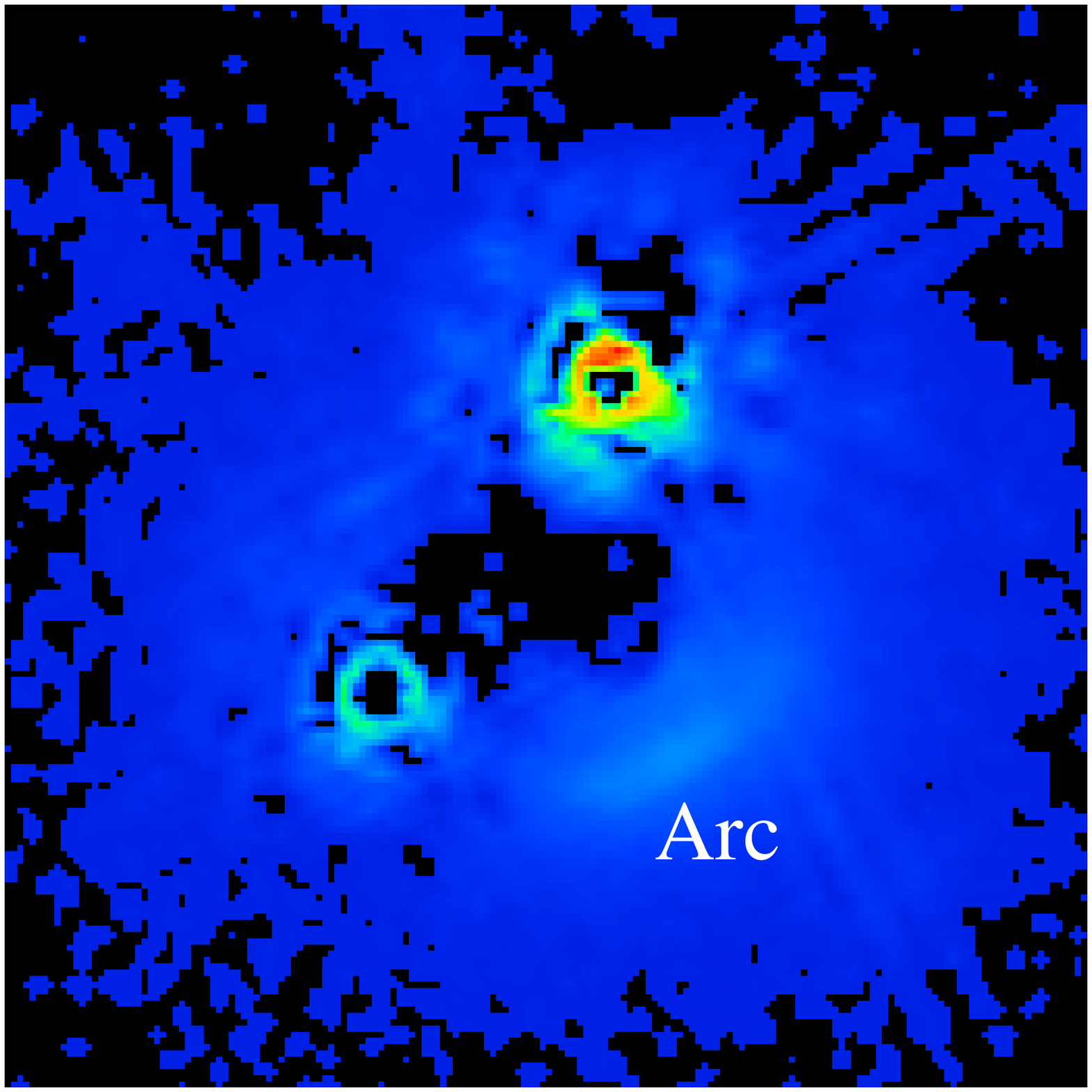}
   \caption[Image-plane models]{The image-plane light profile
     subtraction.  From top to bottom, the rows are the
     \emph{HST/WFPC2} $V$-band, $I$-band and \emph{HST/NICMOS}
     $H$-band images, respectively. From left to right, the columns
     show the \emph{PSF} models, the original images, the convolved
     image models, and the residual images after the subtraction,
     respectively. The intensity is plotted on logarithmic scale with
     pseudo colours. The pixel ranges are adjusted to show optimal
     contrast.  The primary and the secondary lensed images, the main
     lens galaxy, the perturber galaxy, the diffuse arc and the
     resolved quad are labelled as A, B, G, X, Arc and R1, R2, R3
     respectively. }
   \label{fig:galfit}
\end{figure*}

We composed the \emph{NICMOS} $H$-band and \emph{WFPC2} $I$- and
$V$-band images into the RGB channels and formed a pseudo-colour image
as seen in Fig.~\ref{fig:rgb}. From the RGB image, we can notice there
are a noticeable amount of emissions around the lens galaxy and the
X-galaxy, especially in the $I$- and $H$- band images. The intrinsic
emissions from the galaxies are blended with the spread of the lensed
light from both A and B images.  We have utilized the {\sc
  tinytim}\footnote{The Hubble Space Telescope PSF generating program
  by John Krist at STSCI.}~\citep{krist.aspc.93} to generate the
\emph{PSF} of \emph{WFPC2} and \emph{NICMOS}, then convolve and
subtracted the models in the \emph{HST} images retrieved from the
Hubble Legacy Archive (\emph{HLA}). The image-plane model fitting is
carried out with the {\sc Galfit} devised
by~\citet{peng.02.aj,peng.10.aj}. The foreground galaxies were fitted
against the S\'{e}rsic model and the background quasar images were
fitted by the \emph{PSF} kernel. Obvious the primary image was not
representable by a simple \emph{PSF} due to saturation, and there is a
tendency for the X-galaxy to fit the spurious emissions spread from
the lensed images with a bigger effective radius. So we masked a
significant amount of pixels around the galaxies including all lensed
images and the surrounding diffuse emission to reduce the \emph{PSF}
effect. The {\sc galfit} image models and residuals of tri-band data
are shown in Fig.~\ref{fig:galfit}. We here list the newly measured
photometry and derived mass ratios in Table~\ref{tab:bratio}. Their
error estimates are the projected covariances as quoted in {\sc
  Galfit}~\citep{peng.02.aj}. It is noticeable that the new mass ratio
between the host galaxy and the satellite is consistently around 2
across three bands, rather than ranging broadly from 3 to 5 in the
previous measurements. This leads to different estimations about the
X-galaxy's total mass and conceivably affects the lens
modelling. Though it is arguable whether the optical luminosity of the
lensing galaxy is compromised with uncertainty due to dust extinction,
we note there the positional angle of fitted S\'{e}rsic profiles of
the host galaxy and the perturber here in Table~\ref{tab:photpar}.

\begin{table}
   \centering
   \caption[Photometric mass ratios]{Photometry and derived mass ratios 
       between lens and X-galaxy from the \emph{HST} archived images. 
       Here we take the conventional STmag system for \emph{HST} observations.}
   \label{tab:bratio}
   \begin{tabular}{lrrr} 
      \hline
             &  $V$ &  $I$ & $H$  \\
      \hline
       G mag   & 20.88$\pm$0.17 & 20.50$\pm$0.09  & 19.12$\pm$0.02   \\
       X mag   & 23.25$\pm$0.38 & 22.14$\pm$0.76  & 20.52$\pm$0.14   \\
      $b'_G/b'_X$ &  2.98$\pm$0.58 &  2.13$\pm$0.74  &  1.90$\pm$0.12   \\
     \hline
   \end{tabular}
\end{table}

\begin{table}
   \centering
   \caption[Photometric position angles]{Fitted light profile parameters of 
     S\'{e}rsic profiles of the galaxies in three \emph{HST} bands. 
     Values are shown in ($R_e$, $n$, $q$, $\phi_q$) tuples. 
     The tuples below each parameter tuple shows the quoted covariance errors.
     Note that the ellipticity $e$ equals to 1-$q$ and the position angles is 
     measured from north to east.}
   \label{tab:photpar}
   \begin{tabular}{ccc}
      \hline
          &   Host galaxy  &    X-galaxy      \\
      \hline
      $V$ & (0.44'', 0.89, 0.37, -54.24$^{\circ}$)     & ~(0.17'', 0.62, 0.58, -46.85$^{\circ}$) \\
          & $\pm$(0.08'', 0.13, 0.03,   1.65$^{\circ}$) & $\pm$(0.06'', 0.39, 0.11,  12.27$^{\circ}$) \\
      $I$ & (0.35'', 1.39, 0.46, -51.75$^{\circ}$)     & ~(0.29'', 1.68, 0.63, -52.95$^{\circ}$) \\
          & $\pm$(0.03'', 0.12, 0.01,   1.64$^{\circ}$) & $\pm$(0.23'', 1.00, 0.10,  13.48$^{\circ}$) \\
      $H$ & (0.29'', 1.72, 0.53, -51.38$^{\circ}$)     & ~(0.17'', 1.14, 0.85, -65.60$^{\circ}$) \\
          & $\pm$(0.01'', 0.05, 0.01,   1.30$^{\circ}$) & $\pm$(0.03'', 0.25, 0.05,  15.62$^{\circ}$) \\
     \hline
   \end{tabular}
\end{table}

\citet{rusin.02.mn} had already mentioned the detected diffuse
emission to the west of image B in the $I$-band image. To
quantitatively estimate whether this extended structure will play a
role in the overall lens modelling, we need more observational
evidence. As we see in \emph{HST} multiband images, the diffuse west
to image B is barely visible in $V$-band image, when we go to longer
wavelengths like I and H bands, it starts to be noticeable. The diffuse
emission roughly runs along the tangential direction around the
X-galaxy which may form an arc shape. In the $I$-band image, the arc
shape is resolved into three faint blobs which resemble a lensing
case of the cusp-crossing quad. Comparing the {\sc clean}ed images to
the original drizzled images, we can see that some emissions are
restored around the arc region. It could be possible that the
arc-shaped emission is purely produced by surrounding interstellar
dust, however the recent spectroscopic study on the damped Ly$\alpha$
absorption together with the metallicity estimation from the X-ray
observation~\citep{dai.20.mnras} found no adequate evidence for the
dust existence and thus disfavoured this scheme.

\subsection{\emph{VLBI} observations and data reduction}\label{sec:vlbi}

Precisely measured observables of lensed images are crucial for lens
modelling, especially when the substructures are to be identified
therein. The Very Long Baseline Interferometry (\emph{VLBI}) is
currently the most powerful method to get highly resolved images for
extragalactic radio sources. Since the light extinction is entirely
negligible at centimetre wavelengths, the positional and flux
constraints obtained from radio interferometry can be used to
determine the lens model with more accuracy compared to those in the
optical bands. High-resolution \emph{VLBI} images thus constrain the
lens model the most and have been utilized to detect deviations from a
smooth mass
model~\citep{ros.00.a&a,bradac.02.a&a,biggs.04.mn,spingola.18.mn},
galactic central profile~\citep{zhang.07.mn,quinn.16.mn}, Hubble
constant~\citep{biggs.03.mn,wucknitz.04.mn1,wucknitz.04.mn2}, jet
proper motion~\citep{spingola.19.a&a} and new lens discovery
survey~\citep{spingola.19.mn}.

\subsubsection{C-band VLBI observations}\label{sec:5ghz}

There have been several \emph{VLBI} observations carried out on the
target of B1152+199 in recent years. Hence, we collected and reduced
the archived 5-GHz data from the \emph{VLBA}\footnote{The {\em Very
    Long Baseline Array} is a radio interferometric network run by the
  {\em National Radio Astronomy Observatory (NRAO)}} and the
\emph{EVN}\footnote{{\em European VLBI Network} is coordinated by the
  {\em Joint Institute for VLBI in Europe (JIVE)}} in order to obtain
milli-arcsecond-scale astrometry of the dual-image lensing
systems. The data set includes the \emph{VLBA} observation (BB133, PI:
Biggs) from 2001, the global-\emph{VLBI} observation (GJ010, PI:
Jackson) from 2003 and the High-Sensitive Array (\emph{HSA})
observation (BW084, PI: Winn) from 2005. All data were recorded with
four contiguous 8-MHz intermediate frequencies (\emph{IF}) and
correlated with a channel resolution of 0.5~MHz and a time resolution
of 2~s. The global-\emph{VLBI} observation (GJ010) was recorded only
with left circular polarization (\emph{LCP}), while the other two were
obtained with dual circular polarization. The effective
on-source integration time of BB133, GJ010 and BW084 was 3.4, 7.7 and
3.5 hours respectively.

\begin{figure*}
\centering
\includegraphics[width=7cm]{b1152_bb133u_a.ps}
\includegraphics[width=7cm]{b1152_bb133u_b.ps}\\
\includegraphics[width=7cm]{b1152_gj010u_a.ps}
\includegraphics[width=7cm]{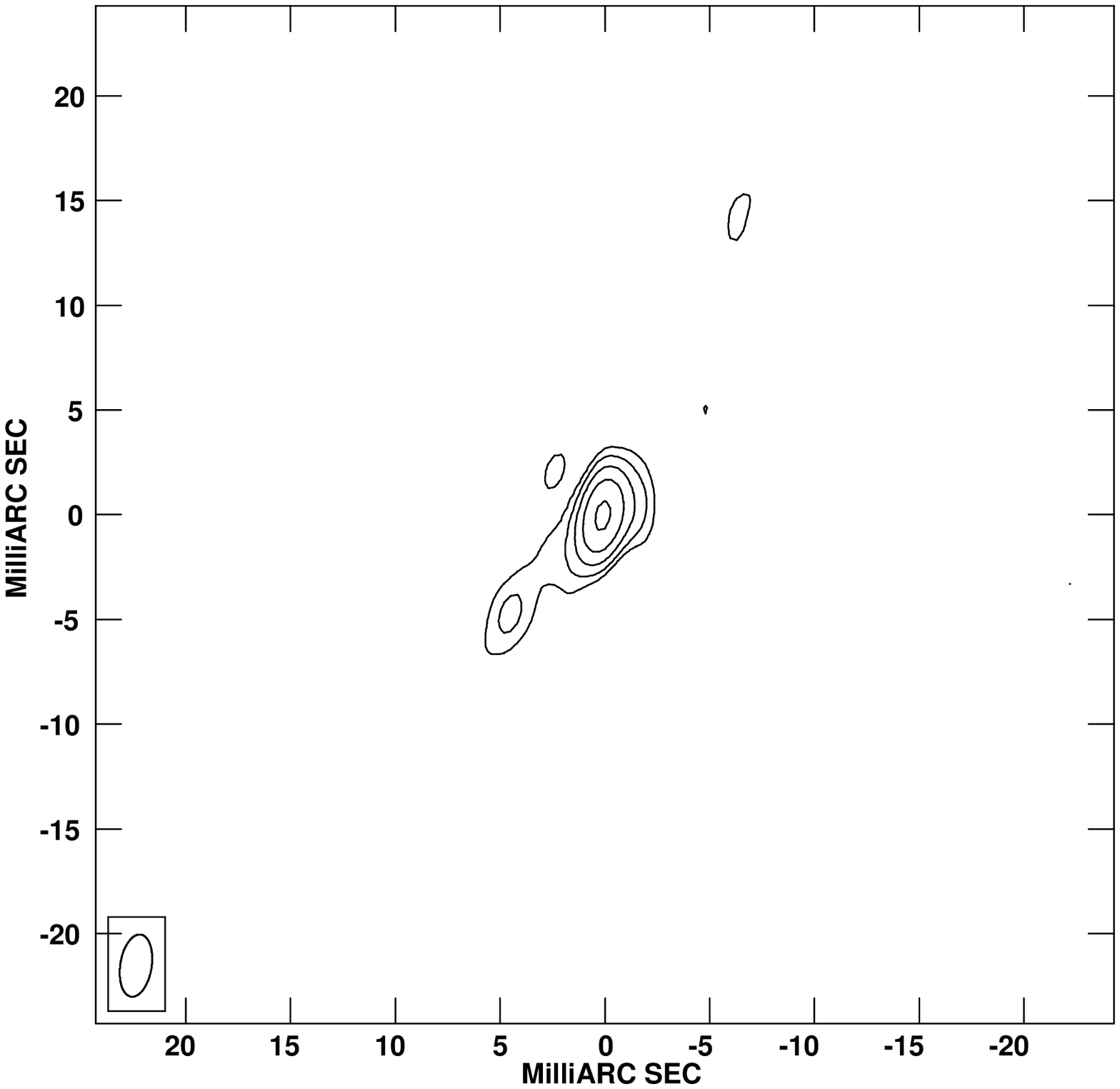}\\
\includegraphics[width=7cm]{b1152_bw084u_a.ps}
\includegraphics[width=7cm]{b1152_bw084u_b.ps}
\caption[Three-epoch \emph{VLBI} observations]{ Three-epoch 5-GHz
  \emph{VLBI} observations of B1152+199. From top to bottom, the A and
  B images are of observations BB133 (2001), GJ010 (2003) and BW084
  (2005), respectively. The data have been uniformly weighted. The
  restoring beam size is 3$\times$1.5~mas$^2$ at a position angle of
  $-9^\circ$. Contours in the map are plotted at multiples of -1, 1,
  2, 4, 8, 16, 32, 64, 128, 256, 512, 1024, 2048 $\times$ $3\sigma$
  where $\sigma$ is the local RMS noise. The off-source background
  noise of each observation is 75, 31 and 21 \mujybm respectively.
  The marginally resolved jet components in image A are labelled from
  the core as A1, A2, A3, and the unresolved jet in image B is labelled ad
  B$_{jet}$.}
\label{fig:5ghz3epo}
\end{figure*}

To cross-check the observations, we re-reduced the visibility data of
the three epochs to obtain high-dynamic-range images.  The data
calibration was carried out with {\sc aips}\footnote{The Astronomical
  Image Processing System distributed by NRAO (National Radio
  Astronomy Observatory).}. Standard procedures for data inspection
and flagging, as well as calibration of amplitude, phase and bandpass,
and multi-source splitting, have been applied to the three sets of VLBI
data using appropriate {\sc aips} tasks. The self-calibration and
deconvolution procedures were carefully used in each case, especially
when mapping the jet in image B. To maintain the highest resolution
and suppress extended features, we {\sc clean}ed the uniformly
weighted data before self-calibration. The three-epoch {\sc
  clean}ed images are shown in Fig.~\ref{fig:5ghz3epo}. All images
are restored to the same 3$\times$1.5~mas$^2$ beam, and the root mean
squared (RMS) map noise was {\sc clean}ed down to the noise level
respectively. The jet in image A has been resolved into at least two
components, while in image B there is only one distinguishable
component. With the highest sensitivity and resolving power of the
\emph{HSA}, we can see that there is a third jet component just barely
resolved between the outermost and innermost components in image
A. There were up to three jet components resolved in image A, while
there was only one unresolved jet emission protruding out in image
B. The discrete jet components were model-fitted in the image plane
with {\sc aips} task {\sc jmfit} and the errors were estimated by the
beam size over signal-to-noise ratio, as shown in
Table~\ref{tab:cxobs}.

It is noticeable that the contour lines of the jet in image B show
a slight bending trend, even when the contour lines are raised over
100~\mujybm. The slight bending impression is mainly guided by outer
contours, so it might come from a non-circular restoring beam and the
presence of any unresolved asymmetric structures. To investigate the
fidelity of observing a bent jet, we use a {\em flip-subtraction}
method to check the collimation of a jet (Section~\ref{sec:flip}). The
result shows the jet bending is merely marginal at the first contour
level. \citet{metcalf.02.apj} has elaborately shown that this subtle
deflection of the jet can be a strong indication of the existence of local
substructures. Undoubtedly, it is possible to add a local compact
perturber to account for the local image morphology without changing
the overall lensed image properties. Recently \citet{asadi.20.mn} has
also confirmed that the jet bending in image B is non-significant with
higher-frequency (22~GHz and 8.4~GHz) \emph{VLBI} observations.

\subsubsection{X-band global-\emph{VLBI} observation}\label{sec:8ghz}

\begin{figure*}
\centering
\includegraphics[width=7cm]{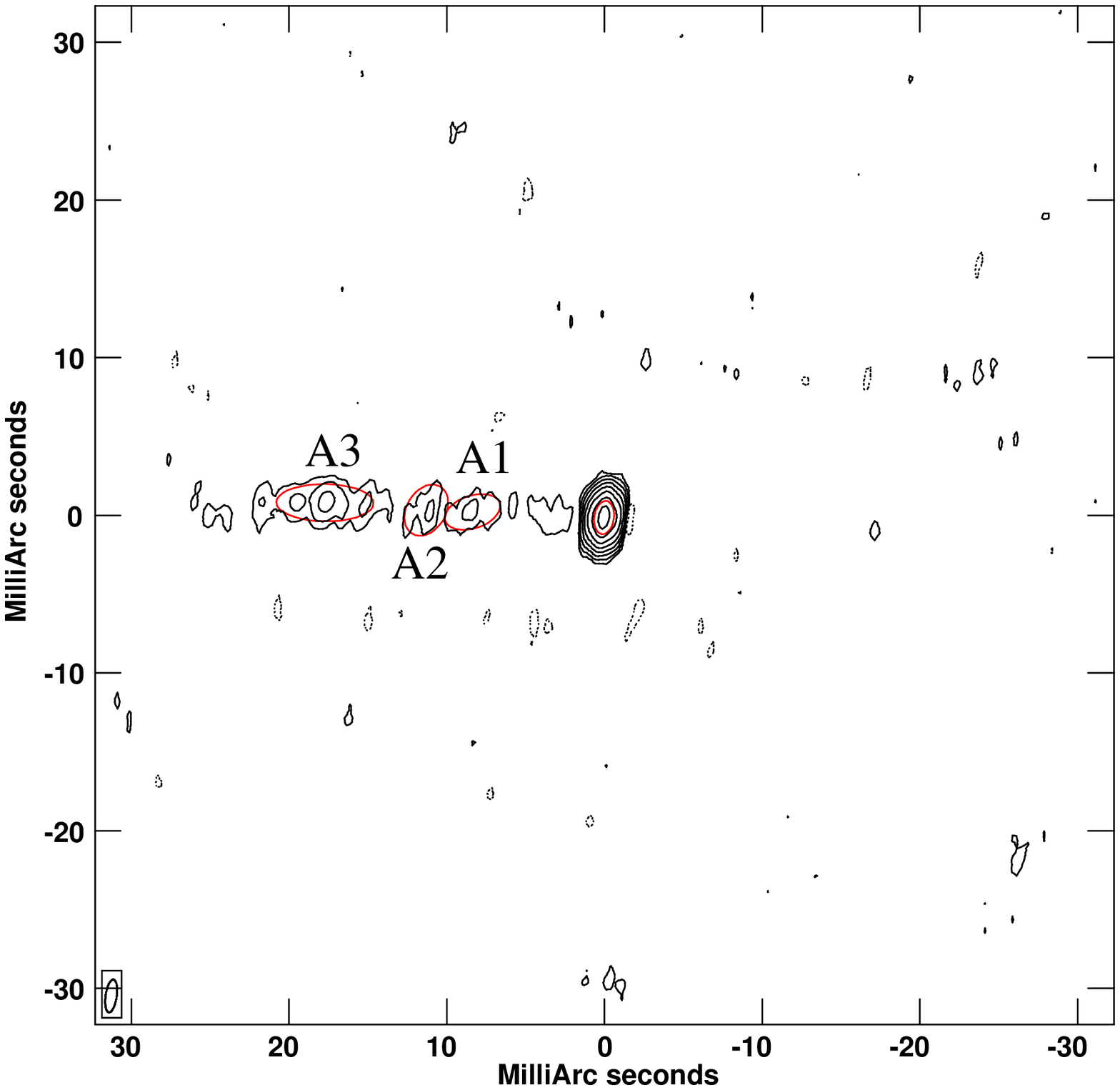}
\includegraphics[width=7cm]{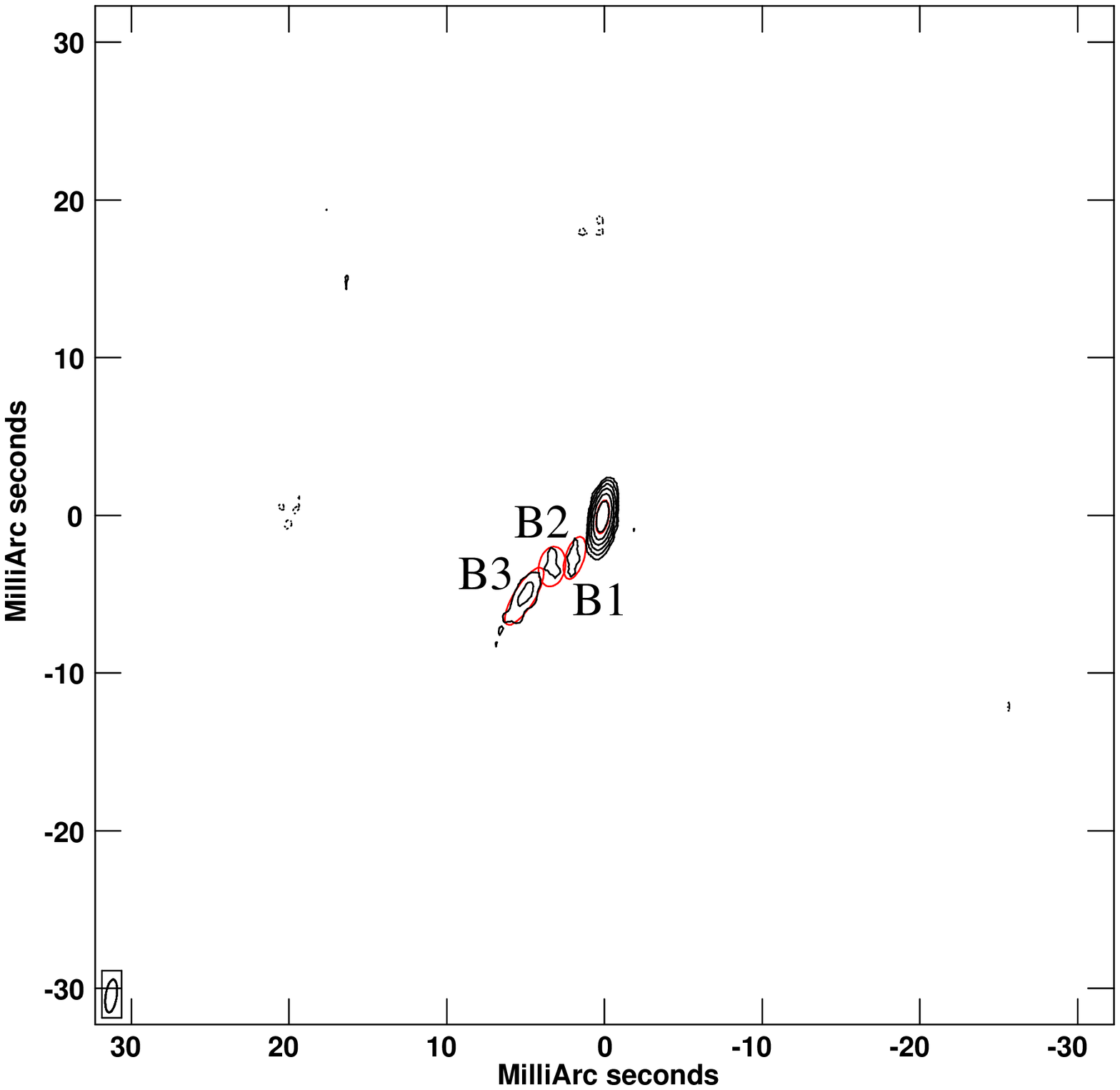}
\caption[X-band \emph{VLBI} observations]{ 8.4-GHz \emph{VLBI}
  observations of B1152+199. The restoring beam size is
  2.1$\times$0.7~mas$^2$ at a position angle of $172^\circ$. Contours
  in the map are plotted at multiples of -1, 1, 2, 4, 8, 16, 32, 64,
  128, 256, 512, 1024 $\times$ $3\sigma$ where $\sigma$ is the local
  RMS noise ($\sim$34~\mujybm). The red circles indicate the fitted
  image models of individual components. The jet components are
  sequentially labelled outward from each core as A1, A2, A3, and B1,
  B2, B3 respectively. }
\label{fig:8ghzimg}
\end{figure*}

\begin{figure*}
\centering
\includegraphics[height=6cm]{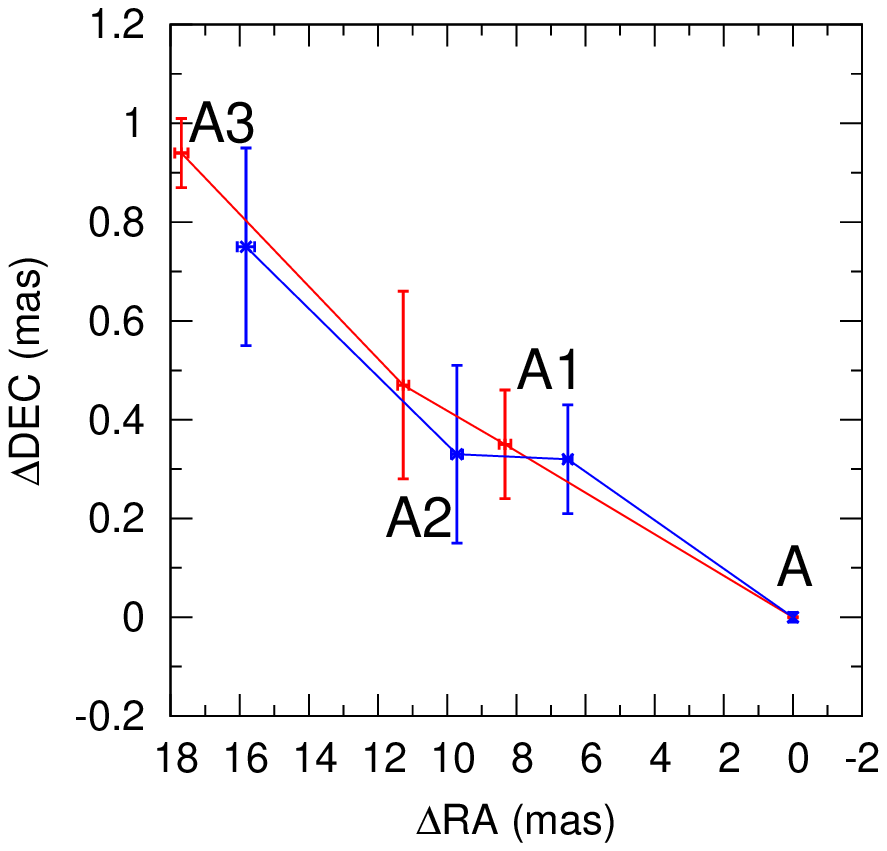}
\includegraphics[height=6cm]{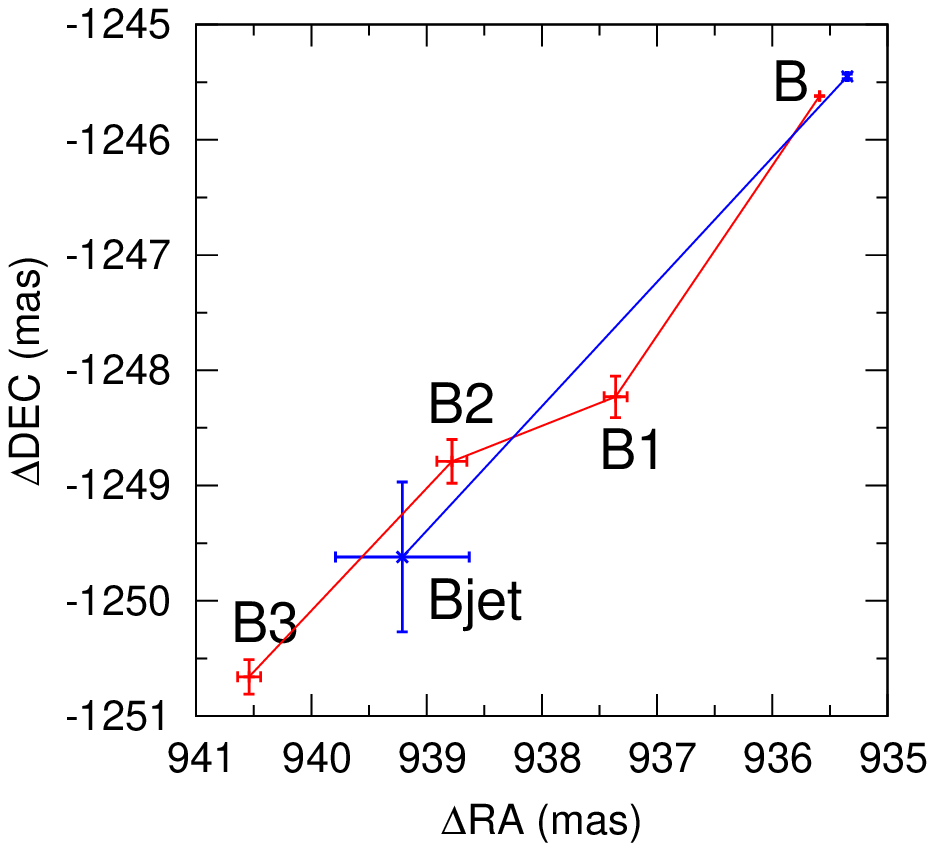}
\caption[C/X-band image positions]{ Jet component positions obtained
  from 5-GHz \emph{HSA} and 8.4-GHz \emph{VLBI} observations. The
  blue dots are measured from the 5-GHz data and the red dots are
  measured from the 8.4-GHz data. The dots are connected by
  polylines to indicate the direction of the jet stream. }
\label{fig:cxobs}
\end{figure*}

The 8.4-GHz global-\emph{VLBI} observation (GA036, PI: Asadi) of
B1152+199 from 2015 was carried out by \citet{asadi.20.mn} . It was a
full-track observation with 22 telescopes from both \emph{EVN} and
\emph{VLBA}. The effective on-source integration time of GA036 was
around 11 hours. In the original paper of \citet{asadi.20.mn}, the
radio jet in image B was resolved into 3 discrete components, which
match to the jet components in image A. However, the authors only used
the image for morphological comparison between the curvature
translations in the A and B images. To extract the observables from
each component as mass modelling constraints, we re-reduced the data
and produced maps of the A and B images from the archived data. The
data came out from the correlator as two passes of pipeline-calibrated
$uv$ data. We found the phase centers of the two passes offset and
used {\sc aips} task {\sc uvfix} to rectify it and combined the two
data set together with task {\sc dbcon}. Multiple rounds of
self-calibrations of solution intervals from two hours down to one
minute were applied to solve the phase error and achieve the highest
dynamic range. The self-calibration task we used is {\sc scimg}. The
uniform weighting leads to a mottled feature as seen in
\citet{asadi.20.mn}, so we use the Briggs weighting scheme with a
robustness parameter equal to 4~\citep{briggs.95.phd}. The images were
restored with a 2.1$\times$0.7~mas$^2$ beam and the background noise
is {\sc clean}ed down to 29~\mujybm. The {\sc clean}ed images are
shown in Fig.~\ref{fig:8ghzimg} and the observables measured from
image-plane model-fitting with a 2-dimensional Gaussian are shown in
Table~\ref{tab:cxobs}. The {\sc clean}ing and image-plane
model-fitting are carried out with the {\sc aips} task {\sc imagr} and
{\sc jmfit}. We can notice from Fig.~\ref{fig:cxobs} that the jet
position angle seems not changed over a decade.

\begin{table}
   \centering
   \caption[C/X-band observables]{ The observable constraints 
     of B1152+199 from the 5-GHz \emph{HSA} and 8.4-GHz \emph{VLBI} observations. 
     The components are fitted by a 2-dimensional Gaussian model in the image plane. The flux
     is measured as the integral intensity of the component. A and B are the core
     images, A1, A2, A3 and B1, B2, B3 are the model-fitted jet components in Fig.~\ref{fig:8ghzimg}, while B$_{jet}$ denotes the unresolved jet component in Fig.~\ref{fig:5ghz3epo}. }
   \label{tab:cxobs}
   \begin{tabular}{cccc}
    \hline
    HSA  &  Flux  &    RA offset    &    DEC offset   \\
   (5~GHz)  &    (mJy)  &      (mas)      &       (mas)     \\
    \hline
    A     &  33.41 $\pm$ 0.20 &  0.00 $\pm$ 0.01 & 0.00 $\pm$ 0.01  \\
    A1    &   3.91 $\pm$ 0.41 &   6.51 $\pm$ 0.12 & 0.32 $\pm$ 0.11  \\
    A2    &   1.92 $\pm$ 0.56 &   9.72 $\pm$ 0.16 & 0.33 $\pm$ 0.18  \\
    A3    &   8.73 $\pm$ 1.15 &  15.82 $\pm$ 0.25 & 0.75 $\pm$ 0.20  \\
    B     &  11.91 $\pm$ 0.21 & 935.35 $\pm$ 0.01 & -1245.45 $\pm$ 0.02  \\
    B$_{jet}$ & 5.07 $\pm$ 2.01 & 939.21 $\pm$ 0.58 & -1249.62 $\pm$ 0.65  \\
    \hline
   g-VLBI &   Flux  &    RA offset    &    DEC offset   \\
   (8.4~GHz)  &    (mJy)  &      (mas)      &       (mas)     \\
    \hline
    A  &  33.95 $\pm$ 0.07 &  0.000 $\pm$ 0.001 & 0.000 $\pm$ 0.001  \\
    A1 &   1.24 $\pm$ 0.16 &  8.33 $\pm$ 0.17 & 0.35 $\pm$ 0.11  \\
    A2 &   1.14 $\pm$ 0.18 & 11.27 $\pm$ 0.16 & 0.47 $\pm$ 0.19  \\
    A3 &   3.63 $\pm$ 0.29 & 17.69 $\pm$ 0.19 & 0.94 $\pm$ 0.07  \\
    B  &  10.78 $\pm$ 0.07 & 935.590 $\pm$ 0.001 & -1245.620 $\pm$ 0.003  \\
    B1 &   0.48 $\pm$ 0.11 & 937.36 $\pm$ 0.10 & -1248.23 $\pm$ 0.18  \\
    B2 &   0.58 $\pm$ 0.13 & 938.78 $\pm$ 0.13 & -1248.79 $\pm$ 0.19  \\
    B3 &   1.31 $\pm$ 0.16 & 940.54 $\pm$ 0.10 & -1250.66 $\pm$ 0.15  \\
    \hline
   \end{tabular}
\end{table}

\section{Lens mass modelling}

\subsection{Modelling of \emph{VLBI} data}\label{sec:parmod}

A typical softened power-law ellipsoid (\emph{SPLE}) model with
elliptical symmetry can be written as a two-dimensional convergence:
\begin{equation}
    \kappa(x,y) = \frac{b^{2-\alpha}}{2(s^2+x^2+y^2/q^2)^{1-\alpha/2}},
\label{eq:siequ}
\end{equation}
where $\alpha$ is the power-law index, and $\alpha=1$ for the
isothermal case; $b$ is the mass scaling factor, i.e. the so-called
Einstein radius for circular symmetry; $s$ is the core size (for
\emph{SIE} $s$=0); and $q$ is the axial ratio. It has been clearly
shown from the optical survey from the Sloan Lens Advanced Camera for
Survey (\emph{SLACS})~\citep{koopmans.06.apj}, the lensing galaxy most
likely has an isothermal mass profile. Besides, the core parameter is
proved to be insensitive to outer image constraints of a dual-image
lens~\citep{zhang.07.mn}. Thus, in our modelling, we fixed $\alpha=1$
and $s=0$ to check in detail to see how the positional observables
gradually constrained the primary singular isothermal ellipsoid model.

The \emph{HST/WFPC2} $I$-band and \emph{HST/NICMOS} $H$-band image
have both already shown the presence of a lensing galaxy as well as
the perturber galaxy X. As such, we defined an object function with
complete constraints from observables, including the discrepancies
between the observed and predicted image positions and fluxes, the
galaxy positions and the mass ratio of the perturber galaxy to the
host galaxy, to minimize in the image plane as:

\begin{equation}
\begin{aligned}
    \chi^2 = & \sum _{i=A,B,A_1,B_1,...}\left[ \frac{(x'_i-x_i)^2}{\sigma^2_{xi}} + \frac{(y'_i-y_i)^2}{\sigma^2_{yi}} + \frac{(f'_i-f_i)^2}{\sigma^2_{fi}} \right] \\
   & + \sum_{g=G,X}\left[ \frac{(x'_{g}-x_{g})^2}{\sigma^2_{xg}} + \frac{(y'_{g}-y_{g})^2}{\sigma^2_{yg}} \right] \\
   & + \frac{(b_1/b_2-r_b)^2}{\sigma^2_{r_b}},
\end{aligned}
\label{eq:chisum}
\end{equation}
where the observed image positions and fluxes are noted as ($x'_i$,
$y'_i$, $f'_i$), and the observed galaxy positions are noted as
($x'_g$, $y'_g$); all the modelled counterparts are noted without the
prime symbol. The mass ratio derived from photometry is also used as a
constraint, which is reflected in the total $\chi^2$. The ratio
between the mass scales of the lensing galaxy ($b_1$) and the X-galaxy
($b_2$) is noted as $r_b$. The denominators are the observational
errors, noted as $\sigma$ with a subscript corresponding to their
nominators respectively. The $\chi^2$ in the image plane explicitly
defines the variance between the observational and theoretical
data. The constraints from the multiple jet components are logically
complete. So other lensing properties like image parities, jet
curvatures and position angles can be derived from the alignment of
the jet components. We use the direct observables rather than
  the derived constraints in our study because the derived
constraints can have degeneracy and introduce
ambiguity. For example, the same positional angle can be formed with a
reverse alignment of the components which have opposite parity. If
we consider the number of degrees of freedom (\emph{NDF}
\footnote{The \emph{NDF} calculation: \enskip $NDF=N_{Cons}-N_{Para}$ \\
  Para: \enskip \undertext{$b$ \, ($e$, $\phi_e$) \, ($\gamma$, $\phi_\gamma$)}{$\times N_{lens}$} \undertext{($x_s$, $y_s$, $f_s$)}{$\times N_{src}$} \\
  Cons: \enskip \undertext{($x_i$, $y_i$, $f_i$)}{$\times N_{img}\times N_{src}$} \undertext{($x_g$, $y_g$)}{$\times N_{lens}$} \undertext{$r_b$}{$\times (N_{lens}-1)$} \\
  N.B. the lensing galaxy positions are considered as observable
  constraints in this paper.}=$\nu$), the reduced $\chi^2$
($\bar{\chi}^2$=$\chi^2/\nu$) is approaching 1, if the extent of
the match between observations and estimates is in accord with the
error variance.

\begin{figure*}
\centering
\includegraphics[width=7cm]{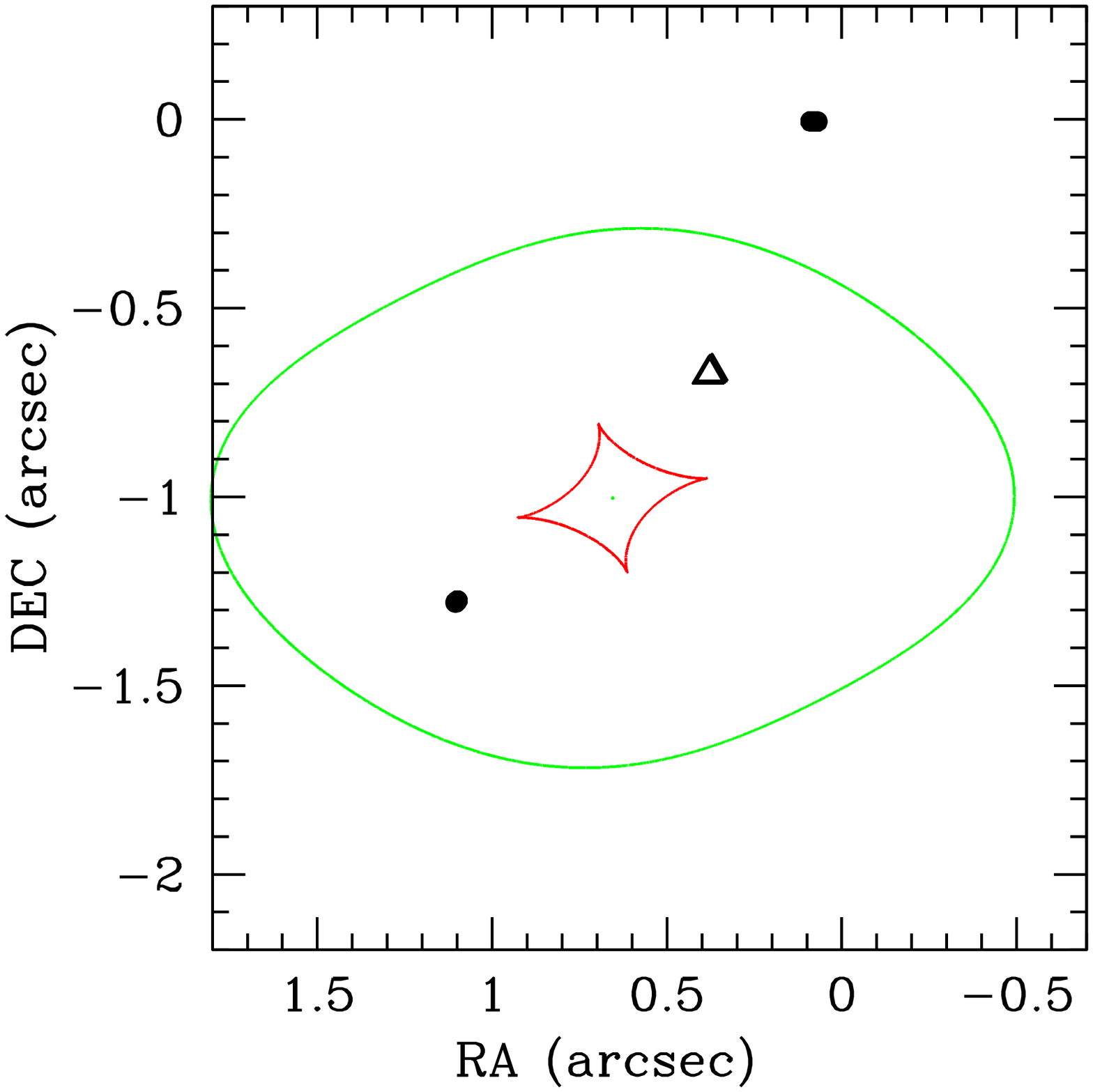}
\includegraphics[width=7cm]{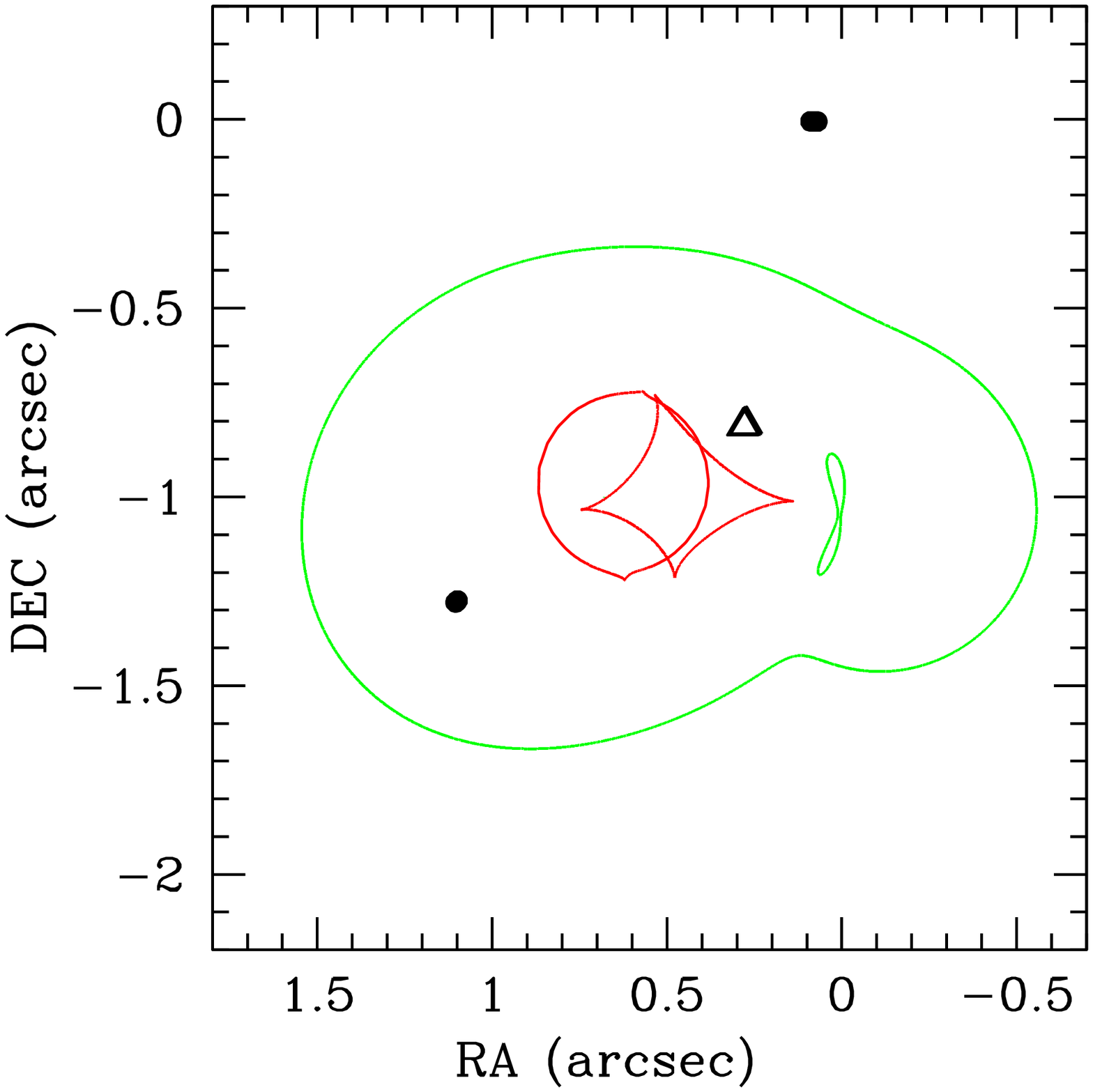}
\caption[Critical lines of power-law models]{ The critical curves and
  caustics from the power-law mass modelling results. The left panel
  is for the power-law ellipsoid only, the right panel is for the
  power-law ellipsoid with an \emph{SIS} X-galaxy perturber. Green
  lines are the critical curves, and red lines are the caustics. Black
  dots and triangles denote the images and sources respectively. }
\label{fig:powcrit}
\end{figure*}

\begin{figure*}
   \centering
   \includegraphics[width=5.5cm]{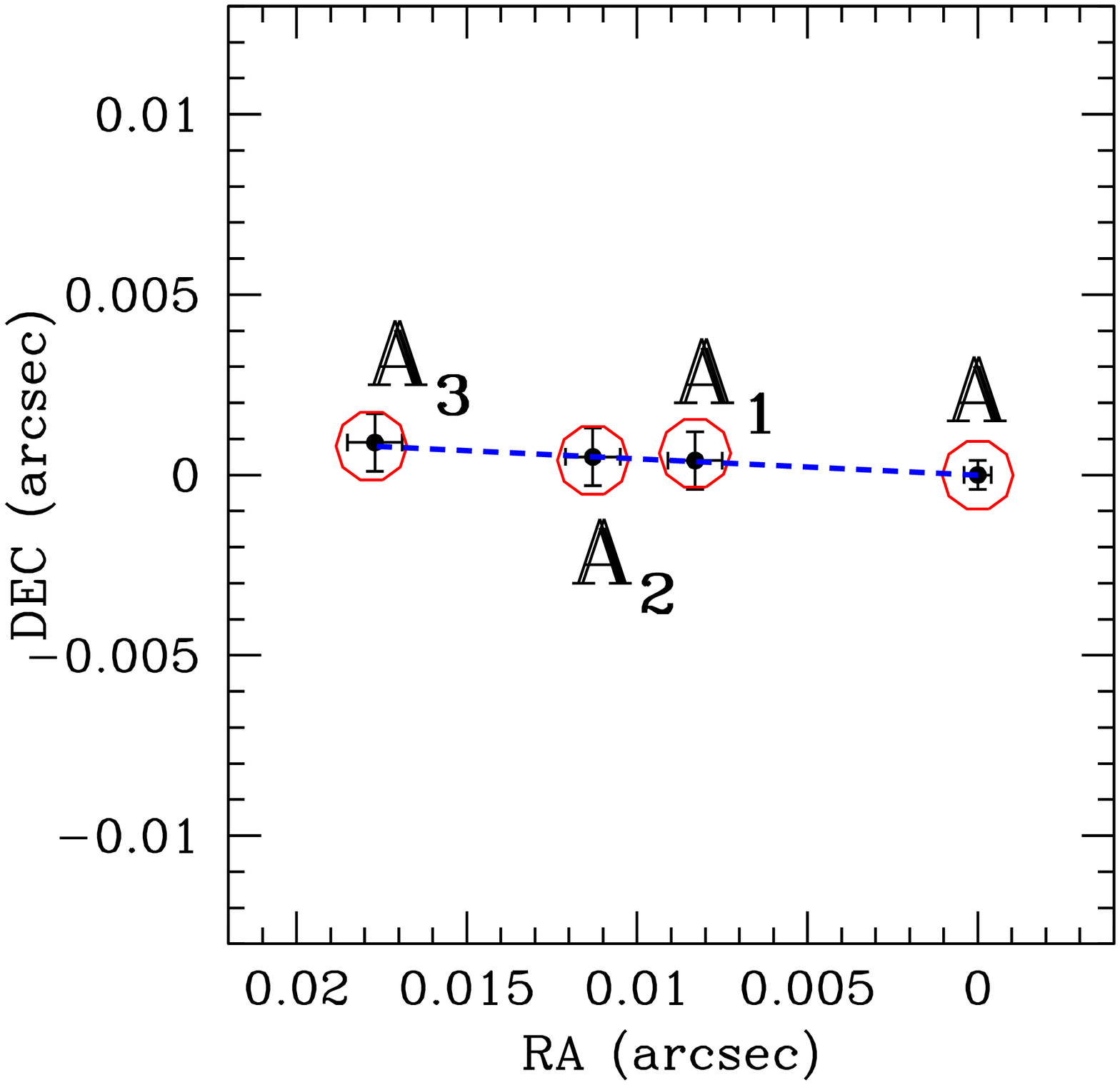}
   \includegraphics[width=5.5cm]{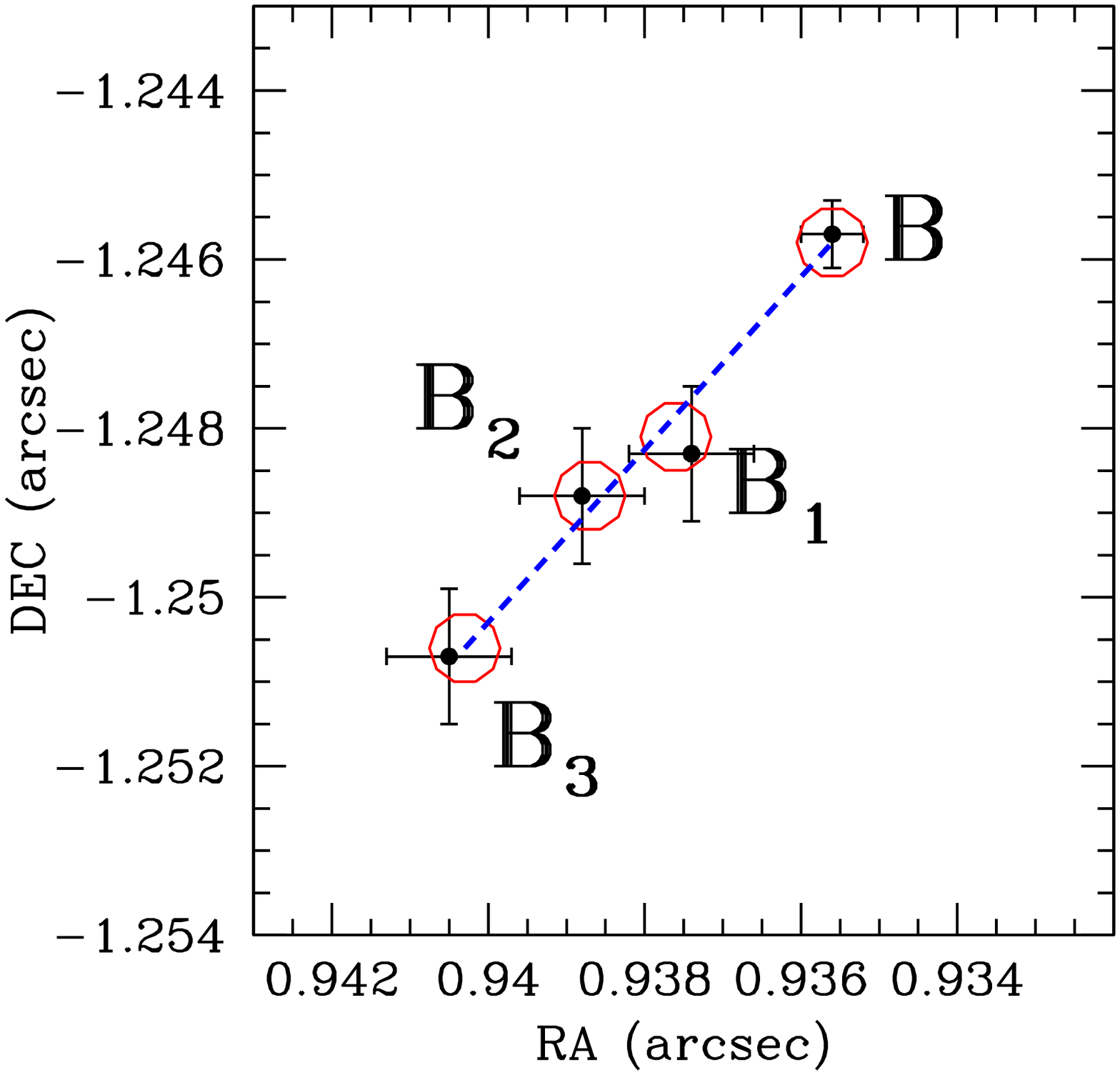}
   \includegraphics[width=5.5cm]{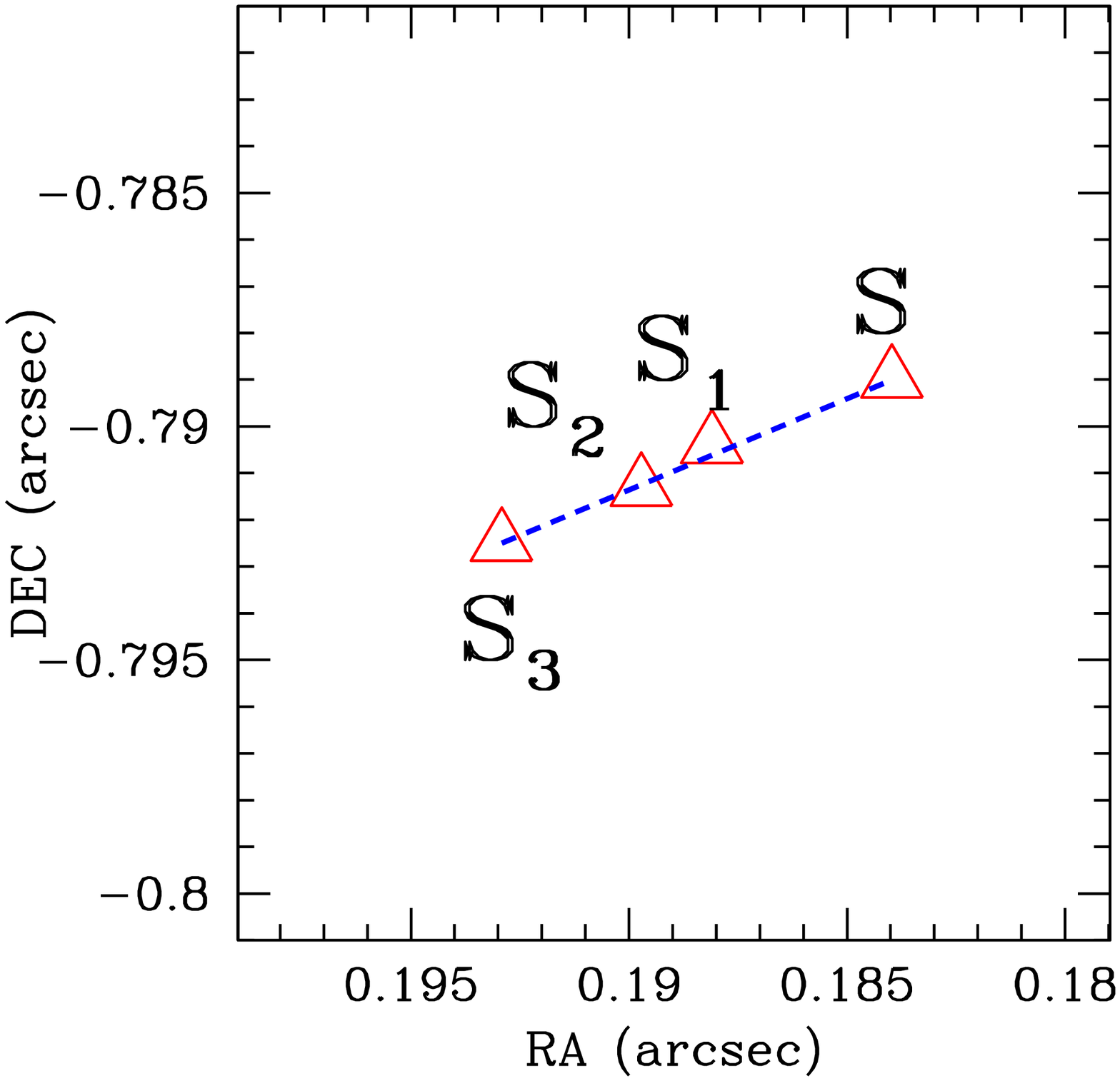}
   \caption[SPLE+X model results of 8GHz AB3]{Predicted image and
     source positions from the optimal \emph{SPLE+SIS} model with
     8.4-GHz \emph{VLBI} multi-component constraints. The observed
     image positions are denoted as black dots with isotropic error
     bars, while the predicted image positions are denoted as red
     circles. The source positions are denoted as red triangles. The
     blue dashed lines indicate the collimated alignment of the
     modelled components in the image plane and the source plane. The
     error bars have been magnified 4 times for visual convenience.}
   \label{fig:splex}
\end{figure*}

The X-band global-VLBI observation currently provides the highest
resolution of this lens system. As the jet components in both image A
and image B were resolved at 8.4~GHz, it offers very strong positional
and flux constraints to the mass model.  The modelling was carried out
with the comprehensive modelling program {\sc gravlens} a.k.a. {\sc
  lensmodel} developed by \citet{keeton.01.astro}. Since the radially
aligned jet components in image B are sensitive to the radial mass
profile, we have tried to optimize the mass model using \emph{SIE} and
\emph{SPLE} plus external shear, with and without a second perturber
respectively. The optimized model parameters are shown in
Table~\ref{tab:powmod}. We can notice that, the presence of the
X-galaxy as a second perturber does give a better fit, while if the
slope of the power-law model is allowed to vary, the fit improves but
not significantly. The optimal power-law profiles indicate the mass
distribution is steeper than isothermal. With the presence of a second
perturber, the profile can be steeper, as the mass distribution
is sort of compensated by the perturber. The critical curves and
caustics of the optimized power-law models are shown in
Fig.~\ref{fig:powcrit}. We can see from the curve configurations that
there are degeneracies between ellipticity, external shear and the
present mass of the X-galaxy. As the goodness of fit indicated, the
X-galaxy is needed to fit the constraints from the three resolved
jet components in the 8.4-GHz \emph{VLBI} images. The best fit is
given by a power-law model with an \emph{SIS} perturber, with a
reduced $\chi^2$ close to unity. The predicted image and source
configurations are shown in Fig.~\ref{fig:splex}. As we can see, all
modelled jet components are aligned linearly very well, except the
B$_1$ component has a slight offset from the alignment.

\begin{table}
   \centering
   \caption[Table of the power-law models]{The optimized parameters of
     the isothermal and power-law mass models. 
     \emph{SPLE} is the softened power-law ellipsoid. 
     \emph{SIS} is the singular isothermal spheroid. 
     $b'$ is the modified mass scale (refer to~\ref{sec:trans}). 
     $(x,y)$ is the mass profile centroid, while the subscripts 
     $g$ and $X$ denote the host galaxy and the X-galaxy respectively. 
     $e$ is the ellipticity and $\phi_e$ is its position angle.
     $\gamma$ is the external shear and $\phi_\gamma$ is its position angle. 
     $\alpha$ is the power-law slope. The \emph{NDF} is noted as $\nu$. }
   \label{tab:powmod}
   \begin{tabular}{lll} 
    \hline
    SIE + $\gamma$ \\
    \hline
    $b'$ = 0.7536    & & $\chi^2$ = 21.73 \\
    $x_g$ = 0.5396  & & $\nu$ = 10\\
    $y_g$ = -0.9645 \\
    $e$ = 0.4596    \\
    $\phi_e$ = 84.48 \\
    $\gamma$ = 0.1238 \\
    $\phi_\gamma$ = 5.9306 \\
    \hline
    SIE + SIS + $\gamma$ \\
    \hline
    $b'$ = 0.5516    & $b'_X$ = 0.2518  & $\chi^2$ = 13.87 \\
    $x_g$ = 0.5609  & $x_X$ = -0.0564  & $\nu$ = 12 \\
    $y_g$ = -0.9752 & $y_X$ = -1.0284  & \\
    $e$ = 0.4513    \\
    $\phi_e$ = 58.64 \\
    $\gamma$ = 0.0826 \\
    $\phi_\gamma$ = -45.02 \\
    \hline
    SPLE + $\gamma$ \\
    \hline
    $b'$ = 0.7530    & & $\chi^2$ = 11.68 \\
    $x_g$ = 0.5321  & & $\nu$ = 9\\
    $y_g$ = -0.9792 \\
    $e$ = 0.5006    \\
    $\phi_e$ = -87.53 \\
    $\gamma$ = 0.1012 \\
    $\phi_\gamma$ = 20.18 \\
    $\alpha$ = 0.8031 \\
    \hline
    SPLE + SIS + $\gamma$ \\
    \hline
    $b'$ = 0.6136    & $b'_X$ = 0.2328  & $\chi^2$ = 11.26 \\
    $x_g$ = 0.5538  & $x_X$ = -0.0572  & $\nu$ = 11 \\
    $y_g$ = -0.9767 & $y_X$ = -1.0260  & \\
    $e$ = 0.2410    \\
    $\phi_e$ = 46.26 \\
    $\gamma$ = 0.1000 \\
    $\phi_\gamma$ = -83.44 \\
    $\alpha$ = 0.6716 \\
    \hline
   \end{tabular}
\end{table}

\begin{figure*}
   \centering
   \includegraphics[width=\textwidth]{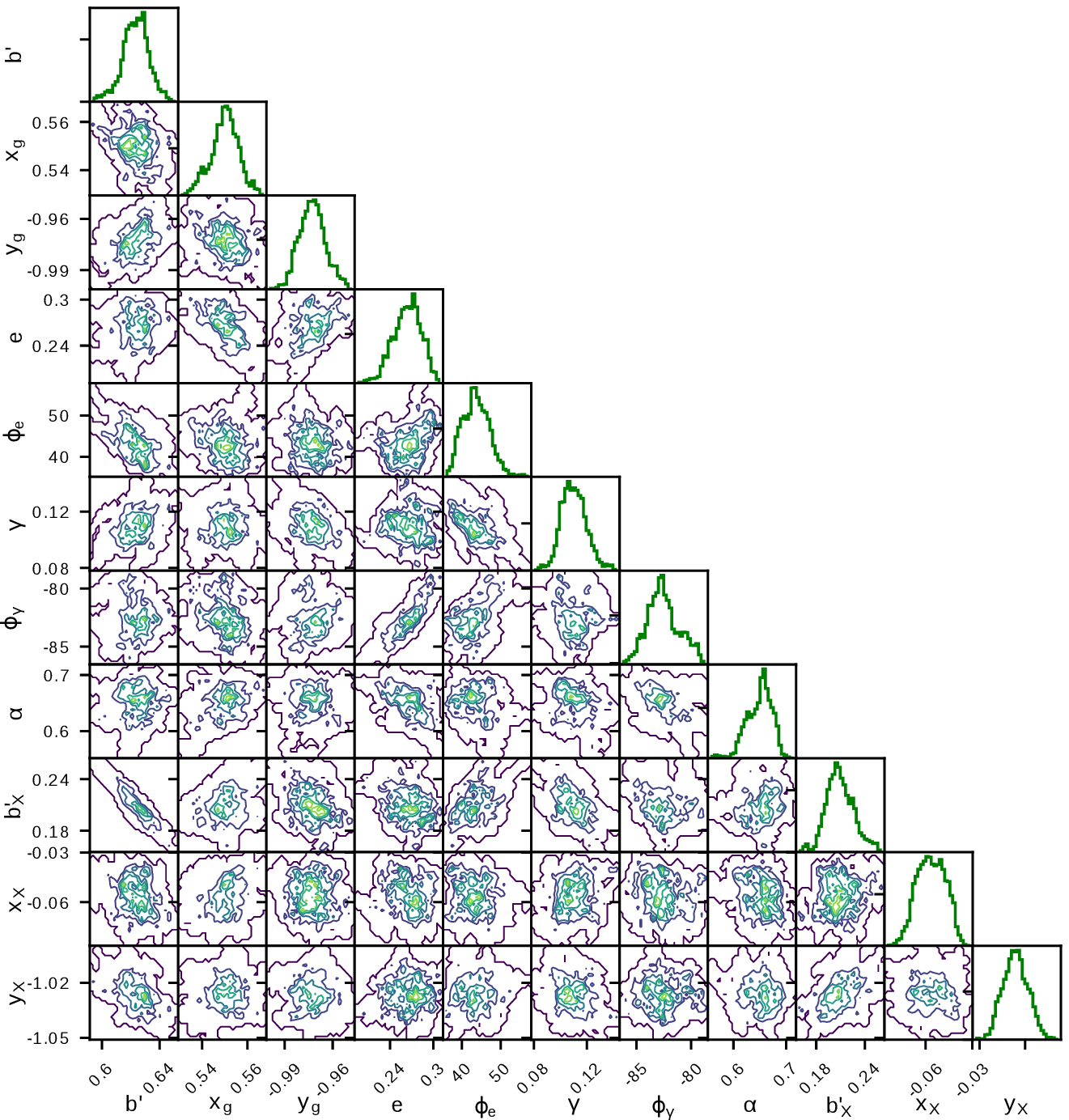}
   \caption[MCMC trials of lens parameters]{ Lens parameter estimation
     with \emph{MCMC} realizations for the optimal \emph{SPLE} with
     the \emph{SIS} X-galaxy perturber. The histograms lying on the
     diagonal show the marginalized posterior distributions of each
     lens parameter. The two-dimensional plots show the joint
     posterior densities of all parameter pairs. The contours are
     binned into five intervals. The parameter notations are the same
     as those in Table~\ref{tab:powmod}. }
   \label{fig:mcpox}
\end{figure*}

To estimate the uncertainty of optimized parameters, we deploy a
Markov chain Monte Carlo (\emph{MCMC}) realization with 13 adaptive
random walks and a maximum step of 10000 through the multi-parameter
space. Uniform priors are used for input uncertainties of all
observables. The multivariate probability distributions of the lens
parameters for the optimal SPLE+SIS+$\gamma$\footnote{To avoid the
  symbol overload confusion, the external shear and power-law index
  use the slant ($\gamma$) and upright ($\upgamma$) Greek letters
  respectively.} model are shown in Fig.~\ref{fig:mcpox}. The
marginalized posterior distributions of lens parameters are
illustrated as the histograms along the diagonal. Other panels show
the joint posterior densities of all coupled parameters and resemble
the realistic covariances.

\subsection{Modelling of \emph{HST} data}\label{sec:tromy}

The photometric measurements showed a significant existence of the
X-galaxy. The derived photometric mass of this companion galaxy is
nearly half of the primary lens' mass and its impact on the lens
system should be checked with lens modelling. Since the discrete
blobs and the arc-shape emission are detected in the \emph{WFPC2}
$I$-band and \emph{NICMOS} $H$-band images respectively, we need the
mass modelling to check its consistency with a dual-lens system. 

We carried out the modelling with the {\sc lenstronomy} program, by
using the entire emission from the extended images as constraints. It
is a multi-purpose Python package to model strong gravitational lenses
which is presented in \citet{birrier.18.pdu} and is based on
\citet{birrier.15.apj}. It does light profile decomposition onto a
shapelet basis rather than direct pixelation, which gives it advantages
to be used in describing an arbitrary profile within a limited parameter
space. Since every pixel in the image is helping to constrain the
model, we can investigate a bigger parameter space. 

So we did combinatorial modelling trials, by assuming the lens mass
model of the host and the companion X-galaxy as a model pair of any
two of the following mass model pairs: (\emph{SIE}, \emph{SIS}),
(\emph{SIE}, \emph{SIE}), (\emph{SPEP}, \emph{SPEP}) and (\emph{SPEMD},
\emph{SPEMD}), where \emph{SIS} denotes a singular isothermal spheroid
mass, \emph{SIE} denotes a singular isothermal ellipsoid mass,
\emph{SPEP} denotes a softened power-law potential, and \emph{SPEMD}
denotes a smooth power-law mass density. The softening core or
smoothing scale was set to approach zero to act as the singular
case. All trials were carried out both with external shear and without
it. The central position and ellipticity of a mass profile and its
light profile are set to be consistent during optimization. The
power-law potential and mass density profiles were also tested with
varying slopes. Since the source must have a non-delta light
profile to contribute the extended emissions in lens modelling, in the
optimization, the source profiles were set to S\'{e}rsic
ellipses. Besides, the lensed images were used as position constraints
but they were assigned S\'{e}rsic profiles to optimize the light
modelling. However, we found it difficult to converge to the same
optimal parameters as obtained from the \emph{VLBI} data modelling with
{\sc gravlens}. This is because the light profiles of the lenses and
source need to be optimized simultaneously in {\sc lenstronomy}, the
increased parameter space leads to non-unique solutions especially
when the model constraining strength is weak.

\begin{figure*}
	\includegraphics[width=0.98\textwidth,left]{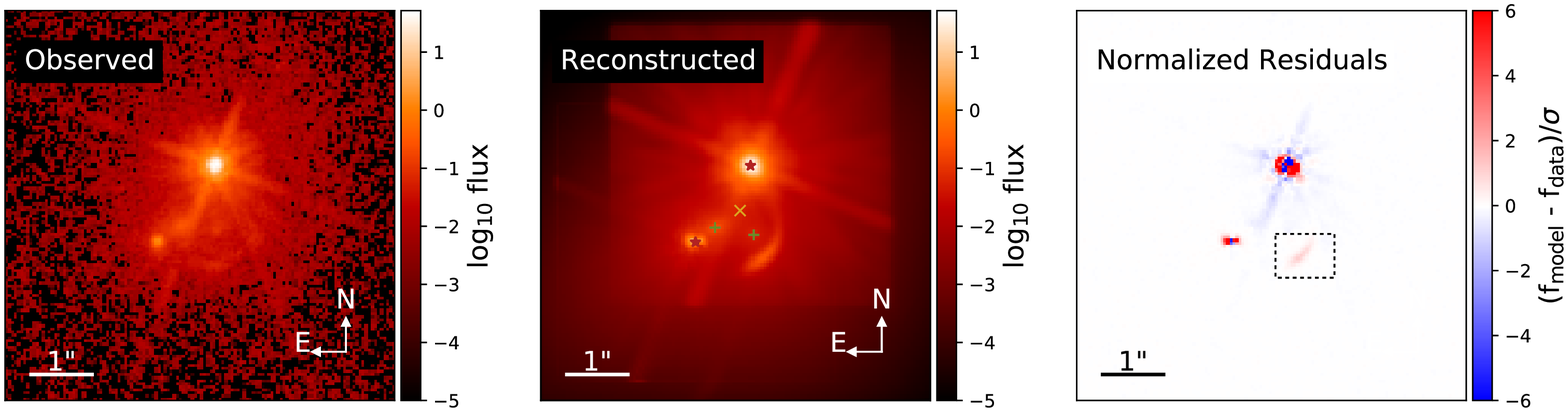}\\
	\includegraphics[width=0.98\textwidth,left]{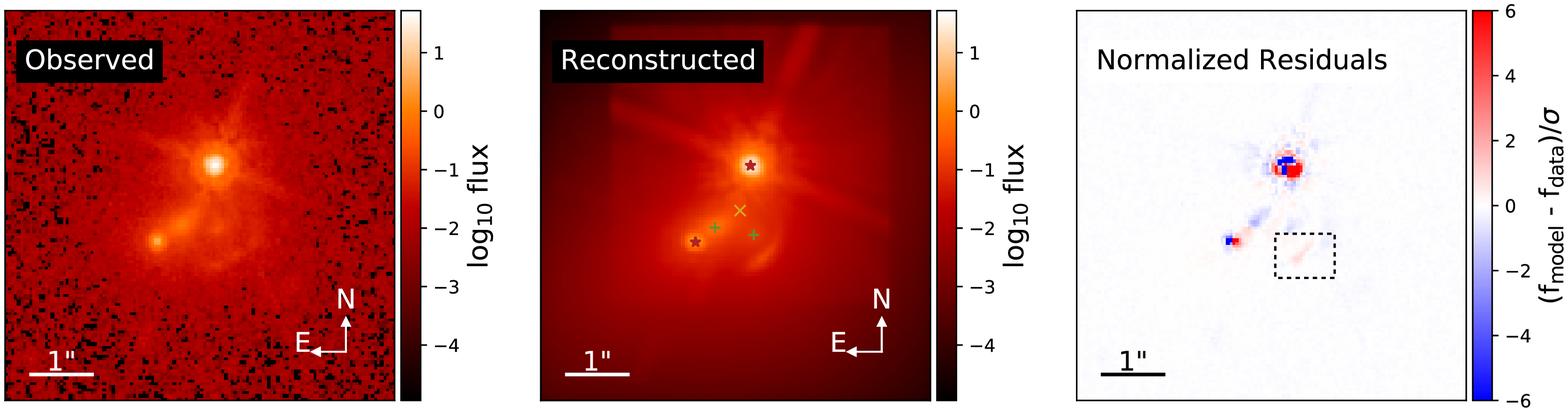}\\
	\includegraphics[width=0.98\textwidth,left]{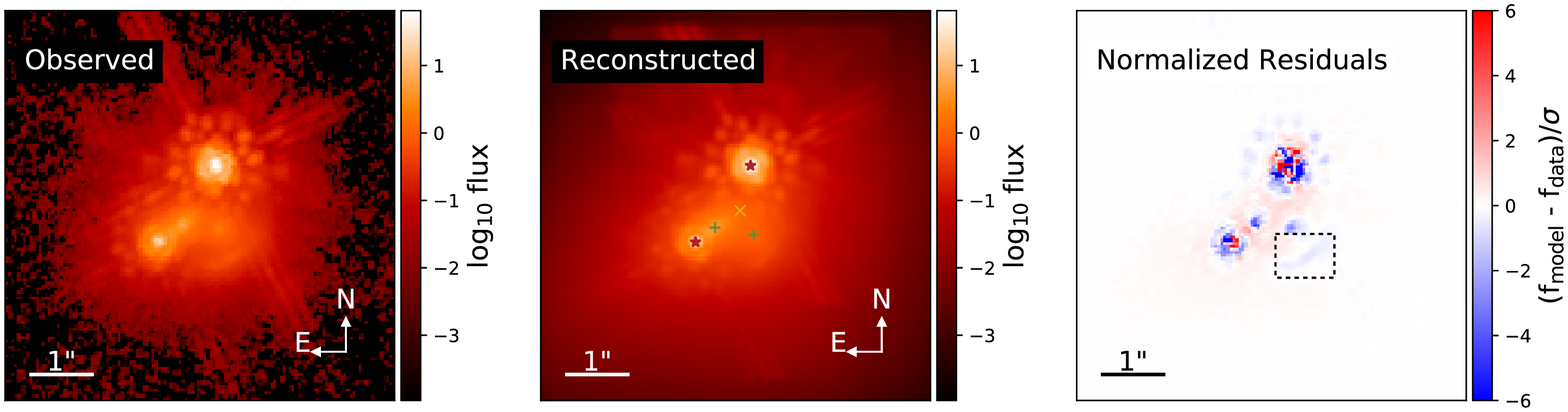}\\
	\includegraphics[width=\textwidth,left]{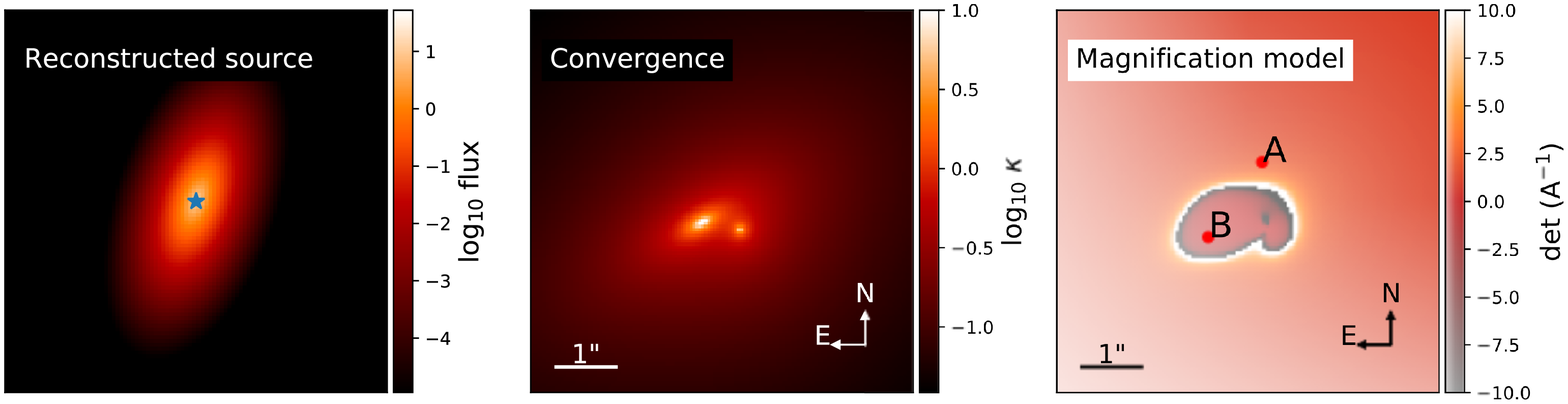}
        \caption[Image model subtraction 1]{ The optimal models of the
          image, sources, lensing convergence and magnification. The
          top three rows are for the \emph{HST} $V$-band, $I$-band and
          $H$-band data, respectively. The bottom two rows show the
          source, mass and light models. In the reconstructed image
          panel, the red stars, green pluses and yellow crosses denote
          the observed image positions, the modelled lens positions
          and source positions respectively. In the reconstructed
          source panel, the blue stars denote the source light centroid
          positions. The dashed boxes indicates the area used for arc
          residual statistics.  }
    \label{fig:tromy1}
\end{figure*}

\begin{figure*}
        \includegraphics[width=0.98\textwidth]{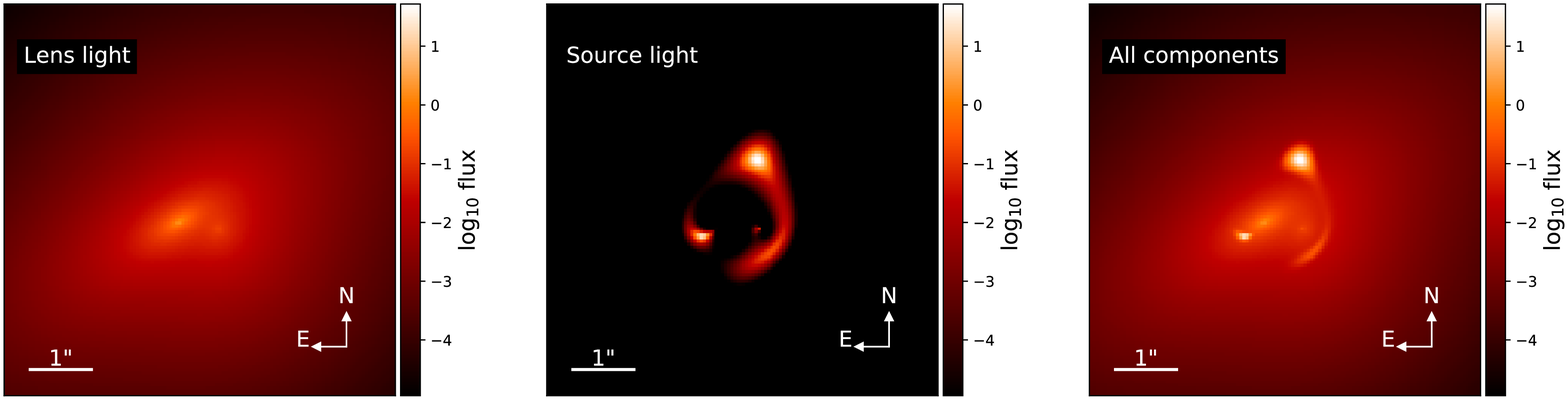} \\
        \vspace{1mm}
        \includegraphics[width=0.98\textwidth]{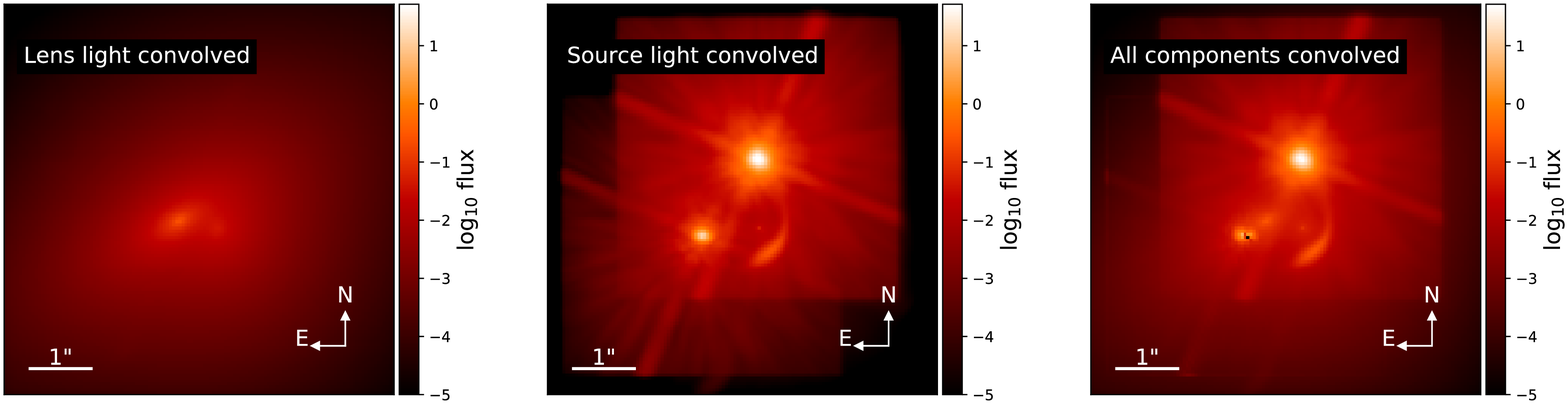} \\
        \vspace{1mm}
        \includegraphics[width=0.98\textwidth]{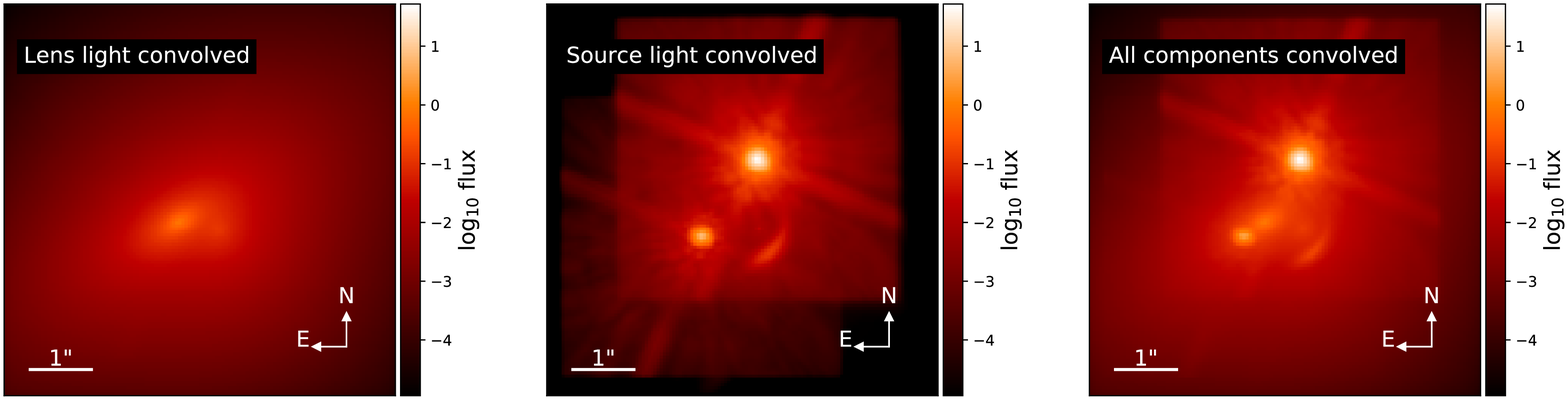} \\
        \vspace{1mm}
        \includegraphics[width=0.98\textwidth]{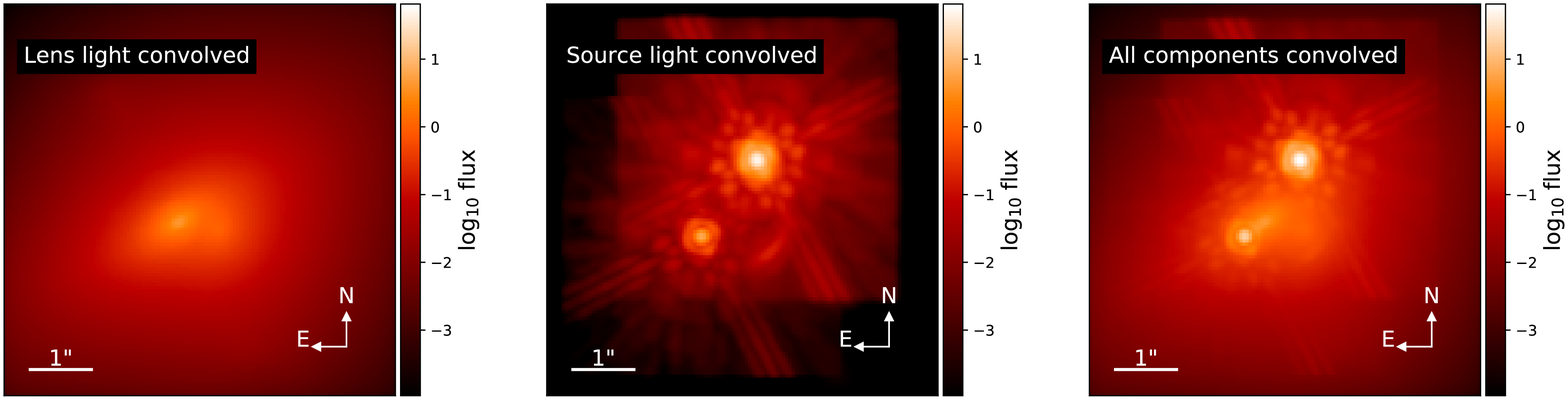}
        \caption[Light model convolution]{ The optimal light models and
          convolved images of the lenses and lensed source. The top
          row shows the light models of the lens and lensed source.
          The bottom three rows are convolved images for the
          \emph{HST} $V$-band, $I$-band and $H$-band data,
          respectively. }
    \label{fig:tromy2}
\end{figure*}

As the lensing objects are the same at the radio and optical bands,
the same lens model must reproduce both observations. So we imported
the optimal {\sc gravlens} model for the \emph{VLBI} 8.4-GHz
observation to {\sc lenstronomy}, and kept the optimized parameters
confined within a certain threshold, to see how similar to the
\emph{HST} images that the positions and fluxes of can be
reproduced. To keep the dimensionality of the parameter space low, we
assumed there is only one background source. The active galactic
nucleus (AGN) contributes to the compact emissions in both radio and
optical bands, while its host galaxy contributes to the extended
emissions in the optical band. All tri-band \emph{HST} data were used
for simultaneous optimization. The optimized results are shown in
Fig.~\ref{fig:tromy1} and \ref{fig:tromy2}, the statistics and
optimized model parameters are listed in Table~\ref{tab:resid} and
Table~\ref{tab:trompar}. Ideally, the reduced $\chi^2$ should be close
to unity. From the reconstructed source light in the image plane, as
shown in Fig.~\ref{fig:tromy2}, we can see the Einstein ring around
the X-galaxy is distorted by the primary lens and the detected diffuse
emission is actually a part of the ring. We can notice that the
optimal lens model slightly overfitted the \emph{HST} $V$- and
$I$-band data. The average residuals over three bands at the targeted
diffuse arc area have been reduced to about 0.7$\sigma$. Taking the
source position as a free parameter, we found that the optimized
source positions with the radio and optical observations are slightly
offset as shown in Table~\ref{tab:srcpos}.

\begin{table}
   \centering
   \caption[Lenstronomy residuals]{ The statistics of the optimal modelling
     with tri-band \emph{HST} data. $\bar{\chi}^2$ is the reduced $\chi^2$. 
     $\sigma$ is the standard deviation of the residual noise and the residuals 
     have been normalized by $\sigma$. The arc box means the area in the dashed
     boxes in Fig.~\ref{fig:tromy1}. }
   \label{tab:resid}
   \begin{tabular}{cccc} 
      \hline
         &  & \multicolumn{2}{c}{Residuals ($\sigma$)} \\
         \cline{3-4}
         \noalign{\vskip 1mm}
         &  $\bar{\chi}^2$  & Full image  & Arc box \\
      \hline    
      $V$-band  &  0.81  &  0.90  & 0.27  \\
      $I$-band  &  0.67  &  0.82  & 0.55  \\
      $H$-band  &  1.02  &  1.01  & 1.25  \\
      \hline
   \end{tabular}
\end{table}

We can see that with a secondary lens - the X-galaxy, the
reconstructed tri-band images can practically resemble arc-shaped
emission at the observed location. This implies the diffuse emission
are highly likely from the lensed background source rather than
intrinsically from the host galaxy, such as a spiral arm.  From
Table~\ref{tab:photpar} and \ref{tab:trompar}, we can see the fitted
S\'{e}rsic index tends to be non-flattish but not significantly far
from a flat index ($n\sim1$) of a disk galaxy over a bulge-dominated
system. Recently \citet{mao.17.nast} detected differential
polarization and Faraday rotation in the two lensed images, which
suggests a fairly significant axisymmetric magnetic field in the host
galaxy. So, the possibility for the non-smooth light structure of the
host galaxy to interfere with the lens system has not totally been
ruled out.

\begin{table}
   \centering
   \caption[Lenstronomy model parameters 1]{ The optimized parameters of 
     lens mass, lens light and source light profiles with \emph{HST} image 
     constraints. $(e_1, e_2)$ are the orthogonal components of the eccentricity
     related to the ellipticity and its position angles as: 
     $e_1=(1-q)\cos 2\phi_q/(1+q)$, $e_2=(1-q)\sin 2\phi_q/(1+q)$. 
     $(\gamma_1, \gamma_2)$ are the orthogonal components of the external shear 
     related to its position angles as: 
     $\gamma_1=\gamma\cos 2\phi_\gamma$, $\gamma_2=\gamma\sin 2\phi_\gamma$. 
     The length unit of $R_e$ and $(x,y)$ is in arcsec and $Amp$ is 
     the amplitude parameter in normalized counts. The coordinates' origin is
     at the image center. $\upgamma$ is the power-law index. }
   \label{tab:trompar}
   \begin{tabular}{ccccccccc} 
      \hline
       Mass   &  $\theta_E$  & $e_1$ & $e_2$  & $x$   & $y$   & $\upgamma$ \\
      \noalign{\vskip 1mm}    
      G   & 0.630 & 0.182 & -0.291 & 0.337 & -0.360 & 2.328 \\
      X   & 0.233 &       &      & -0.299 & -0.480  & \\
           & $\gamma_1$ & $\gamma_2$ \\
      \noalign{\vskip 1mm}
      $\gamma$ & -0.048 & -0.040 \\
      \hline
       Light  & $Amp$ & $R_e$ & $n$ & $e_1$ & $e_2$ & $x$ & $y$ \\
      \noalign{\vskip 1mm} 
      G  &  211.070 & 0.630  & 2.503 & 0.182 & -0.291 &  0.337 & -0.360 \\ 
      X  & 53.138 & 0.716  & 1.530 &       &      & -0.299 & -0.480 \\
      S  & 4.464e4 & 0.047  & 1.095 & -0.274 & -0.209 & -0.074 & -0.082 \\ 
     \hline
   \end{tabular}
\end{table}

\begin{table}
   \centering
   \caption[Predicted source positions]{ The predicted background source 
     position. The coordinates' origin is at the A image core. 
     The errors are projected covariance estimated from \emph{MCMC} trials. }
   \label{tab:srcpos}
   \begin{tabular}{lrr} 
      \hline
         &  Radio (VLBI) & Optical (HST) \\
      \hline    
      $\Delta$RA ($''$)  &  0.184$\pm$0.020 &  0.180$\pm$0.006 \\
      $\Delta$DEC ($''$) & -0.789$\pm$0.012 & -0.709$\pm$0.009 \\
      \hline
   \end{tabular}
\end{table}


\subsection{Comparison between models for optical and radio datasets}

Ideally, the optical models for optical and radio data should converge
to the same point in the parameter space. However, this only applies
when there is a unique minimization of the model on the same
dimensionality of the parameter space and the constraining strength of
the observables. Our combinatorial trials have shown the chance to get
a consistent model independently from radio and optical data is low.

Fitting to the light profiles in \emph{HST} images utilizes
information from all pixels. Besides the lens mass model, the light
profiles of the lens and the source need to be constrained at the same
time. The increased observable constraints are competing with the
expanded parameter space then. Due to the degeneracy between model
parameters, the optimization will face more local minima when the
parameter space increases. Advanced algorithms like particle swarm
optimization (PSO) hardly mitigate this. Intuitively, different image
features imply different optimal models. Unlike the radio emissions,
in a galaxy lens, the optical emissions are easily affected by many
factors, like the blending between foreground and background
emissions, the dust absorptions and even the instrumental
deficiency. As we have noticed, the diffuse emission looks different
in the \emph{HST} $I$-band and $H$-band images. Compared with the
arc-shape emission in the $H$-band image, the discrete blobs in the
$I$-band image give an impression of a quad-image configuration. If we
fit each band individually, it is easy to reach the modelling
inconsistency. Provided with multi-band data, it is better to optimize
the model over the whole bands simultaneously to avoid the discrepancy
between individual bands. Nevertheless, the different image features
at different bands still need to be explained, such as, by noise or
resolution limits. The different appearances of the diffuse emission
in \emph{HST} images are probably due to the instrumentation effect,
as the $I$-band image has a lower signal-to-noise ratio but a higher
resolution than the $H$-band image.

Unlike the \emph{HST} image modelling, the modelling of \emph{VLBI}
data utilize accurate positional and flux constraints from the
resolved jet components in radio images. Given only the mass model of
the lens to be constrained, the \emph{VLBI} data can help to determine
the mass model parameters more precisely with the image-plane
optimization, especially with the multi-image multi-component lens
examples, like B1152+199. Since the lens model is the same, a
practical approach is to use the high-resolution radio observables to
constrain the lens mass model first and feed the optimal model to the
optical optimization to get the optimal light models of the lenses and
source second.

The parametric modelling has shown that the X-galaxy does improve the
modelling, given the tight constraints from the 8.4-GHz global
\emph{VLBI} observation. If the pow-law profile is let to vary, the
optimal mass model of the primary lens tends to have a
steeper-than-\emph{SIE} slope. When the X-galaxy perturber is added,
its mass profile becomes even steeper to compensate for the mass
profile extension due to the presence of the X-galaxy. This is
consistent with the previous investigation done by~\citet{rusin.02.mn}
and \citet{auger.08.mn}.

\section{Discussions}

Here we propose some topics on the lenses, images, sources, mass and
light models with regard to the observations and lens modelling, that
need to be discussed as follows.

\subsection{Contribution of the X-galaxy}

The arc-shaped emission detected in \emph{HST/NICMOS} $H$-band data
conjured up an image of secondarily lensed arc in the lens system. Its
convergence centre coincides with the location of the surrounded
X-galaxy. Our detailed mass modelling with both \emph{HST} $V$-, $I$-
and $H$-band observations have shown that the X-galaxy does contribute
to the optical morphology in the lensed optical images. That means the
diffuse emission in \emph{HST} images is actually the detection of the
lensing effects of the companion galaxy.

As we can see from the optical modelling, the diffuse arc is mainly
contributed by the extended structure from the host galaxy of the
source. So there is no direct lensing effect of the X-galaxy detected
in the radio observation. The lensed images are well resolved and
serval discrete sub-components can be identified in the 8.4-GHz global
\emph{VLBI} observation. So the observables from those three resolved
jet components of this dual-image lens add up to the constraining
strength and make it possible to find the subtle contribution of the
X-galaxy in fitting to the constraints. Our parametric mass modelling
shows that the X-galaxy is required to achieve an optimal fitting to
all the observables, especially when handling the misalignment of the
jet components in image B. If the radial profile is left as a free
parameter, the optimal mass model of the primary lens turns to have a
steeper profile as the presented \emph{SIS} perturber can contribute
part of the extended mass distribution in the whole lens system. This
could be explained merely as a modelling degeneracy if the direct
lensing effects of companion lens have not been detected. In other
words, the up-to-date optical and radio observations testify the role
of the X-galaxy in this lens system is not ignorable.

\subsection{Evidence for a bent jet}

B1152+199 is a particularly controversial lensing system, which first
provided evidence for substructure lensing because of a bending jet in
one of the lensed images and recently was found to not have such
change in the jet direction. The possible bent radio jet in one of the
lensed images was seen as evidence for a local perturbation in the
mass model by CDM substructure~\citep{metcalf.02.apj}. This has
motivated a number of recent observations like \citet{asadi.20.mn}.
The 8.4-GHz global \emph{VLBI} observation has
resolved the sub-components in the lensed images with a position
accuracy down to about 0.1~mas, as seen in Fig.~\ref{fig:8ghzimg} and
\ref{fig:splex}. We may notice that the sub-components line up very
well in each image, except that there is a very slight misalignment
for component B$_1$ in image B. The orthogonal distance between
B$_1$ and the collimation line is about 0.5~mas. The optimized lens
model also reflects this small offset in its predicted image
positions. However, it is arguable whether this small offset is enough to
depict a bent jet. As asserted by \citet{asadi.20.mn}, it exhibits no
apparent milliarcsecond-scale curvature. Nevertheless, our modelling
has shown that the lens model can easily tackle this slight
misalignment within the given measurement errors, as long as the
X-galaxy perturber is considered in the mass model, i.e., no jet
curvature or extra substructure is needed for that. This is different
from \citet{metcalf.02.apj}'s approach. We have provided a toy model
to produce an evidently bent jet with an SIE+SIS+$\gamma$ lensing
configuration in Section~\ref{sec:bender}.

\subsection{Source structures}

As the diffuse emission is considered as a secondary lensed feature,
it is natural to speculate there is a secondary lensed source, as
either independent or attached to the primary one by a certain
structure. Assuming that the background source is an AGN abiding in an
elliptical host galaxy, our optical modelling shows that the extended
structure of the host galaxy spread over the caustics and forms the
lensed arc. The radio observations see the radio emissions from the
AGN and the optical observations see the optical emissions from both
the AGN and the host galaxy. 

There are also other possibilities, such as that the compact radio
quasar just has an extended companion galaxy to contribute to the
diffuse lensed features with itself not being so extended. We should
point out that, either the case of an extended host galaxy or the case
of an extended companion, is merely a projection along the line of
sight. The source components could be in different source planes but
coincidentally align along the line of sight, as a discovered case
by~\citet{gavazzi.08.apj}. We show the modelling of a two-source case
in Section~\ref{sec:twosrc}.

The light centroid of the background source is roughly 80~mas offset
from the position of the radio core as seen in
Table~\ref{tab:srcpos}. This corresponds to a distance scale of
$\sim$625~pc in the source plane\footnote{A flat $\Lambda$CDM
  cosmological model is used ad hoc as $\Omega_m=0.3$,
  $\Omega_\Lambda=0.7$ and $H_0=72$ kms$^{-1}$Mpc$^{-1}$.}. It is not
uncommon to find radio-optical offsets in AGNs~\citep{orosz.13.a&a},
even in the (\emph{CLASS}) lens samples~\citep{skipper.18.mn,
  spingola.20.mn}. In the gravitational lens modelling, the source
position is self-constrained by ray-tracing image positions back from
the image plane to the source plane. The positional observables in the
image plane build up a self-relative reference frame. Our lens
modelling is proven to be an independent method to measure the
relative offset without resorting to an absolute celestial reference
frame. Since the offset between the radio and optical core positions
of AGNs may indicate a profound evolutionary scenario of the central
black hole and the dark matter halo in the host galaxy, the
gravitational lensing thus becomes an effective research tool on this
topic.

It needs to be mentioned that in the optimized two-source
configuration, the optimized source light profile becomes flatter,
while the X-galaxy light profile becomes steeper; the
radio-optical offset of the primary source is then lowered to
$\sim$60~mas. Since the secondary source mainly contributes to the
extended emissions along the Einstein ring which blends with the
emissions from the lens and the bright images of the lensed AGN, its
presence does not violate the image number theorem. We can notice from
Fig.~\ref{fig:tromy2} that, one extra central image is blended with
the X-galaxy.

\subsection{Inner power-law slope}

The $\Lambda$CDM halo density profile, as applied to clusters, is
shallower than isothermal near the centre and steeper than isothermal
near the virial radius~\citep{navarro.04.mn}. While down to the
galactic scale, the statistical results from the Sloan \emph{SLACS}
survey~\citep{koopmans.06.apj} shows that the inner regions of lens
galaxies exhibit isothermal density profiles. The optimal lens model
obtained from the 8.4-GHz \emph{VLBI} data modelling has inferred a
steeper-than-isothermal inner slope of the primary lens galaxy, which
is also verified by the optical modelling with tri-band \emph{HST}
data. Its inconsistency with the statistical results implies the lens
galaxy is at a different stage of evolution. It has been found that
the lens system with a companion is more likely to have steeper
density profiles~\citep{auger.08.mn}. From the lens modelling point of
view, in a nearly-isothermal lens system, if a companion is included
in the modelling, the optimization will steepen the inner slope of the
primary lens to compensate for the added extended mass profile by the
companion. This modelling degeneracy can be broken when the extra
lensing effects due to the companion lens are detected. The detection
of the diffuse arc in the lens B1152+199 has put the claim of a
steeper inner slope on a solid foothold. \citet{dobke.07.mn} used
N-body simulation to demonstrate that the steepening of the density
profile is due to the temporary interactions within galaxy groups and
will return to isothermality eventually.

\subsection{Quad, arc or ring}

Though the current modelling restores the optical observation to the
greatest extent, we still notice that there are subtle differences
between the observed diffuse features and the reconstructed
emissions. The source light reconstruction in the image plane does
show a fractured ring. After the convolution with the \emph{PSF}s,
only the diffuse arc remains visible since other parts of emissions
are totally buried in the glows of the two quasar images. As limited
by the sensitivities and resolutions of current observations, there is
no compact core resolved from the diffuse emission, so it is hard to
tell whether the lensing configuration is a quad, arc or
ring. Nevertheless, we can notice that the shape of the fractured
ring does resemble a quad configuration, which verified our naive
speculation from the \emph{HST} $I$-band image. The only difference is
that the model resembles a fold-crossing rather than a cusp-crossing
case. As the arc is composed of two extended images (R2, R3) and other
two images (including R1) (refer to Fig.~\ref{fig:galfit}) are shifted
close the primarily lensed images A and B, due to the presence of the
primary lens.

If the diffuse arc can be resolved into discrete images, it will
definitely add up to the positional observables and help to constrain
the lens model to another level. The high-sensitivity infrared
observation with the upcoming James Webb Space Telescope (\emph{JWST})
can definitely have a deeper insight into this lens system.

\section{Conclusions}

In this paper, we reinvestigated the archived \emph{HST/WFPC2}
$V/I$-band and \emph{HST/NICMOS} $H$-band data as well as the 5-GHz
and 8.4-GHz \emph{VLBI} data of the target lens \emph{CLASS}
B1152+199. Our re-analysis and lens modelling of the archival data,
has shown that improved observations are important for astronomers to
understand this lens system.

The resolved radio sub-components in lensed images give more direct
constraints on the parametric lens model and the image-plane
optimization gives more precise estimations on the model
parameters. Currently, the 8.4-GHz global-\emph{VLBI} observation now
offers the highest resolution of the discrete jet components in the A
and B images. The collimation of the jet components in image B
dismisses the previous debate over the jet
bending~\citep{asadi.20.mn}. Actually, the jet bending can be proved
non-significant with lower resolution data from 5-GHz observation, as
shown in Section~\ref{sec:flip}. Nevertheless, a slight misalignment
within 0.5~mas of a B jet component can still be detected. Meanwhile,
the model fitting of resolved components gives the strongest
up-to-date constraints on the macro mass model. Especially, the radial
mass profile may be inferred from the radially aligned jet
components. So we tested the observables with \emph{SIE} and
\emph{SPLE} models plus external shear both with and without an
\emph{SIS} perturber. The optimal models tell us that the companion
X-galaxy is needed for better optimization and a
steeper-than-isothermal mass profile is thus preferred.

The photometric measurements from tri-band images give a consistent
luminosity ratio between the host galaxy and the X-galaxy, which is
different from the previous study by \citet{rusin.02.mn}. This leads to a
different estimation of the mass ratio between two galaxies, which
affects the mass model considerably. Our attention was brought to the
diffuse emission that was detected in the \emph{HST} $H$-band and
$I$-band images around the X-galaxy. A careful inspection has revealed
that the diffuse arc-shaped emission in the $H$-band image is resolved
into three faint blobs in the $I$-band image. We utilized the shapelet
modelling program {\sc lenstronomy} to investigate the possibility
of the diffuse emission as a secondary lensed feature by the presence
of the X-galaxy. Optical modelling is very useful to reconstruct
the morphology of the observed image. However, as there are light
profiles of the lens and the source to be constrained at the same
time, the expanded parameter space makes its determination of the mass
model less effective. So we imported the optimal lens model from the
radio modelling to the optical modelling as a reference model to keep
the optimizations consistent. The optimization with optical data shows
that a power-law ellipsoid with external shear, plus an isothermal
spheroid companion, can reproduce the optical features observed in
\emph{HST} images.

An under-constrained lens system leads to parameter degeneracy, so
that a subtle observed effect may be explained by many
possibilities. It is risky to draw conclusions without considering all
the components detected in the lens system. The X-galaxy and diffuse
optical emissions were detected from the first discovery of the
B1152+199 lens~\citep{rusin.02.mn}. However, it is until the
sub-components are fully resolved by the 8.4-GHz globally \emph{VLBI}
observation that the X-galaxy can finally find its role in this lens
system. So a high-sensitivity full-track global \emph{VLBI} follow-up
observation is still the most powerful tool to probe the mass
distribution of those radio lenses.

Though it is favoured by current mass modelling that the dubious
diffuse emission around the X-galaxy in \emph{HST} images are lensed
features from the source, the diffuse emission may still play an
important role in further determining the light and mass profiles of
the lens and source when a higher-sensitivity-and-resolution
observation is achieved. Hence future investigations of this specific
lens system with more powerful or technically advanced ground-based or
space-based optical telescopes are needed. As shown in the work of
\citet{lagattuta.12.mn}, the Keck adaptive optics can provide better
constraints than the \emph{HST}. Proved by the mass modelling, the
X-galaxy, as a massive companion satellite of the host galaxy, does
play a consequential role in the overall mass model. This kind of
companion galaxies are not rare in \emph{CLASS} lenses, such
identified samples as B1608+656~\citep{myers.95.apj},
B2108+213~\citep{mckean.05.mn} and B1359+154~\citep{myers.99.aj}. As a
matter of fact, we don't even know the redshift of the
X-galaxy~\citep{rusin.02.mn}, not even mention the redshift of the
diffuse emission. The spectrographs of very faint galaxies
($\sim$30~mag) have been successfully taken by the \emph{HST} slitless
grism and Keck-I Multi-Object Spectrometer For Infra-Red Exploration
(\emph{MOSFIRE})~\citep{schmidt.14.apj}. Even the detection of the
arc-blob emission in another \emph{HST} filter will give a crude
first-order inference on the photometric redshift with the
colours. Yet the redshifts of the primary lensing galaxy and the
X-galaxy have not been confirmed the same until the present.

\section*{Acknowledgements}

MZ is grateful to the anonymous referee for the constructive and
informative suggestions which helped to improve the paper. We also
thank Simon Birrier for his practical insider advice about the {\sc
  lenstronomy} code's behaviours. This work was supported by the
National Science Foundation of China (12173078 and 11773062) and the
West Light Foundation of Chinese Academy of Sciences
(2017-XBQNXZ-A-008). The \emph{VLBA} is an instrument of the National
Radio Astronomy Observatory, a facility of the National Science
Foundation operated by Associated Universities, Inc. The European
\emph{VLBI} Network is a joint facility of independent European,
African, Asian, and North American radio astronomy institutes. The
scientific results from data presented in this publication are derived
from the following EVN project codes: BB133, GJ010, BW084 and
GA036. This research used observations made with the NASA/ESA Hubble
Space Telescope, obtained from the data archive at the Space Telescope
Science Institute. STScI is operated by the Association of
Universities for Research in Astronomy, Inc. under NASA contract NAS
5-26555.


\section*{Data availability}

The data underlying this article are available in the Hubble Legacy
Archive at https://hla.stsci.edu and the NRAO Science Data Archive at
https://archive.nrao.edu.



\bibliographystyle{mnras}
\bibliography{refs}




\appendix
\section{Extra materials}

\subsection{Jet flip-subtraction}\label{sec:flip}

\begin{figure*}
\centering
\includegraphics[width=11cm]{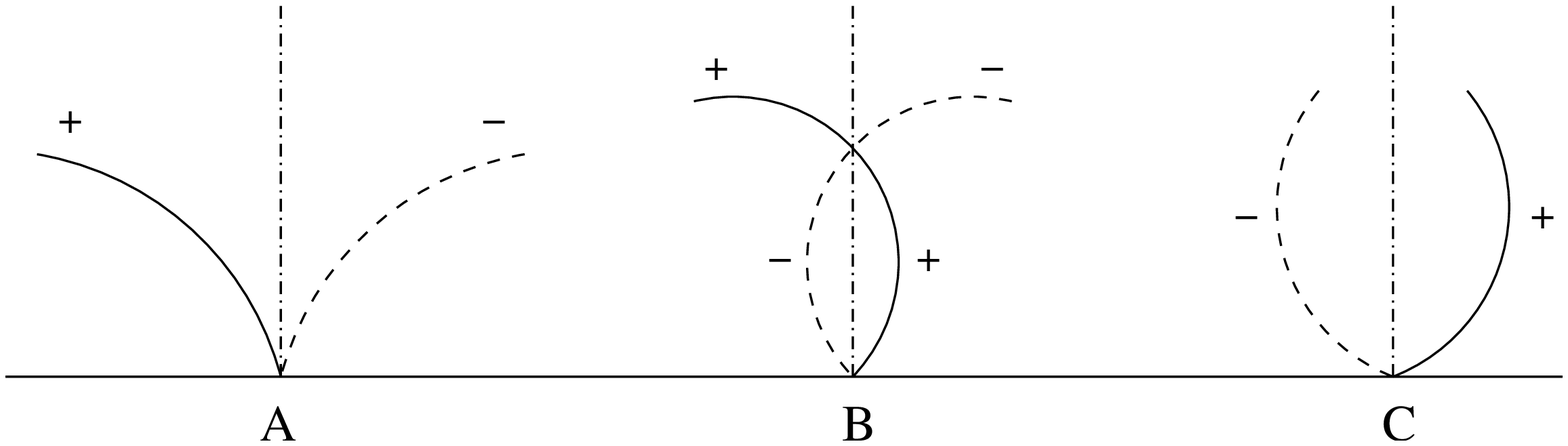}
\caption[Flip-subtraction of a bent jet]{ Flip-subtraction of a bent
  jet. The solid lines are the bent jet and dashed lines are the
  negative mirror image of the jet after flip-subtraction. A, B and C
  show the effects of choosing different flipping axes. A real bent jet
  cannot be cancelled along any flipping axis. }
\label{fig:flipsub}
\end{figure*}

\begin{figure*}
\centering
\includegraphics[width=15cm]{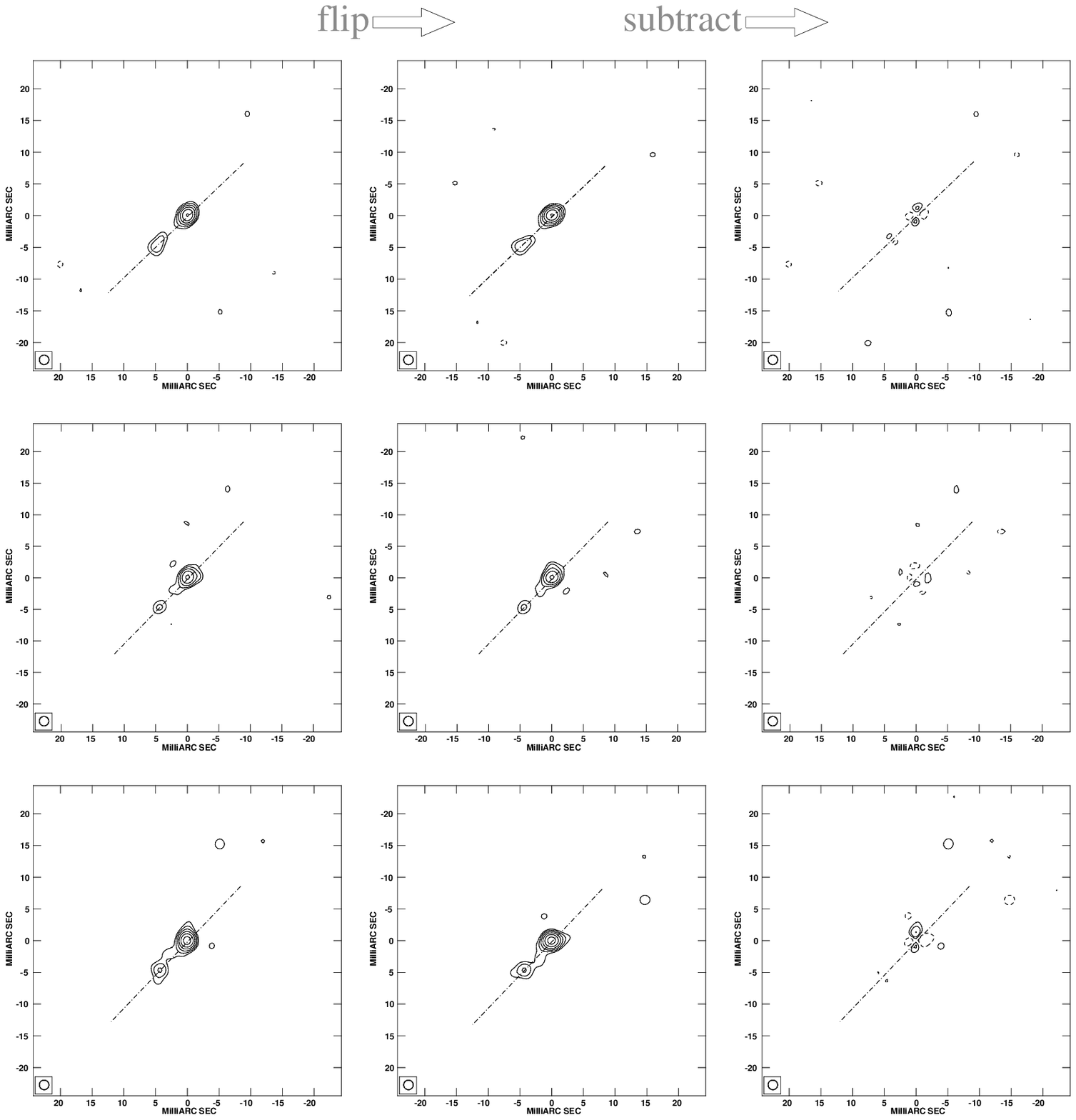}
\caption[Circular beam restored jet structure in image B]{ Three-epoch
  \emph{VLBI} observations and flip-subtractions of image B restored
  with a circular beam of 1.5$\times$1.5~mas$^2$ . From top to bottom,
  the A and B images correspond with observations BB133 (2001), GJ010
  (2003) and BW084 (2005), respectively. The data have been uniformly
  weighted. Contours in the map are plotted in multiples of -1, 1, 2,
  4, 8, 16, 32, 64, 128, 256, 512, 1024, 2048 $\times$ $3\sigma$,
  where $\sigma$ is the local RMS noise ($\sim 100\mu$Jy). The images
  in the middle column are flipped along the jet direction, while the
  image in the right column is the residual image after the
  subtraction of the original image from the flipped image. }
\label{fig:1152bjet}
\end{figure*}

For collimated jet emission, if we flip the image along the
collimation axis and subtract mirrored image from the original one, we
will find the flux along the collimation axis cancelled. However, if a
jet is really bent, its flux cannot be cancelled by flip-subtraction
no matter how the collimation axis is chosen. The systematics of the
residuals with mirrored parities should be detected for a
flip-subtracted bent structure, see Fig.~\ref{fig:flipsub}. So, rather
than using an elliptical beam, we chose a circular beam to restore
image B to check for any unbiased shape perception
(Fig.~\ref{fig:1152bjet}).  The radius of the circular beam is chosen
as the minor axis of the original elliptical beam calculated from $uv$
coverage. Along the major axis direction, the image is thus slightly
super-resolved, which may help to trace unresolved components. The
collimating axis was determined by the peak positions of the image
components. We can see that, when restored with a circular beam, the
jet in image B from \emph{VLBA} and global-\emph{VLBI} observations
looks less bent than that restored with the elliptical beam shown in
Fig.~\ref{fig:5ghz3epo}. Though a slightly noticeable bend can be
seen from the contour connection between the core and the jet
component in the \emph{HSA} observation, after flip-subtraction, the
marginal jet bending is only at a $3\sigma$ detection level.

\subsection{Toy model to produce a bent jet}\label{sec:bender}

\begin{figure*}
   \centering
   \includegraphics[width=5cm]{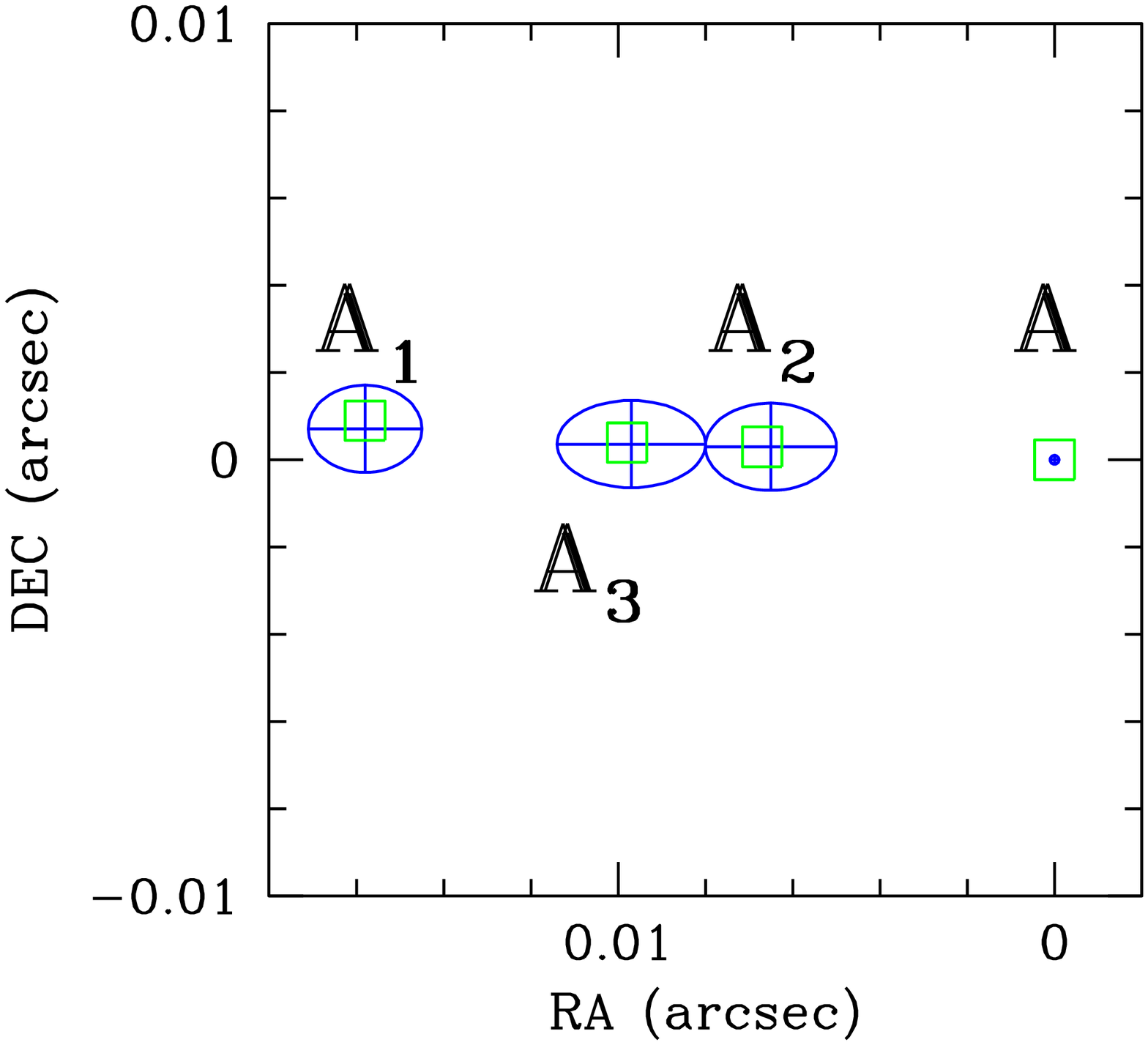}
   \includegraphics[width=5cm]{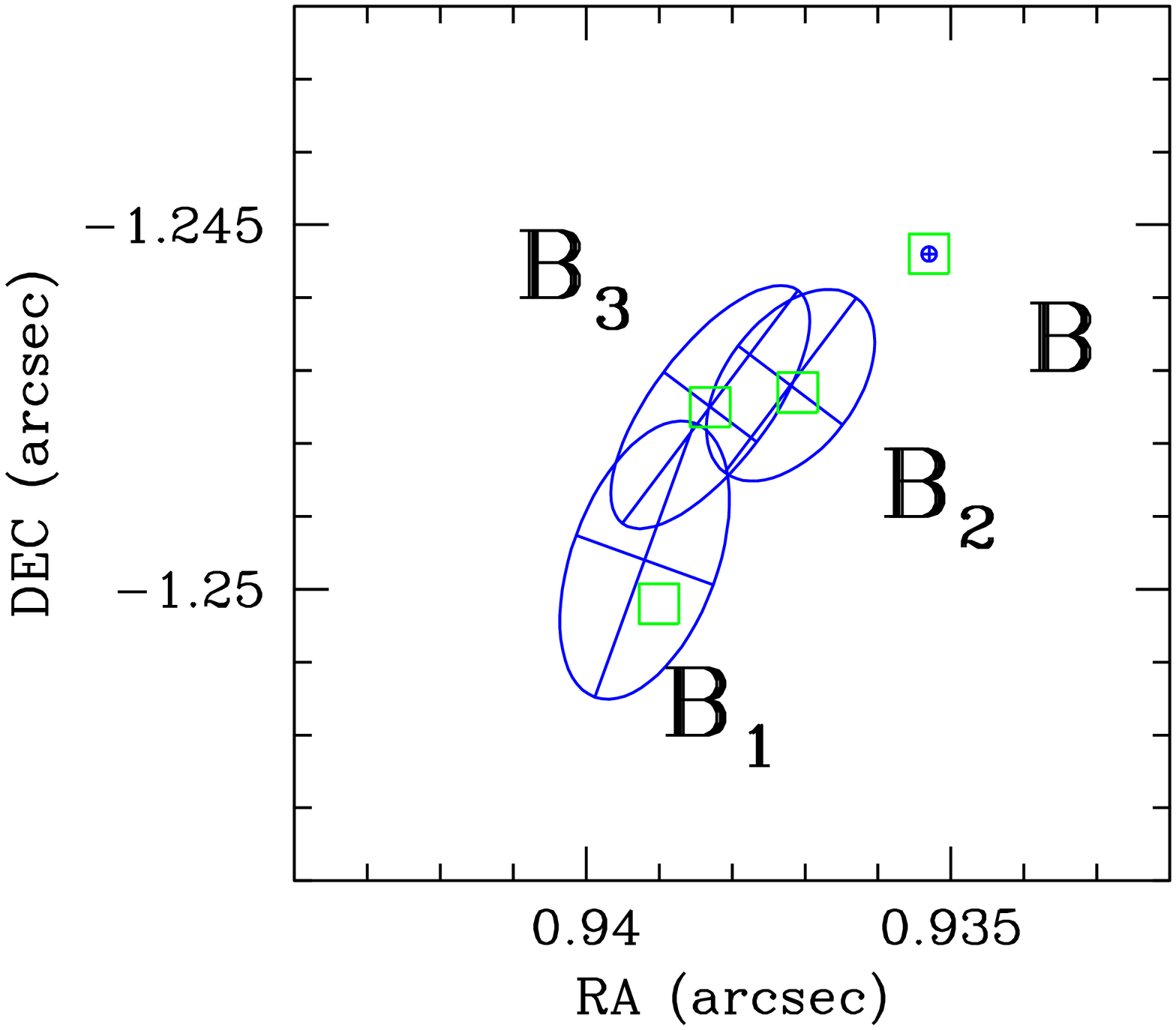}
   \includegraphics[width=5cm]{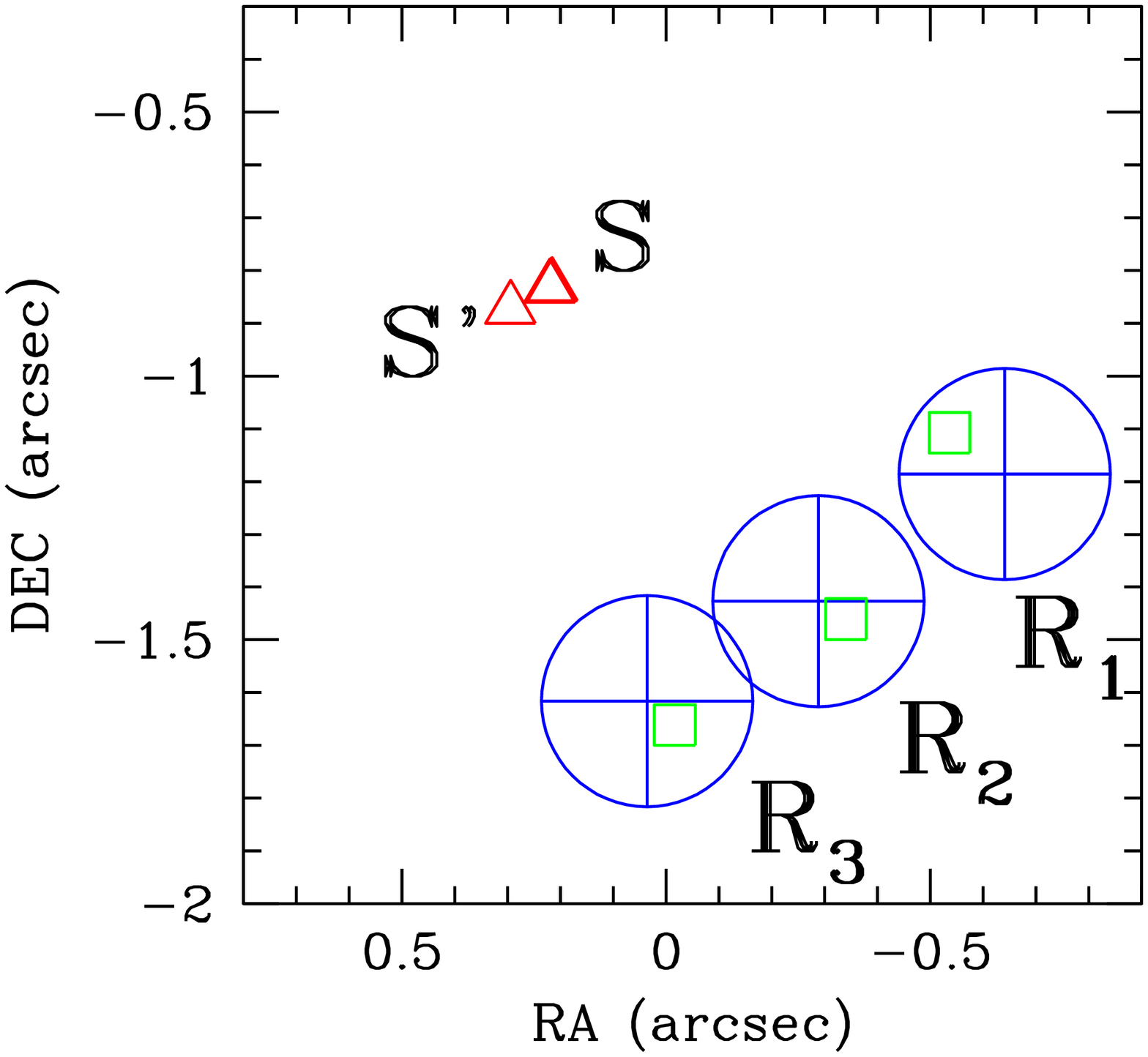}
   \caption[Modelling of AB3+quad]{ Predicted image and source
     positions from the optimal model with multi-component and
     quad-image constraints. The observed image positions are denoted
     as blue error ellipses, while the predicted image positions are
     denoted as green squares. The source positions are denoted as red
     triangles. The S' denotes the secondary source. }
   \label{fig:ab3quad}
\end{figure*}

Here present a toy model to show that with the companion galaxy as a
secondary lens, the macro lens model can produce a bent jet without
any extra substructure local to the jet. The exaggerated error
ellipses are used to give a greater relaxation for the lens model to
move around the image positions. We can see from the middle panel in
Fig.~\ref{fig:ab3quad}, merely an SIE+SIS+$\gamma$ model is adequate
enough to produce a bent jet if the lensing strength of the companion
is emphasized. The diffuse emission as a lensing effect can be treated
as a partial Einstein ring, an arc or a quad-image case.

\subsection{Parameter translation between {\sc gravlens} and {\sc lenstronomy}}\label{sec:trans}

We summarize here the parameter conventions of the two modelling
software packages and the conversion between them to facilitate
cross-check and clarify the use of symbols. The {\sc gravlens} Alpha
model's parameter set is \{ $b'$, ($e$, $\phi_e$), ($\gamma$,
$\phi_\gamma$), $\alpha$ \}, while the {\sc lenstronomy} \emph{SPEMD}
model's parameter set is accordingly \{ $\theta_E$, ($e_1$, $e_2$),
($\gamma_1$, $\gamma_2$), $\upgamma$ \}.
\\

\noindent Convergence:
\begin{align}
\kappa(x,y) &= \frac{1}{2} {b^{2-\alpha} \over {(x^2+y^2/q^2)^{(2-\alpha)/2}}} \\
            &= \frac{1}{2} {b'^{2-\alpha} \over {[(1-\epsilon)x^2+(1+\epsilon)y^2]^{(2-\alpha)/2}}} \\
            &= \frac{3-\upgamma}{2} {\theta_E^{\upgamma-1} \over (q x^2+y^2/q)^{(\upgamma-1)/2}}
\end{align}
\\
\noindent Mass scale (1-dimensional):
\begin{align}
& \frac{b'}{b} = \sqrt\frac{2 q^2}{1+q^2} \\
& \frac{\theta_E}{b'} = \sqrt\frac{1+q^2}{2q} \frac{1}{\alpha^{1/(2-\alpha)}} \\
& \frac{\theta_E}{b} = \frac{\sqrt q}{\alpha^{1/(2-\alpha)}} 
\end{align}
\\
\noindent Ellipticity:
\begin{align}
& q^2 = (1-\epsilon)/(1+\epsilon) \\
& e = 1-q \\
& e_1 = \frac{1-q}{1+q} \cos{2\phi_e}, \quad e_2 = \frac{1-q}{1+q} \sin{2\phi_e}
\end{align}
\\
\noindent External shear:
\begin{equation}
\gamma_1 = \gamma \cos{2\phi_\gamma}, \quad \gamma_2 = \gamma \sin{2\phi_\gamma} 
\end{equation}
\\
\noindent Power-law index:
\begin{equation}
\upgamma = 3 - \alpha
\end{equation}

\subsection{Model case with two background sources}\label{sec:twosrc}

\begin{figure*}
	\includegraphics[width=0.98\textwidth,left]{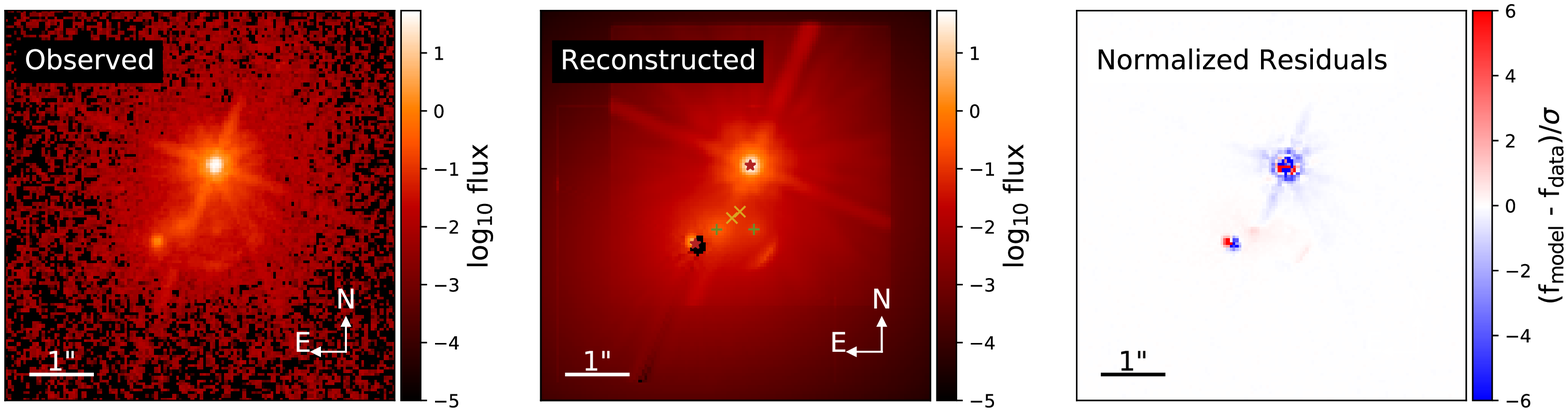}\\
	\includegraphics[width=0.98\textwidth,left]{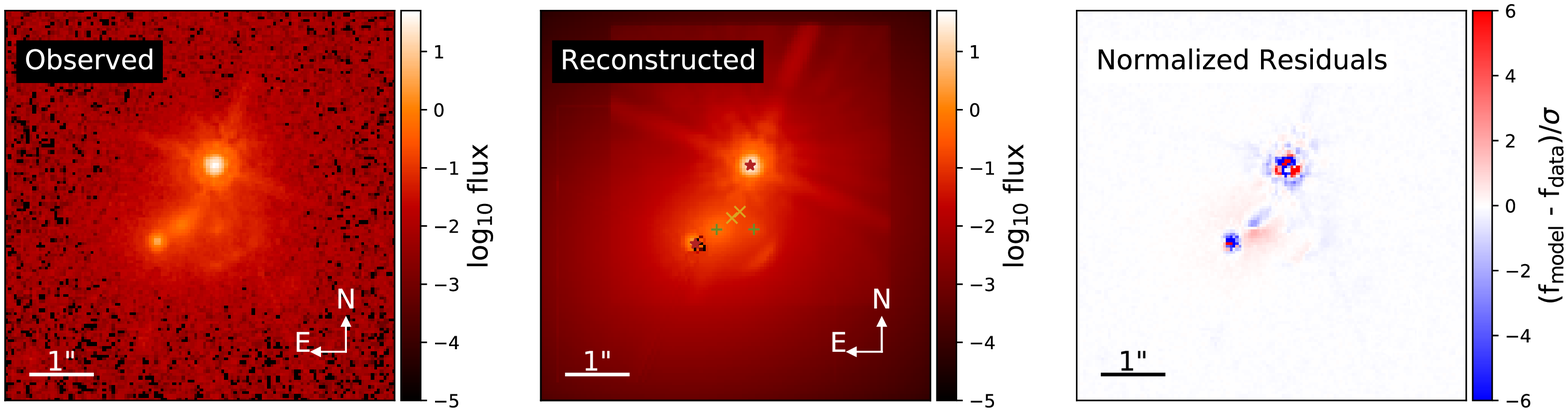}\\
	\includegraphics[width=0.98\textwidth,left]{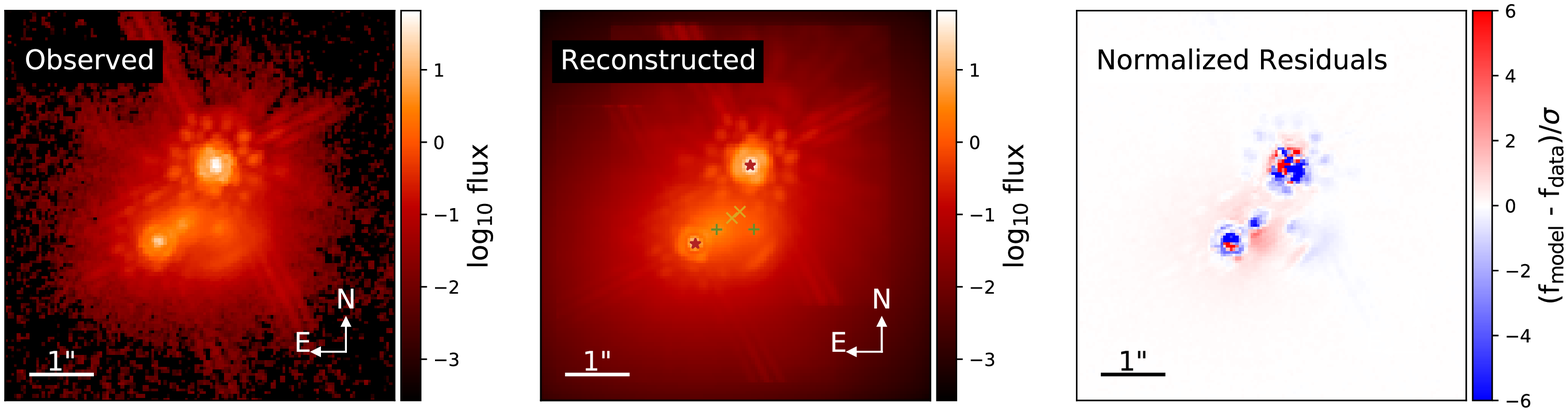}\\
	\includegraphics[width=\textwidth,left]{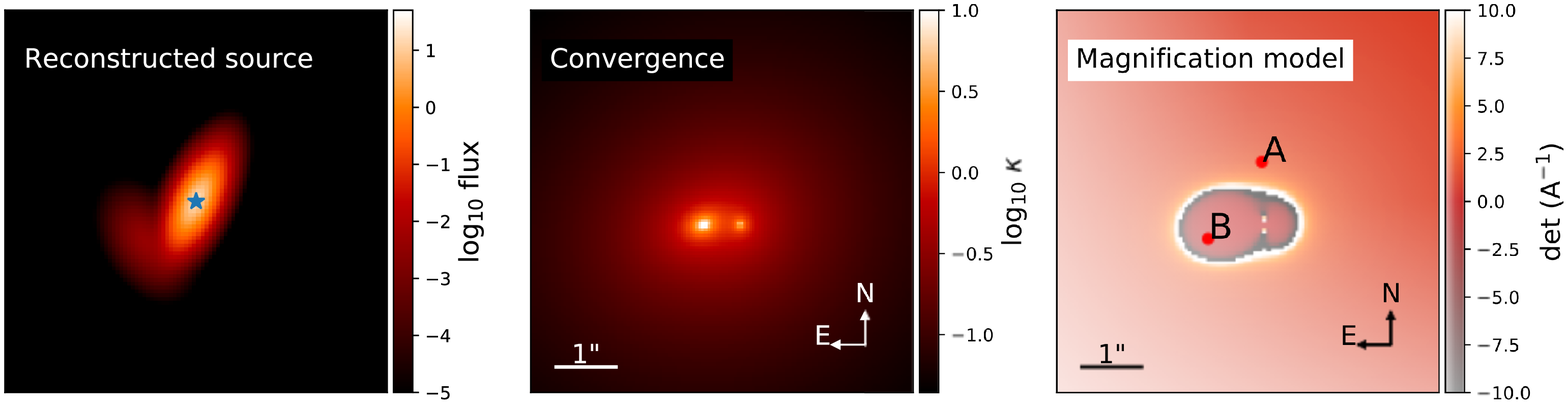} \\
	\includegraphics[width=0.973\textwidth,left]{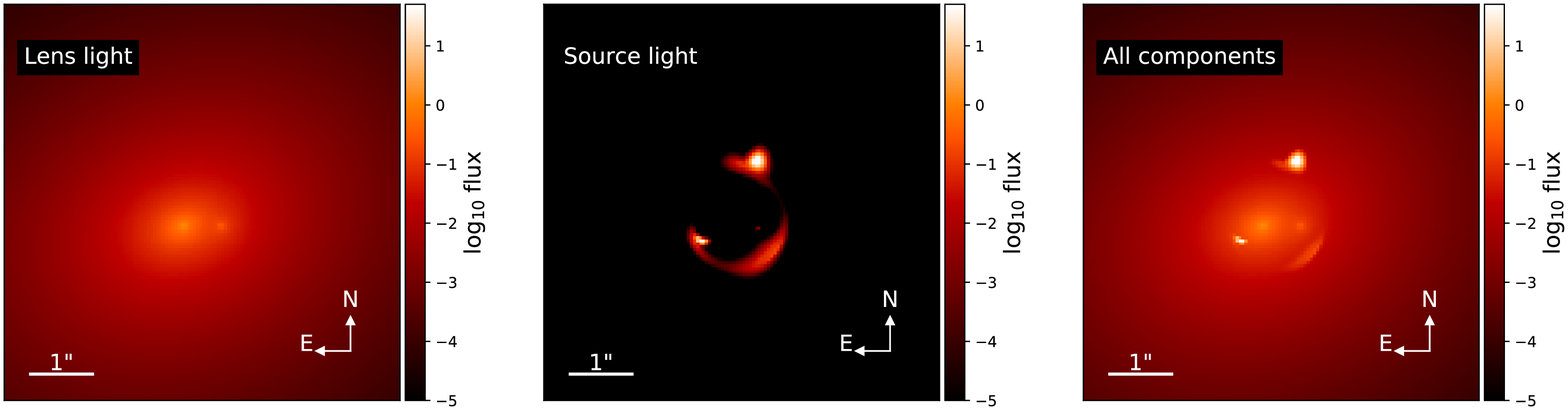}
        \caption[Image model subtraction 2]{ The optimal models for the
          two-source configuration. The top three rows are for the
          \emph{HST} $V$-band, $I$-band and $H$-band data,
          respectively. The bottom two rows show the source, mass and
          light models. In the reconstructed image panels, the red
          stars, green pluses and yellow crosses denote the observed
          image positions, the modelled lens positions and source
          positions respectively. In the reconstructed source panel,
          the blue stars denote the source light centroid positions. }
    \label{fig:twosrc}
\end{figure*}

Here we present an example model with two lensed background sources
which is optimized with tri-band \emph{HST} images (see
Fig~\ref{fig:twosrc}). Since the secondary source also contributes to
forming the Einstein ring, the primary source needs not to be so extended
as shown in Fig.~\ref{fig:tromy1} and \ref{fig:tromy2}. The optimized model parameters are shown in Table~\ref{tab:trompartw}

\begin{table}
   \centering
   \caption[Lenstronomy model parameters 2]{ The optimized parameters of 
     lens mass, lens light and source light profiles with \emph{HST} image 
     constraints for a two-source case. The symbol notations are the same as
     in Table~\ref{tab:trompar}. }
   \label{tab:trompartw}
   \begin{tabular}{ccccccccc} 
      \hline
       Mass   &  $\theta_E$  & $e_1$ & $e_2$  & $x$   & $y$   & $\upgamma$ \\
      \noalign{\vskip 1mm}    
      G   & 0.627 & 0.112 & -0.086 & 0.310 & -0.393 & 2.328 \\
      X   & 0.233 &       &       & -0.302 & -0.391  & \\
           & $\gamma_1$ & $\gamma_2$ \\
      \noalign{\vskip 1mm}
      $\gamma$ & -0.094 & -0.007 \\
      \hline
       Light  & $Amp$ & $R_e$ & $n$ & $e_1$ & $e_2$ & $x$ & $y$ \\
      \noalign{\vskip 1mm} 
      G  &  178.803 & 0.698  & 2.111 & 0.112 & -0.086 &  0.310 & -0.393 \\ 
      X  &  79.702 & 0.279  & 2.161 &       &       & -0.302 & -0.391 \\
      S1  & 1.434e5 & 0.042  & 0.712 & -0.258 & -0.309 & -0.070 & -0.100 \\ 
      S2  & 240.880 & 0.057  & 0.592 & -0.090 &  0.203 &  0.060 & -0.204 \\ 
     \hline
   \end{tabular}
\end{table}


\bsp	
\label{lastpage}
\end{document}